\documentclass[aip,
amsmath,amssymb,
 reprint,
]{revtex4-1}

\usepackage{dcolumn}
\usepackage{bm, color}
\usepackage{hyperref}
\usepackage{natbib}
\usepackage{booktabs,tabulary}
\usepackage{multirow}
\usepackage[export]{adjustbox}
\usepackage{float}
\usepackage[caption=false]{subfig}
\usepackage{array}
\usepackage{cases}
\usepackage{makecell}
\usepackage{physics}
\usepackage{yhmath}

\usepackage{enumitem}

\newcommand{\spD}[1]{\fn{\tilde{\chi}_{_V}}{#1}}

\usepackage{soul}
\setstcolor{red}

\setul{0}{0.3ex}

\newcommand{\R}{\mathbb{R}}

\newcommand{\fn}[2]{\mathinner{#1\mathopen{\left(#2\right)}}}
\newcommand{\vect}[1]{{\bf #1}}
\newcommand{\E}[1]{\left\langle#1\right\rangle}
\newcommand{\Order}[1]{\fn{\mathcal{O}}{#1}}

\newcommand{\Smax}{S_{\max}}
\newcommand{\rmin}{r_{\min}}
\newcommand{\phimax}{\phi_{\max}}
\newcommand{\fail}{p_\mathrm{fail}}

\usepackage[colorinlistoftodos,shadow,textwidth=18mm]{todonotes}

\begin{document}

\preprint{AIP/123-QED}

\title{Ultradense Sphere Packings Derived From Disordered Stealthy Hyperuniform  Ground States}

\author{Jaeuk Kim}
\affiliation{Princeton Materials Institute, Department of Physics, Department of Chemistry, Princeton University, Princeton, NJ 08544, USA}
\author{Salvatore Torquato}
\affiliation{Department of Chemistry, Department of Physics, Princeton Materials Institute, Program in Applied and Computational Mathematics, Princeton University, Princeton, NJ 08544, USA}
\email[]{torquato@princeton.edu}

\date{\today}

\begin{abstract}

Disordered stealthy hyperuniform (SHU) packings are an emerging class of exotic amorphous two-phase materials endowed with novel optical, transport, and mechanical properties.
Such packings of identical spheres have been created from SHU ground-state point patterns via a modified collective-coordinate optimization scheme that includes a soft-core repulsion, besides the standard ``stealthy" pair potential.
To explore maximal ranges of the packing fraction $\phi$, we investigate the distributions of minimum pair distances as well as nearest-neighbor distances of ensembles of SHU point patterns without and with soft-core repulsions in the first three space dimensions as a function of the stealthiness parameter $\chi$ and number of particles $N$ within a hypercubic simulation box under periodic boundary conditions.
Within the disordered regime ($\chi<0.5$), we find that the maximal values of $\phi$, denoted by $\phimax(\chi,d)$, decrease to zero on average as $N$ increases if there are no soft-core repulsions.
By contrast, the inclusion of soft-core repulsions results in very large $\phimax(\chi,d)$ independent of $N$, reaching up to $\phimax(\chi,d)=1.0, 0.86, 0.63$ in the zero-$\chi$ limit and decreasing to $\phimax(\chi,d)=1.0, 0.67, 0.47$ at $\chi=0.45$ for $d=1,2,3$, respectively.
We obtain explicit formulas for $\phimax(\chi,d)$ as functions of $\chi$ and $N$ for a given value of $d$ in both cases with and without soft-core repulsions.
In two and three dimensions, our soft-core SHU ground-state packings for small $\chi$ become configurationally very close to the corresponding jammed hard-particle packings created by fast compression algorithms, as measured by their pair statistics.
As $\chi$ increases beyond $0.20$, the packings form fewer contacts and linear polymer-like chains as $\chi$ tends to $1/2$.
The resulting structure factors $\fn{S}{k}$ and pair correlation functions $\fn{g_2}{r}$ reveal that soft-core repulsions significantly alter the short- and intermediate-range correlations in the SHU ground states.
We show that the degree of large-scale order of the soft-core SHU ground states increases as $\chi$ increases from 0 to 0.45 for $d=2,3$.
We also compute the spectral density $\spD{k}$, which can be used to estimate various physical properties, including electromagnetic properties, fluid permeability, and mean survival time, of SHU two-phase dispersions. 
Our results offer a new route for the discovery of novel disordered hyperuniform two-phase materials with unprecedentedly high density.

\end{abstract}


\maketitle

\section{Introduction}
\label{sec:intro}

Hyperuniform many-particle systems in $d$-dimensional Euclidean space $\R^d$ are characterized by vanishing (normalized) density fluctuations at large length scales, i.e.,
the structure factor $S({\bf k})$ tends to zero in the limit $\mathinner{\abs{\vect{k}}\to0}$.\cite{torquato_local_2003}
This classification includes all perfect crystals and quasicrystals as well as special disordered systems.
Thus, the hyperuniformity concept generalizes the traditional notion of long-range order in many-particle systems to not only include all perfect crystals and perfect quasicrystals, but also the aforementioned exotic amorphous varieties.
Disordered hyperuniform systems are an emerging class of correlated amorphous states of matter of increasing interest due to the fact that they are poised between a crystal
and an isotropic liquid and hence have a ``hidden order" on large length scales that is not apparent at small length scales.\cite{torquato_hyperuniform_2018}
During the last decade, a variety of disordered hyperuniform states have been identified that can be achieved via either equilibrium and nonequilibrium routes as classical or quantum-mechanical states,\cite{hexner_hyperuniformity_2015,jadrich_consequences_2016,chremos_hidden_2018,wang_hyperuniformity_2018,zhuravlyov_comparison_2022,oppenheimer_hyperuniformity_2022,onishi_topological_2024} which are often endowed with novel transport, electromagnetic,
and mechanical properties; see Refs. \onlinecite{torquato_hyperuniform_2018, torquato_extraordinary_2022} and references therein.

We focus on an important subclass of such disordered hyperuniform systems, called disordered stealthy hyperuniform (SHU) system, defined by an {\it ensemble-averaged} structure factor that is exactly zero for a range of wavevectors around the origin up to a positive wavenumber $K$, i.e.,
\begin{equation}
S({\bf k})=0 \quad \mbox{for} \quad 0 \le \abs{\bf k} \le K.
\label{S-stealthy}
\end{equation}
This condition implies that there is no single scattering down to intermediate wavelengths of the order of $2\pi/K$.\cite{uche_constraints_2004, zhang_ground_2015}
It has been shown that disordered SHU configurations are highly degenerate classical ground states 
of particles that interact with certain bounded long-ranged pair potentials,\cite{torquato_ensemble_2015} as detailed later.

Generally, a numerically obtained ground-state configuration depends on the space dimension $d$, number of particles $N$ within the fundamental cell, the {\it stealthiness parameter} $\chi$, initial particle configuration, the shape of the fundamental cell, and particular optimization technique employed.
Here, the stealthiness parameter $\chi$ is a measure 
of the size of this ``exclusion region" in which $S(k)$ is identical to zero (for $k\ge K$) relative to the total of degrees of freedom $d(N-1)$ in the system; see Sec. \ref{sec:algorithm} for a precise definition.
One important property of the constructed ground states across dimensions ($d=2,3$) is that they are, counterintuitively, disordered (statistically isotropic without long-range order), stealthy, hyperuniform, and highly degenerate for $\chi<1/2$, but there is a phase transition to crystal structures for $\chi>1/2$.\cite{uche_constraints_2004, batten_classical_2008,torquato_ensemble_2015} 
Such a phase transition can occur because the dimensionality of the SHU configuration space per particle, $d_c$, decreases linearly with $\chi$ as $d_c=d(1-2\chi)$ in the thermodynamic limit.\cite{torquato_ensemble_2015}

Mapping such particle configurations to networks enabled the discovery of the first disordered
dielectric networks to have large isotropic photonic band gaps comparable in
size to photonic crystals.\cite{florescu_designer_2009}
This computational study led to the design and fabrication of disordered cellular solids with the predicted
photonic bandgap characteristics for the microwave regime, enabling unprecedented
free-form waveguide geometries that are robust to defects
not possible with crystalline structures.\cite{aeby_fabrication_2022}
Subsequently, SHU network solids were shown to possess
novel wave propagation, transport, and elasticity characteristics,
including low-density cellular materials with nearly optimal effective
electrical conductivities and elastic moduli,\cite{torquato_multifunctional_2018} photonic polarizer,\cite{zhou_ultrabroadband_2020} photonic
bandgaps that potentially persist in the infinite-volume limit,\cite{klatt_wave_2022} and high quality-factor optical cavity.\cite{granchi_nearfield_2022}

A systematic study of the transport, geometrical and topological
properties of 2D and 3D disordered SHU packings was
carried out in Ref. \onlinecite{zhang_transport_2016}. It was shown there that one can convert SHU ground-state point configurations into corresponding two-dimensional (2D) and three-dimensional (3D) SHU packings of ``identical spheres'' across a range of packing fractions by decorating the points with nonoverlapping spheres.
Subsequently, a variety of investigations demonstrated the novel physical properties of SHU packings, including
wave transparency,\cite{kim_multifunctional_2020,romero-garcia_stealth_2019,kim_effective_2023,froufe-perez_bandgap_2023,alhaitz_experimental_2023} maximal absorption,\cite{bigourdan_enhanced_2019} tunable localization and
diffusive regimes,\cite{froufe-perez_band_2017, sgrignuoli_subdiffusive_2022} enhanced solar cell efficiency,\cite{tavakoli_65_2022,merkel_stealthy_2023} phononic
properties,\cite{gkantzounis_hyperuniform_2017,gkantzounis_freeform_2017,romero-garcia_wave_2021,rohfritsch_impact_2020,cheron_experimental_2022,zhuang_vibrational_2024} directional wave extraction,\cite{castro-lopez_reciprocal_2017,gorsky_engineered_2019} designs of terahertz quantum cascade laser,\cite{deglinnocenti_hyperuniform_2016} Luneberg lenses with reduced
backscattering,\cite{zhang_experimental_2019} extraordinary phased arrays,\cite{christogeorgos_extraordinary_2021,tang_hyperuniform_2023} metamaterials suppressing backscattering,\cite{zhang_hyperuniform_2021,kuznetsova_stealth_2021}
and optimal sampling array of 3D ultrasound imaging.\cite{tamraoui_hyperuniform_2023}

Why do disordered SHU systems have superior physical properties among isotropic amorphous states of matter? 
Unlike all other disordered systems, they possess characteristics of crystals from intermediate to infinitely large scales while being isotropic.\cite{torquato_hyperuniform_2018} 
Specifically, in disordered SHU systems, there can be no single scattering from intermediate to infinite wavelengths [see Eq. \eqref{S-stealthy}] and ``holes" (i.e., spherical regions empty of particle centers) are strictly bounded with maximal hole size on the order of the mean nearest neighbor distance,\cite{zhang_can_2017,ghosh_generalized_2018} which are also properties of crystals.
This is to be contrasted with typical disordered many-particle systems in $\R^d$ in which the probability of finding a hole of arbitrarily large size in the thermodynamic limit is non-vanishing.\cite{torquato_hyperuniform_2018,zhang_can_2017}

We have previously used the {\it collective-coordinate} optimization scheme to generate numerically such SHU ground state with isotropic potentials
in one-, two- and three-dimensional Euclidean spaces\cite{uche_constraints_2004,batten_classical_2008}
as well as with anisotropic potentials.\cite{martis_exotic_2013, torquato_hyperuniformity_2016}
In the simplest setting of isotropic interactions, we previously examined pairwise additive potentials 
$v({\bf r})$ that are bounded and integrable such that their
Fourier transforms ${\tilde v}({\bf k})$ exist; see Ref. \onlinecite{torquato_ensemble_2015} and references therein.
If $N$ identical point particles reside in a fundamental region $\mathfrak{F}$ of volume $v_\mathfrak{F}$ in $\R^d$ 
at positions ${\bf r}^N \equiv {\bf r}_1, \ldots, {\bf r}_N$ under periodic
boundary conditions, the total potential energy can be expressed
in terms of ${\tilde v}({\bf k})$ as follows:
\begin{equation}
\Phi({\bf r}^N) = \frac{\rho}{2} \sum_{\bf k\neq 0} {\tilde v}({\bf k}) \fn{\mathcal{S}}{\vect{k}},
\label{eq:pot}
\end{equation}
where $\vect{k}$ in the sum refers to the reciprocal lattice vectors of $\mathfrak{F}$, $\rho\equiv N/v_\mathcal{F}$ is the number density, and $\fn{\mathcal{S}}{\vect{k}} = |\sum_{j=1}^{N} \exp(-i{\bf k \cdot r}_j)|^2/N$ is the structure factor for a single configuration.
\footnote{The total energy expression (\ref{eq:pot}) actually also contains the sum $\sum_{\bf k} {\tilde v}({\bf k})$,
but this is a structure-independent constant that is set to zero here without
loss of generality.} 
Note that $\fn{S}{k} \equiv \E{\fn{\mathcal{S}}{\vect{k}}}$ is the ensemble average of $\fn{\mathcal{S}}{\vect{k}}$, given in Eq. \eqref{eq:pot}, in the thermodynamic limit.
The crucial idea is that if ${\tilde v}({\bf k})$ is defined to be bounded and positive with support in the radial interval $0 \le |{\bf k}| \le K$ and if the initial positions of the particles are displaced so 
that the structure factor $\fn{\mathcal{S}}{\vect{k}}$ is driven to its minimum value of zero for all wavevectors where ${\tilde v}({\bf k})$ has support, i.e.,
as specified by condition (\ref{S-stealthy}),
then it is clear from relation \eqref{eq:pot}
that the system must be at its ground state or global energy minimum.\footnote{More generally, stealthy configurations are ground states that
 minimize $S(k)$ to be zero at other sets of wave vectors, not
necessarily in a connected set around the origin and hence nonhyperuniform, specific examples of which were investigated
in Ref. \onlinecite{batten_classical_2008}. 
The collective-coordinate optimization technique has been used to target more general forms
of the structure factor for a prescribed set of wave vectors such that $S({\bf k})$
is not minimized to be zero in this set (e.g., power-law forms and positive constants).\cite{batten_classical_2008, zhang_perfect_2016} 
There, the resulting configurations are the ground states of interacting many-particle systems with 2-,3- and 4-body interactions.}
Thus, one can quantify the accuracy of the resulting SHU point patterns in terms of the {\it distance to stealthiness} $\Smax$:\cite{morse_generating_2023}
\begin{align} \label{eq:Smax}
 S_{\max}\equiv \max_{0<k\leq K} \fn{S}{k}.
\end{align}
Various optimization techniques have been employed
to find the globally energy-minimizing configurations within an exceedingly small numerical 
tolerance from zero within the exclusion region of $\Smax \sim 10^{-20}$;\cite{uche_constraints_2004, batten_classical_2008, martis_exotic_2013, shih_fast_2024} see also a very recent study that provides even greater precision.\cite{morse_generating_2023}\footnote{Reference \citenum{morse_generating_2023} reports techniques that reduce the deviations from zero within the exclusion region (distance to exact stealthiness) by a factor of approximately $10^{30}$ compared to the previous ones described in Refs. \cite{uche_constraints_2004,batten_classical_2008,martis_exotic_2013}, even for much larger system sizes.} 


\begin{figure}[h]
 \includegraphics[width=0.45\textwidth]{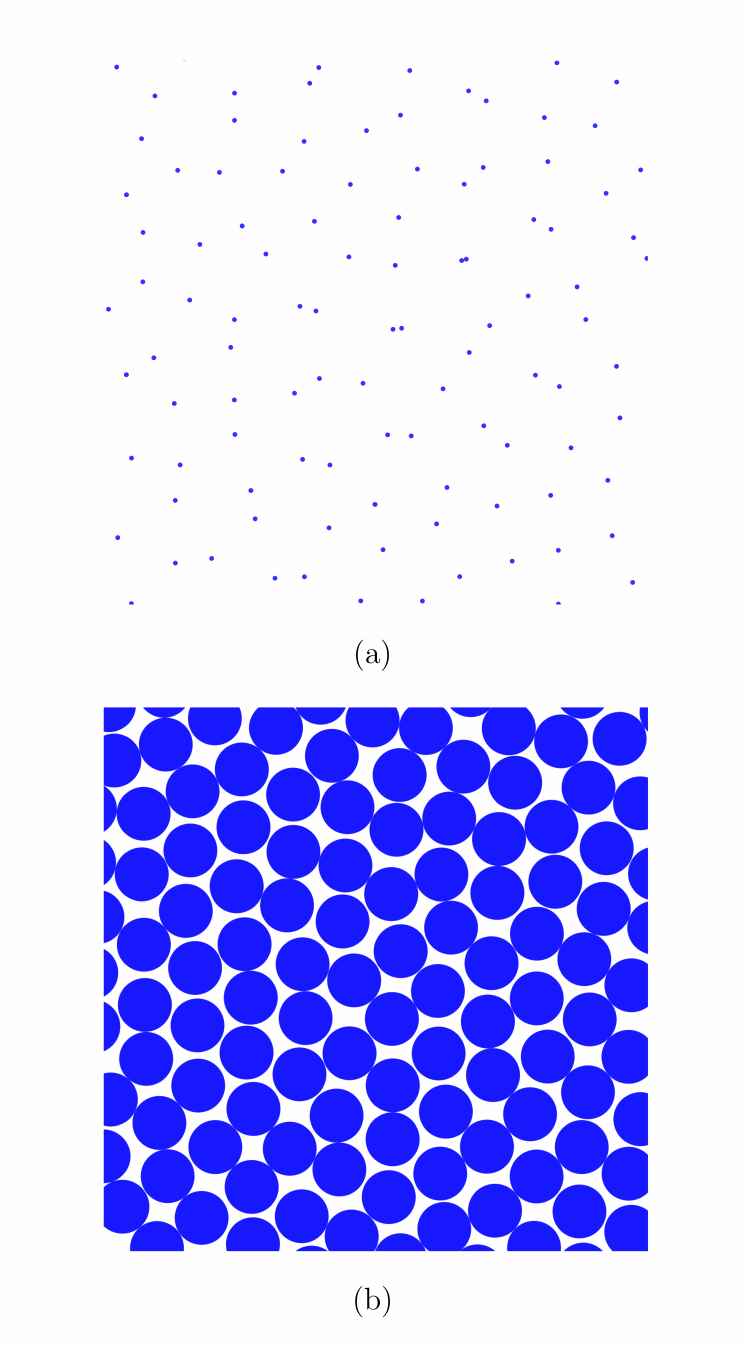}
 \caption{
 Portions of two representative images of 2D SHU packings with their maximal packing fractions that are derived from 2D SHU ground-state point patterns with $\chi=0.30$ and $N=400$ associated with the standard potential \eqref{eq:pot}, shown in the top panel (a), and with the modified potential \eqref{eq:Phi}, shown in the bottom panel (b).
 The packing with soft-core repulsion shown in (b) is ultradense compared to that without it shown in (a).
 Specifically, the maximal packing fractions $\phimax(\chi,d)$ of (a) and (b) are $6.05\times 10^{-3}$ and $0.77$, respectively.
 In each panel, particles are decorated with blue nonoverlapping disks of identical radii.
\label{fig:schem}}
\end{figure}

A primary aim of this paper is the creation of SHU sphere packings that are substantially denser than those that can be achieved via the standard collective-coordinate optimization protocols using the potential \eqref{eq:pot}.
We carry out this task by using a modified collective-coordinate optimization scheme that includes short-range soft-core repulsions, in addition to the standard `stealthy' potential \eqref{eq:pot}.
This modified scheme biases the search for point patterns within the SHU ground-state manifold by penalizing configurations containing close particle pairs.

While the standard collective-coordinate optimization protocols using the potential \eqref{eq:pot} can yield
SHU sphere packings with high accuracy $S_{\max} \lesssim \Order{10^{-20}}$, the range of packing fractions is notably limited and depends on the space dimension $d$, $\chi$ value, and $N$.\cite{zhang_transport_2016, zhang_can_2017, middlemas_nearestneighbor_2020}
Specifically, for the one-dimensional (1D) SHU packings without soft-core repulsions at $\chi < 1/2$, it has been barely possible to obtain $\phi > 0.05$ even for the relatively small system size $N=100$.\cite{zhang_can_2017, middlemas_nearestneighbor_2020}
In two and three dimensions ($d=2,3$), SHU packings barely obtain $\phi > 0.05$ for $\chi < 0.3$, but the range of achievable packing fraction increases as $\chi$ increases from $0.3$ to $0.5$: $\phi < 0.5$ for $d=2$ and $\phi < 0.25$ for
$d=3$.\cite{zhang_transport_2016,zhang_can_2017}
Thus, the SHU packings generated without soft-core repulsions were mainly considered for the 2D and 3D cases in a narrow range of packing fractions (e.g., $\phi \leq 0.15$).\cite{zhang_transport_2016, froufe-perez_band_2017, froufe-perez_bandgap_2023, alhaitz_experimental_2023}
Some previous attempts to increase $\phi$ of SHU packings for $d=1$ (Ref. \onlinecite{romero-garcia_wave_2021}) and $d=2$ (Refs. \onlinecite{gkantzounis_hyperuniform_2017, rohfritsch_impact_2020, merkel_stealthy_2023}) resulted in very small system sizes [i.e., $N=\Order{10}$] or high values of $S_{\max}$ [i.e., $S_{\max} \sim \Order{10^{-5}}$], the latter of which are not truly stealthy hyperuniform.
These packings are not suited for studying the novel properties of SHU packings in the thermodynamic limit because both large sizes and ultrasmall $S_{\max}$ are imperative for this purpose.\cite{klatt_wave_2022, morse_generating_2023}

To overcome such restrictions on the range of the packing fraction $\phi$ for the SHU sphere packings from the standard collective-coordinate method, we previously modified the collective-coordinate optimization technique\cite{kim_multifunctional_2020, kim_effective_2023, kim_theoretical_2024} to include an additional short-ranged soft-core repulsive potential that guarantees a sufficiently large minimum pair distance or, equivalently, larger $\phi$ than the previous method. 
We had previously employed this revised collective-coordinate procedure to generate SHU packings for $d=1$ (Refs. \onlinecite{kim_effective_2023, kim_theoretical_2024}) and for $d=2,3$ (Refs. \onlinecite{kim_multifunctional_2020, kim_theoretical_2024}) with relatively small packing fractions, i.e., $\phi \leq 0.25$.
However, little is known about the range of packing fractions attainable by the soft-core SHU point patterns.

It is of great interest, therefore, to understand the range of achievable packing fractions of SHU ground-state packings with and without the soft-core repulsion as a function of $\chi$, system size $N$, and the radius of the short-range soft-core potential $\sigma$ as well as to quantify their structural characteristics. 
For these purposes, we carry out the following analyses in the first three space dimensions ($d=1,2,3$).
We investigate the minimum-distance distributions $\fn{P}{\rmin;N}$ and the nearest-neighbor distributions $\fn{H_P}{r;N}$ (see Sec. \ref{sec:local} for definitions) for SHU point patterns with and without soft-core repulsions as a function of the parameters $d$, $\chi$, $N$, and $\sigma$.
These analyses determine whether those point patterns can be mapped to SHU packings with a positive packing fraction $\phi$ in the thermodynamic limit, i.e., $N\to \infty$.
When SHU packings are attainable, we determine their maximal values of $\phi$, denoted by $\phimax(\chi,d)$, as a function of $\chi$ at given values of $d$ and $N$.
We also study the pair statistics, including structure factor $\fn{S}{k}$ and pair correlation function $\fn{g_2}{r}$, of these SHU point patterns across length scales as well as the spectral density $\spD{k}$ of the corresponding SHU packings. 

For the SHU point patterns without soft-core repulsions, analysis of their local structural characteristics shows that for a given value of $\chi<0.5$, {\it minimum distance} $\rmin$ of all particle pairs decreases to zero as the system size $N$ increases.
Thus, for the SHU packings with $\chi < 0.5$, maximal achievable packing fractions $\phimax(\chi,d)$ decrease with $N$, and they eventually become zero in the thermodynamic limit (i.e., $N\to \infty$).
We numerically obtain upper bounds on $\phimax(\chi,d)$ as a function of $N$ and the associated probability that a ground-state point pattern cannot achieve $\phimax(\chi,d)$ for $d=2,3$.
By contrast, for SHU packings derived from point patterns with soft-core repulsions, we find that $\phimax(\chi,d)$ is given by $1.00,0.86,0.63$ in the zero-$\chi$ limit and then decreases with $\chi$ to 1.00, 0.67, 0.47 at $\chi=0.45$ for $d=1,2,3$, respectively, insensitive to $N$.

To give a visual sense of the vivid
differences in the maximal packing fractions without and with the soft-core repulsion, we show two respective 2D configurations for $\chi=0.3$ and $N=400$ in Fig. \ref{fig:schem}.
Here, it is seen that the packing with the soft-core repulsion is ultradense, achieving $\phimax(\chi,d)=0.77$, as compared to that without the soft-core repulsion, i.e., $\phimax(\chi,d)=6.05\times10^{-3}$.

When $d=1$, these packings remain integer-lattice packings for all values of $\chi$ we investigated.
We show that for very small $\chi$, the 2D and 3D ultradense SHU packings are configurationally very close to the corresponding jammed hard-particle packings obtained via fast compression algorithms, i.e., polycrystalline-like disk packings and maximally random jammed (MRJ) sphere packings,\cite{torquato_is_2000,donev_unexpected_2005, maher_hyperuniformity_2023} respectively.
These observations imply when $d_c$ (a measure of the cardinality of the infinitely degenerate SHU ground-state manifold set) is maximized (i.e., $d_c\to d$ as $\chi$ tends to zero), this manifold contains such nonequilibrium jammed states.
(In the 3D case, these results are expanded upon in Ref. \onlinecite{torquato_existence_2025}.)
As $\chi$ increases beyond $0.20$, they start to form fewer contacts and linear polymer-like chains with a decreasing mean chain length as $\chi$ tends to $1/2$.
We also obtain approximate formulas for $\phimax(\chi,d)$ for the attainable SHU packings as functions of $\chi$, $N$, $\sigma$, and the number of configurations $n_c$ for a given value of $d$.
We also show that the degree of large-scale order of those SHU packings increases with $\chi$ from 0 to 0.45 for $d=2,3$ by measuring the hyperuniformity order metric $\overline{\Lambda}$.
The evaluation of $\fn{S}{k}$ and $\fn{g_2}{r}$ reveals that the soft-core SHU point patterns can be converted to SHU packings with a wide range of $\phi$ by effectively altering the short- and intermediate-range correlations without sacrificing the accuracy of the long-range stealthy hyperuniform correlations.
We also compute in Appendix \ref{sec:chi} the spectral density $\spD{k}$ of the SHU packings, which can be used to estimate various effective properties of such two-phase dispersions, including the effective dynamic dielectric constant,\cite{kim_effective_2023,kim_theoretical_2024} fluid permeability, and mean survival time.\cite{torquato_predicting_2020}

The rest of the paper is organized as follows.
In Sec. \ref{sec:defs}, we provide precise definitions of the statistical descriptors and the stealthy potential.
In Sec. \ref{sec:algorithm}, we describe simulation details of the standard and modified collective-coordinate optimization schemes. 
In Sec. \ref{sec:local}, we quantify two types of the local structural characteristics of the SHU point patterns. 
Here, we also provide approximation formulas of $\phimax(\chi,d)$ for the attainable SHU packings as functions of $\chi$ and $N$ for $d=1,2,3$; see Sec. \ref{sec:max}.
The pair statistics of SHU point patterns are reported in Sec. \ref{sec:2pt}. Section \ref{sec:conclusion} provides concluding remarks.

\section{Backgrounds and Definitions}
\label{sec:defs}

\subsection{Correlation Functions}
\label{sec:correlation}

A point process in $d$-dimensional Euclidean space $\R^d$ is a spatial distribution of infinitely many particles $\vect{r}_1,~\vect{r}_2,~ \cdots$, which can be described by a microscopic density function $\fn{n}{\vect{x}}$:\cite{torquato_new_2006}
\begin{equation}\label{imp-eq:micro density}
\fn{n}{\vect{x}} = \sum_{i=1}^\infty \fn{\delta}{\vect{x}-\vect{r}_i},
\end{equation}
where $\delta(\vect{x})$ denotes the Dirac delta function in $\R^d$.
The $n$-particle density correlation function $\fn{\rho_n}{\vect{x}^n}$ is defined by
$\fn{\rho_n}{\vect{x}^n} \equiv \E{\fn{n}{\vect{x}_1}\fn{n}{\vect{x}_2} \cdots \fn{n}{\vect{x}_n}}$,
where $\vect{x}^n$ is a shorthand notation of $n$ position vectors $\vect{x}_1,\vect{x}_2,\cdots,\vect{x}_n$, and $\E{\cdot}$ represents an ensemble average.
This function is proportional to the probability density associated with finding $n$ different particles at $\vect{x}^n$.
For a statistically homogeneous point process at a given number density $\rho$, the $n$-particle correlation function depends on relative positions of particles, i.e., $\fn{\rho_n}{\vect{x}^n} = \rho^n \fn{g_n}{\vect{x}_{21}, \cdots, \vect{x}_{n1}}$ with $\vect{x}_{ij} \equiv \vect{x}_j -\vect{x}_i$ for $1\leq i\neq j \leq n$ and $\fn{\rho_1}{\vect{x}} = \rho$. 
The \textit{pair correlation function} $\fn{g_2}{\vect{r}}$ is of special importance in this paper.
For a point process without long-range order, $\fn{g_2}{\vect{r}} \to 1$ as $\abs{\vect{r}}\to \infty$.
If the system is also statistically isotropic, then $\fn{g_2}{\vect{r}}$ is a radial function $\fn{g_2}{r}$, where $r=\abs{\vect{r}}$.

The ensemble-averaged structure factor in the thermodynamic limit is also related to the Fourier transform of the {\it total correlation function} $\fn{h}{\vect{r}} \equiv \fn{g_2}{\vect{r}}-1$, denoted by $\fn{\tilde{h}}{\vect{k}}$, as follows:
\begin{align} \label{eq:Sk}
 \fn{S}{\vect{k}} = 1+\rho \fn{\tilde{h}}{\vect{k}}.
\end{align}
In the absence of long-range order, $\fn{S}{\vect{k}}\to 1$ as $\abs{\vect{k}}\to\infty$.
If the system is statistically isotropic, $\fn{S}{\vect{k}}$ becomes a radial function of wavenumber $k$, i.e., $\fn{S}{k}$.

\subsection{Local Structural Characteristics}
\label{sec:def-local}

We also define two types of local structural characteristics, the minimum-distance distribution $\fn{P}{\rmin;N}$ and nearest-neighbor distribution $\fn{H_P}{r;N}$, of point patterns.
For a point pattern of $N$ particles, a minimum distance $\rmin$ is defined as the minimum of all pair distances, implying that all points are separated by at least $\rmin$.
Thus, the minimum-distance distribution is defined as
\begin{align}
 \fn{P}{\rmin; N} \dd{r} &\equiv \text{probability that the minimum distance} 
 \nonumber \\
&\text{of a point pattern lies}
 \nonumber \\
&\text{between $\rmin$ and $\rmin+\dd{r}$}.\label{eq:min-dist-P}
\end{align}
Similarly, the nearest-neighbor distribution function of point patterns is defined \cite{torquato_new_2006} as
\begin{align}
 \fn{H_P}{r;N}\dd{r} &\equiv \text{probability that the nearest point} 
 \nonumber \\
&\text{to a selected point of the point pattern lies}
 \nonumber \\
&\text{between $r$ and $r+\dd{r}$}.\label{eq:Hp}
\end{align}

For $N$ particles, the minimum-distance distribution $P(r_{\min}; N)$ can be connected to the nearest-neighbor distribution $\fn{H_P}{r;N}$.
Denoting the nearest-neighbor distance of particle $i$ by $r^{(i)}$, we note that $\fn{P}{\rmin;N}$ is associated with an {\it extreme event} (Refs. \onlinecite{salvadori_extremes_2007, gomes_extreme_2015}) of finding the minimum distance $\rmin$ of $N$ values, $r^{(1)},\ldots,r^{(N)}$.
Thus, we can express the complementary cumulative distribution function of $\fn{P}{\rmin; N}$ as 
 \begin{align}
&1 - \int_0^\delta \fn{P}{r_{\min}; N} \dd{r_{\min}}
 \nonumber \\
 =~& \text{Prob. that a $N$-particle configuration}\nonumber \\
&\qquad \text{ has $\rmin > \delta$ or $\phimax \geq \rho v_1(\delta/2)$.} 
 \nonumber \\
 =~& \text{Prob. for finding a $N$-particle configuration}
 \nonumber \\
&\qquad \text{such that $r^{(i)} > \delta$ for all $i=1,\ldots,N$.}
 \nonumber \\ 
 \approx~& [\text{Prob. for finding $r^{(i)} > \delta$}]^{\gamma N} 
\label{eq:step1} \\
 =~& \qty[1 - \int_0^{\delta} \fn{H_P}{r;N} \dd{r}]^{\gamma N}
\nonumber \\
\approx~& \qty[1 - \int_0^{\delta} \fn{H_P}{r;\infty} \dd{r}]^{\gamma N}
\label{eq:step2},
 \end{align}
 where we have assumed that $r^{(i)}$ are independent and identical random variables following $\fn{H_P}{r;N}$ in Eq. \eqref{eq:step1}, and $\fn{H_P}{r;N}$ is independent of $N$ in Eq. \eqref{eq:step2}.
Here, $\gamma=1/2$ is an empirical correction factor to account for the cases of $r^{(i)}=r^{(j)}$ occurring because of the close proximity of two distinct particles $i$ and $j$.
We numerically demonstrate the accuracy of Eq. \eqref{eq:step2} in the \href{...}{supplementary material}.

\subsection{Local Number Variance, Hyperuniformity, and Hyperuniformity Order Metric}
\label{sec:order}

Given a statistically homogeneous point process in $\R^d$ at number density $\rho$, consider sampling the number of points $\fn{N}{R}$ within a $d$-dimensional spherical window of radius $R$.
The local number variance, defined as $\fn{\sigma_N^2}{R}\equiv \E{\fn{N^2}{R}} - {\E{\fn{N}{R}}}^2$, can be obtained from pair correlation function $\fn{g_2}{\vect{r}}$ in direct space or structure factor $\fn{S}{\vect{k}}$ in the Fourier space:\cite{torquato_local_2003}
\begin{align}
\sigma_N^2(R) =& \rho v_1(R) \left[ 1 + \rho \int_{\mathbb{R}^d} h({\bf r}) \alpha_2(r;R) \, d{\bf r} 
\right], \label{eq:nv-D}\\
=& \rho v_1(R)\Big[\frac{1}{(2\pi)^d} \int_{\mathbb{R}^d} S({\bf k}) 
{\tilde \alpha}_2(k;R) d{\bf k}\Big],
\label{eq:nv-F}
\end{align}
where $v_1(R)= \pi^{d/2} R^d/\Gamma(1+d/2)$ is the volume of a $d$-dimensional sphere of radius $R$, $\alpha_2(r;R)$ is the intersection volume of two spherical windows of radius $R$, scaled by $v_1(R)$, whose centers are separated by the distance $r$, and ${\tilde \alpha}_2(k;R)$ is its Fourier transform.
A hyperuniform point process is one in which $\sigma_N^2(R)$ grows slower than the window volume, i.e., $R^d$, for large $R$ 
\begin{align} \label{eq:hu-direct}
\lim_{R\to\infty} \frac{\fn{\sigma_N^2}{R}}{R^d}=0,
\end{align} 
which is equivalent to the hyperuniformity condition in Fourier space, i.e., $\lim_{\abs{\vect{k}}\to 0} S(\vect{k})=0$.\cite{torquato_local_2003,torquato_hyperuniform_2018}

Suppose the structure factor has the following power-law form as $\abs{\vect{k}}$ tends to zero: $\fn{S}{\vect{k}}\sim \abs{\vect{k}}^\alpha~ (\abs{\vect{k}}\to 0)$.
For hyperuniform systems, the {\it hyperuniformity exponent} $\alpha$ is a positive constant, which implies that there are three different scaling regimes (classes) that describe the associated large-$R$ of the number variance:\cite{torquato_local_2003,zachary_hyperuniformity_2009, torquato_hyperuniform_2018}
\begin{align} 
\sigma_N^2(R) \sim 
\begin{cases}
R^{d-1}, \quad\quad\quad \alpha >1 \qquad &\text{(Class I)}\\
R^{d-1} \ln R, \quad \alpha = 1 \qquad &\text{(Class II)}\\
R^{d-\alpha}, \quad 0 < \alpha < 1\qquad &\text{(Class III)}.
\end{cases}
\label{eq:classes}
\end{align}
Classes I and III are the strongest and weakest forms of hyperuniformity, respectively.
Class I includes all periodic systems and stealthy hyperuniform ones, the latter of which is a central interest of this work.

In Class I systems, the degree of order at large length scales has been characterized by the {\it hyperuniformity order metric} $\overline{\Lambda}$ (Refs. \onlinecite{torquato_local_2003,zachary_hyperuniformity_2009}), defined by 
\begin{align} 
 \overline{\Lambda} \equiv& \lim_{L\to \infty} \frac{1}{L}\int_0^L \frac{\fn{\sigma_N^2}{R}}{(R/D)^{d-1}}\dd{R}, 
\label{eq:lambda1} 
\end{align} 
where $D$ is the characteristic length scale of systems.
In Fourier space, Eq. \eqref{eq:lambda1} is rewritten as \cite{torquato_structural_2021b,maher_hyperuniformity_2023}
\begin{align}
 \overline{\Lambda} 
 = & \frac{2^d \rho v_1(D/2) d}{D \pi} \lim_{Q\to\infty} \int_{0}^Q \frac{S(k)-S(0)}{k^2} \dd{k}. \label{eq:lambda2} 
\end{align}
Point patterns arranged by increasing values of $\overline{\Lambda}$ are indeed arranged by increasing structural disorder or, equivalently, decreasing structural order.
The integral in Eq. \eqref{eq:lambda2} converges slowly: 
\begin{align} \label{eq:lambda3}
 \overline{\Lambda}(Q) 
 \equiv &
 \frac{2^d \rho v_1(D/2) d}{D \pi} \int_0^Q \frac{S(k)-S(0)}{k^2} \dd{k}
 \nonumber \\
 \sim & \overline{\Lambda} - c/Q \qquad(Q\to\infty),
\end{align}
where $c$ is a positive constant.
Thus, we estimate $\overline{\Lambda}$ by using the fit function given in Eq. \eqref{eq:lambda3}.
To estimate $\overline{\Lambda}$ using Eq. \eqref{eq:lambda3}, we take $D$ to be the particle diameter.

\subsection{Modified Collective-Coordinate Optimization Scheme}
\label{sec:PHI}

Here, we briefly describe the previous algorithm that we have used to generate disordered SHU sphere packings of moderate packing fractions,\cite{kim_multifunctional_2020, kim_effective_2023, kim_theoretical_2024} which uses a modification of Eq. \eqref{eq:pot} by including a pairwise soft-core repulsion $\fn{u}{r}$ that is bounded and positive with support in the finite range $0\leq r < \sigma$ but zero otherwise:
\begin{align} \label{eq:Phi}
\fn{\Phi}{\vect{r}^N} =\frac{\rho}{2} \sum_{\vect{k} \neq \vect{0}} \fn{\tilde{v}}{\vect{k}} \fn{\mathcal{S}}{\vect{k}} + \sum_{i <j} \fn{u}{r_{ij}}.
\end{align}
Here, we choose the following forms of $\fn{\tilde{v}}{\vect{k}}$ and $\fn{u}{r}$:
\begin{align} \label{eq:v-and-u}
 \frac{\fn{\tilde{v}}{\vect{k}}}{v_0} = \fn{\Theta}{K-\abs{\vect{k}}},
 ~~ 
 \frac{\fn{u}{r}}{\epsilon_0} = \qty(1-\frac{r}{\sigma})^2 \fn{\Theta}{\sigma-r},
\end{align}
where $\Theta(x)$ (equal to 1 for $x>0$ and zero otherwise) is the Heaviside step function, and $v_0$ and $\epsilon_0$ are parameters in the energy unit.
Turning off the stealthy pair potential (i.e., $v_0=0$) makes Eq. \eqref{eq:Phi} the harmonic contact potential that is most commonly employed to study the jamming of soft particles.\cite{ohern_jamming_2003,charbonneau_universal_2012,jin_jamming_21}
We note that exact forms of $\fn{\tilde{v}}{\vect{k}}$ and $\fn{u}{r}$ do not affect the set of possible ground states as far as they satisfy the conditions prescribed in Eqs. \eqref{eq:pot} and \eqref{eq:Phi}.
Specifically, since both sums in Eq. \eqref{eq:Phi} are non-negative, its ground state or global energy minimum with $\fn{\Phi}{\vect{r}^N}=0$, if it exists, must satisfy the stealthy hyperuniform condition \eqref{S-stealthy} and ensure all particle pairs are separated by at least a targeted distance $\sigma$, i.e., $\rmin \geq \sigma$.
Hence, these ground states can be mapped to SHU packings with packing fractions at least $\phi=\rho v_1(\sigma/2)$.
We note that at a specific value of $\chi$, $\phi$ can be increased to its maximum value, $\phimax(\chi,d)$, beyond which the ground state ceases to exist, which can be viewed as a satisfiable-unsatisfiable (SAT-UNSAT) transition.\cite{mezard_satisfiability_2009,franz_universality_2017} 
We numerically determine $\phimax(\chi,d)$ for the SHU packings with soft-core repulsions, as described in Sec. \ref{sec:algorithm}.

As the number of independent $\bf k$ vectors for which $\fn{\mathcal{S}}{\vect{k}}$ is constrained to be zero increases, i.e., as $K$ increases,
the dimensionality of the ground-state configuration manifold decreases.\cite{torquato_ensemble_2015} 
As noted in Sec. \ref{sec:intro}, a measure 
of the size of the ``exclusion region" [specified by the condition (\ref{S-stealthy})] relative to the total of $d(N-1)$ degrees of freedom 
is quantified by the stealthiness parameter $\chi$ in $\mathbb{R}^d$, which is given by
\begin{equation} \label{eq:def-chi}
\chi = \frac{\fn{M}{K}}{d(N-1)},
\end{equation}
where $\fn{M}{K}$ is half the number of $\bf k$ vectors for which $\fn{\mathcal{S}}{\vect{k}}=0$; see Eq. \eqref{S-stealthy}.
In the thermodynamic limit, $\chi$ is inversely proportional to $\rho$ according to the exact relation \cite{torquato_ensemble_2015}
\begin{equation}
\rho \,\chi =\frac{v_1(K)}{2d\,(2\pi)^d}.
\label{rho-chi}
\end{equation}
When $\chi$ is below a critical value ($\chi=1/3$ for $d=1$ and $\chi=1/2$ for $d=2,3$), SHU ground states are highly degenerate and disordered.\cite{zhang_ground_2015}
As $\chi$ (or $K$) increases, the short-range order also increases, leading crystalline ground states to be predominant beyond the critical value.

\section{Algorithm to Generate ultradense Disordered Stealthy Hyperuniform Sphere Packings}
\label{sec:algorithm}

Here, we provide simulation details about the modified collective-coordinate optimization scheme, employing the potentials \eqref{eq:pot} and \eqref{eq:Phi}, respectively, to generate ultradense SHU ground-state packings. 
The initial conditions and the choice of $\fn{\tilde{v}}{\vect{k}}$ and $\fn{u}{r}$ affect the statistics of the ground states obtained.\cite{zhang_ground_2015}
For the purposes of illustration, we begin from random initial conditions in a periodic hypercubic box in $\R^{d}$ and unit number density.
In the pair potentials \eqref{eq:v-and-u} that we employed, we take $v_0 = 1$ as our energy unit and take $\epsilon_0 = 100 v_0$, enabling us to generate valid packings that are ground states in a reasonable amount of computational time for $N=400,4000$ and $n_c=500-5000$, where $n_c$ is the number of ground-state configurations \footnote{For example, on an Intel(R) Xeon(R) CPU (E5-2680, 2.40 GHz), it takes around 15 core-hours to generate one 3D SHU ground-state packing with $\chi=0.45$, $\phi=0.47$, and $N=4000$. It takes around 0.1 core-hours on the same CPU to generate one 3D SHU ground-state packing with $\chi=0.0025$, $\phi=0.63$, and $N=4000$}.
Since we are primarily concerned with disordered states, and the computation cost increases as at least $\Order{\chi N^2}$,\cite{morse_generating_2023} we consider the cases of $0 < \chi\leq 0.45$ for $d=1,2,3$.

We perform the modified optimization scheme, as in Refs. \onlinecite{kim_multifunctional_2020,kim_theoretical_2024}.
However, we use it here to achieve the maximal target packing fractions.
Specifically, for given parameters of $d$, $\chi$, $N$, and $\sigma$, we begin with an initial condition and minimize its potential energy $\Phi$ given in Eqs. \eqref{eq:Phi} and \eqref{eq:v-and-u} via the low-storage Broyden-Fletcher-Goldfarb-Shanno (L-BFGS) algorithm\cite{nocedal_updating_1980, liu_limited_1989}.
The minimization stops when (i) $\Phi < 7\chi \times 10^{-20}$ \footnote{This threshold is close to zero energy of the potentials \eqref{eq:pot} and \eqref{eq:Phi} within the double precision of the machine}, (ii) the number of evaluations exceeds $5\times 10^6$, or (iii) the mean particle displacements are less than $10^{-15}\rho^{-1/d}$.
Among the resulting point patterns, we retain the ground-state point patterns that satisfy condition (i) and discard the others.

For given $d$ and $\chi$, we focus on cases with the maximal packing fraction $\phimax (\chi,d) = \rho v_1(\sigma_{\max}/2)$ and define it as the largest value of the target packing fractions $\phi$ that generates at least 5 ground states from 50 random initial conditions across system sizes $N$; see Table \ref{tab:max-soft} shown later.
This search is done by varying $\phi$ from 0.20 to the packing fraction of the densest lattice packing in increments of 0.01.
With the determined values of $\phimax (\chi,d)$, we obtain $n_c$ ground states; see Table S1 in the \href{...}{supplementary material}.
We note that Ref.\citenum{torquato_existence_2025} provides a more accurate estimation of $\lim_{\chi\to 0^+}\phimax(\chi,d=3)$ by using a finer increment in packing fraction (i.e., 0.001) and a lower target success ratio (i.e., 7\%).

\begin{figure*}[t]
 \begin{center}
 \includegraphics[width=0.40\textwidth]{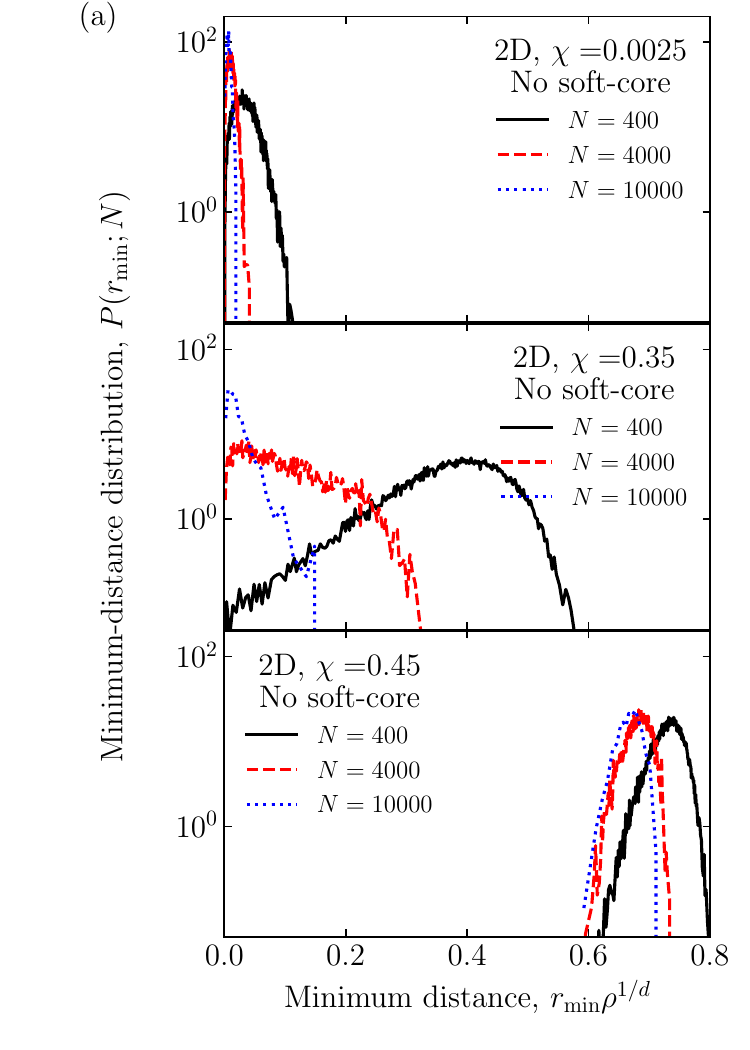}
 \includegraphics[width=0.40\textwidth]{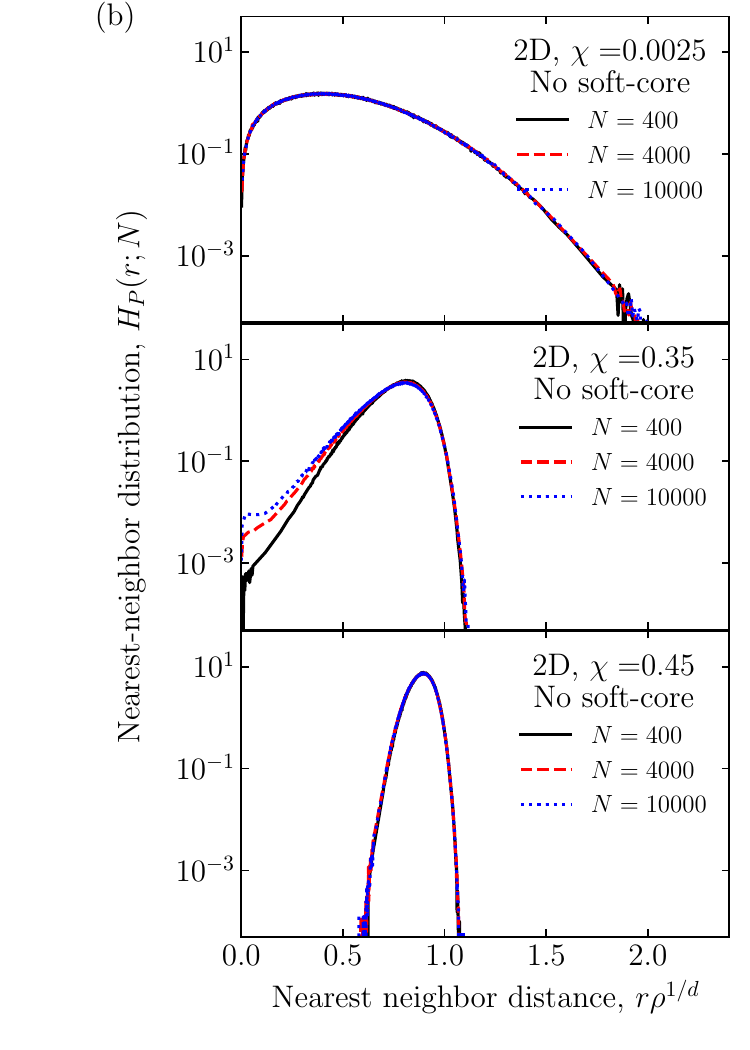}
 \end{center}
 \caption{
 Semi-log plots of (a) minimum-distance distribution $\fn{P}{\rmin;N}$ and (b) nearest-neighbor distribution $\fn{H_P}{r;N}$
 for 2D SHU ground-state point patterns of the standard potential \eqref{eq:pot} for $\chi=0.0025, 0.35, 0.45$.
 Each panel compares the curves with three different system sizes, $N=400,4000,10000$. 
 The analogous plots for $d=1,3$ are reported in the \href{...}{supplementary material}.
 \label{fig:2D-SHU-pure}}
\end{figure*}

Similarly, for purposes of comparison, we determine ground states using the standard potential \eqref{eq:pot} for a range of parameters $d$, $\chi$ and $N$.
The simulation parameters are also listed in Table S1 in the \href{...}{supplementary material}.

\section{Local Structural Characteristics of SHU Ground-State Configurations}
\label{sec:local}

Here, we report numerical results for the minimum-distance distribution $\fn{P}{\rmin;N}$ and nearest-neighbor distribution $\fn{H_P}{r;N}$ of configurations of SHU  ground states 
with $N$ points associated with the standard potential \eqref{eq:pot} and the modified potential \eqref{eq:Phi} defined in Secs. \ref{sec:local-standard} and \ref{sec:local-soft-core}, respectively.
The local structural information contained in these two functions facilitates the determination of the achievable maximal packing fraction $\phimax$ of SHU packings arising from the SHU ground-state point configurations.

Specifically, for a given SHU point pattern, the maximal packing fraction $\phimax(\chi,d)$ is determined by its minimum distance, i.e., $\phimax = \rho v_1(\rmin/2)$.
Thus, the $d$th-order moment of $\fn{P}{\rmin;N}$ leads to the mean value of the maximal achievable packing fraction $\E{\phimax} = \int_0 ^{\infty} \rho v_1(\rmin/2) P(\rmin; N)\dd{\rmin}$.
In addition, $\fn{H_P}{r;N}$ can be employed to analyze the large-$N$ behaviors of $\fn{P}{\rmin;N}$.
While Middlemas and Torquato\cite{middlemas_nearestneighbor_2020} studied $\fn{H_P}{r;N}$ for SHU ground-state configurations of the standard potential \eqref{eq:pot}, these results were not utilized to determine the values of $\phimax$, a quantity of central interest in this paper.

\begin{figure*}[ht]
  \begin{center}
    \includegraphics[width=0.45\textwidth]{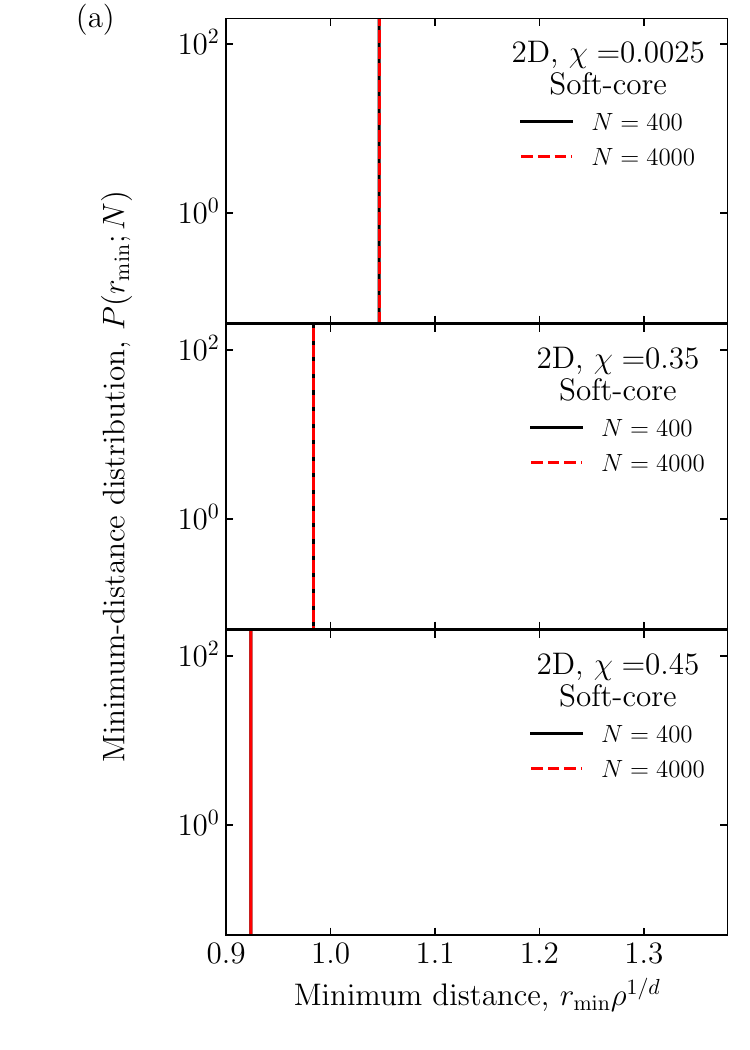}
    \includegraphics[width=0.45\textwidth]{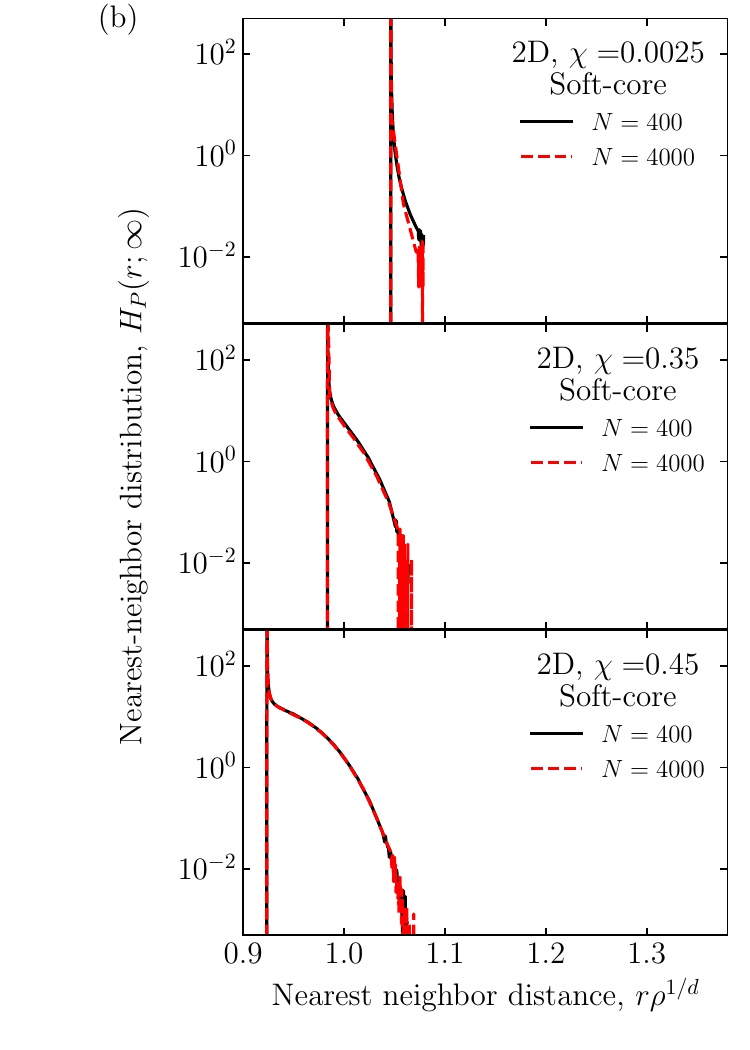}  
  \end{center}
  \caption{
  Semi-log plots of (a) minimum-distance distribution $\fn{P}{\rmin;N}$ and (b) nearest-neighbor distribution $\fn{H_P}{r;N}$ of 2D SHU ground-state point patterns of the modified potential \eqref{eq:Phi} for $\chi=0.0025, 0.35, 0.45$.
  Each panel compares the curves with two different system sizes, $N=400,4000$. 
  The corresponding plots for $d=1,3$ are reported in the \href{...}{supplementary material}. 
  \label{fig:2D-SHU-soft-core}
  }
\end{figure*}

\subsection{SHU point patterns without soft-core repulsions}
\label{sec:local-standard}

For simplicity here, we only present plots of $\fn{P}{\rmin;N}$ and $\fn{H_P}{r;N}$ for 2D SHU point patterns without soft-core repulsions.
The reader can find corresponding 1D and 3D plots in the \href{...}{supplementary material}.
It suffices here to state that while $P(\rmin; N)$ has a narrow range of support around $\rmin=0$ when $\chi<1/2$ for $d=1$, due to the fact that $\fn{H_P}{r;N}$ has large positive values for small $r$, the results of $P(\rmin; N)$ and $\fn{H_P}{r;N}$ for $d=3$ are similar to the 2D cases, as discussed below.

Figure \ref{fig:2D-SHU-pure}(a) shows that for a given $N$, $\fn{P}{\rmin;N}$ has a complicated dependence on $\chi$, i.e., the mean value of $\rmin$ monotonically increases with $\chi$, but its variance is maximized around $\chi=0.35$.
By contrast, for a given $\chi$, the mean value as well as the variance of $\rmin$ (or $\phimax$) monotonically decrease with $N$.
Such a tendency is also observed for $d=1,3$; see the \href{...}{supplementary material}.
It is clearly seen that $\phimax(\chi,d)$ decreases to zero in the thermodynamic limit, when $\chi$ is below a certain threshold value, i.e., $\chi < 0.5$ for $d=1$, $\chi \leq 0.35$ for $d=2$, and $\chi \leq 0.30$ for $d=3$.
In contrast to the quantity $\fn{P}{\rmin;N}$, $\fn{H_P}{r;N}$ is largely independent of $N$ (except at $\chi = 0.35$) as shown in Fig. \ref{fig:2D-SHU-pure}(b).
Thus, we can take $\fn{H_P}{r;\infty}\approx\fn{H_P}{r;N=10000}$, allowing us to predict the large-$N$ behaviors of $\fn{P}{\rmin;N}$ for any values of $\chi$ by using Eq. \eqref{eq:step2}.

Formula \eqref{eq:step2} indicates that it becomes exponentially more likely to encounter a point pattern with a small value of $\phimax=\rho \fn{v_1}{\delta/2}$ as $N$ increases, where the corresponding decay rate constant is equal to the probability that nearest-neighbor distances are larger than $\delta$, i.e., $1-\int_0^\delta \fn{H_P}{r;\infty}\dd{r}.$
Thus, for cases in which $\chi \leq 0.35$, the function $\fn{H_P}{r;\infty}$ has large positive values for small $r$, meaning that the probability of a point pattern having a very small $\phimax$ increases quickly as $N$ increases, as indicated in Fig. \ref{fig:2D-SHU-pure}(b).
Consequently, even for relatively small system sizes of $N \sim 100$, most SHU ground-state packings for $\chi \leq 0.35$ cannot achieve a maximal packing fraction $\phimax \gtrsim 0.05$. 
On the other hand, if $0.40\leq \chi < 0.45$, $\fn{H_P}{r;\infty}$ is negligibly small for small $r$, and thus, most of these SHU packings can have $\phimax \gtrsim 0.10$ when $N\approx 10^4$.
These packings, however, are less likely to have $\phimax \gtrsim 0.10$ as $N$ increases and eventually cannot have a positive value of $\phimax$ in the thermodynamic limit, as demonstrated in Sec. \ref{sec:max}.



\subsection{SHU point patterns with soft-core repulsions}
\label{sec:local-soft-core}

Here, we present the local structural characteristics, as in Sec. \ref{sec:local-standard}, for SHU point patterns with soft-core repulsions.
For simplicity here, we only present the plots of $\fn{P}{\rmin;N}$ and $\fn{H_P}{r;N}$ for 2D cases.
However, in the \href{...}{supplementary material}, we present the analogous plots for $d=1,3$. 
While these plots for $d=1$ are virtually identical to those of integer lattice, the plots for $d=3$ are similar to the 2D cases.

\begin{figure*}[t]
\includegraphics[width=0.45\textwidth]{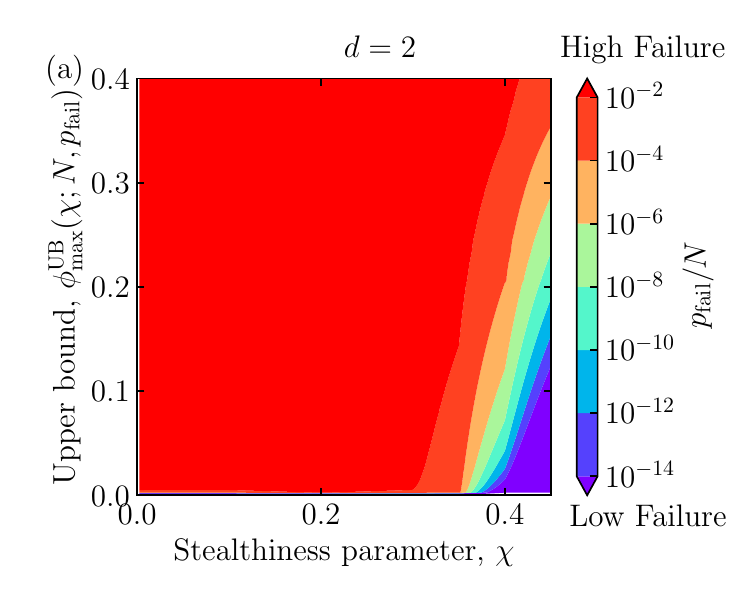}
\includegraphics[width=0.45\textwidth]{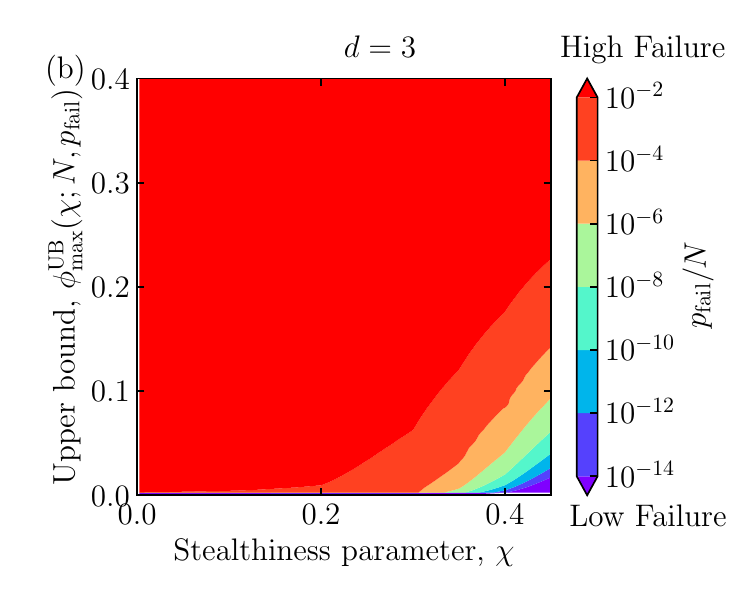}
\caption{ 
Probabilistic upper bounds \eqref{eq:UB} on the maximal packing fraction $\phimax$, denoted by $\phimax^\mathrm{UB}(\chi; N,\fail)$, as a function of stealthiness parameter $\chi$, system size $N$, and the associated probability $\fail$ for SHU ground-state packings without soft-core repulsions for (a) $d=2$ and (b) $d=3$.
For given values of $\chi$ and $\phimax^\mathrm{UB}$, the contours depict the ratio $\fail/N$ of a probability $\fail$ that a ground state cannot obtain $\phimax $ greater than $\phimax^\mathrm{UB}$ to $N$.
The red regions are prohibited in the sense that even for the relatively small system sizes of $N\approx 100$, the likelihood of failing to find a ground-state packing with the corresponding packing fractions is extremely high, i.e., $\fail\gtrsim 1$.
The contours of $\fail/N$ are obtained from Eq. \eqref{eq:UB-} and the simulation data of $\fn{H_P}{r;N}$ with the largest system sizes (i.e., $N=10000$ for $d=2$ and $N=8000$ for $d=3$) if $\fail/N \geq 10^{-4}$.
Otherwise, these contours are obtained from Eq. \eqref{eq:UB-2} and the small-$r$ fitting of $\fn{H_P}{r;N}$; see the \href{...}{supplementary material} for details.
  \label{fig:max-phi_standard}
}
\end{figure*}

The function $\fn{P}{\rmin;N}$ behaves like those in ultradense hard-core systems for $\chi < 1/2$, as shown in Fig. \ref{fig:2D-SHU-soft-core}(a).
In particular, $\fn{P}{\rmin;N}$ is virtually identical to a Dirac delta function centered at $\rmin=\sigma$.
This observation means that pairs of particles cannot be closer than $\sigma$, and every ground-state point pattern has pairs of particles separated by $r=\sigma$, implying that $\phi$ can be as large as $\rho v_1(\sigma/2)$.
In contrast to Fig. \ref{fig:2D-SHU-pure}(a), the plots of $\fn{P}{\rmin;N}$ shown here are independent of $N$, and thus, $\phimax$ depends solely on $\chi$.
The analogous behaviors are observed for $d=1,3$; see the \href{...}{supplementary material}.
For all $\chi$ values, plots of $\fn{H_P}{r;N}$ in Fig. \ref{fig:2D-SHU-soft-core}(b) are also insensitive to $N$ and exhibit common features.
For example, $\fn{H_P}{r;N}$ is identically zero for $r<\sigma$, possesses a Dirac-delta-like peak at $r=\sigma^+$, due to ``contacting'' particles, and then decreases to zero as $r\rho^{1/d}$ increases. 
We deem two particles to be in contact when their separation distance $r$ obeys the inequality, $r<\sigma+0.001\rho^{-1/d}$.
As $\chi$ increases, $\fn{H_P}{r;N}$ tends to have a broader range of support, which is a trend opposite to Fig. \ref{fig:2D-SHU-pure}(b). 
The analogous behavior is observed for $d=3$; see the \href{...}{supplementary material}.

\section{Maximal Packing Fractions}
\label{sec:max}

Here, we provide simulation data for the maximal packing fraction $\phimax(\chi,d)$ for SHU packings of the standard potential \eqref{eq:pot} and ultradense ones using the modified potential \eqref{eq:Phi}. 
We also present corresponding empirical formulas for $\phimax$.

For SHU ground states without soft-core repulsion, the quantity $\phimax$ can take any small positive value determined by an extreme event\cite{salvadori_extremes_2007, gomes_extreme_2015} of finding the minimum distance $\rmin$ of nearest-neighbor distances to $N$ distinct particles, as shown in Fig. \ref{fig:2D-SHU-pure}.
This observation indicates that an increasingly larger number of ground-state point configurations fail to achieve a prescribed packing fraction $\phi$ as the system size $N$ grows, as discussed in Sec. \ref{sec:local-standard}.
The reader is referred to the \href{...}{supplementary material} for some specific values of the fraction of the point configurations that fail to achieve selected values of $\phi$ at various values of $\chi$ and $N$.
Thus, we consider the following probabilistic upper bound on $\phimax$ that depends on $\chi$, $N$, and a probability $\fail$: 
\begin{align}\label{eq:UB}
 \phimax \leq \phimax^\mathrm{UB}(\chi;N,\fail) \equiv \rho \fn{v_1}{\Delta(\chi; N,\fail)/2},
\end{align}
where $\fail$ is the expected probability that one randomly obtained ground-state point configuration with $N$ particles cannot achieve $\phimax$ greater than $\phimax^\mathrm{UB}$, and the diameter $\Delta(\chi; N,\fail)$ is numerically estimated from the following expression for given $d$ and $\chi$ (see Appendix \ref{sec:UB-derivation} for derivation):
\begin{align} \label{eq:UB-}
\frac{2\fail}{N} &\approx 
\int_0^{\Delta(\chi; N,\fail)} \fn{H_P}{r;\infty} \dd{r} .
\end{align}
Here, the values of $\fn{\Delta}{\chi; N,\fail}$ are directly computed from simulation data of $\fn{H_P}{r;N}$ with the largest system sizes (i.e., $N=10000$ for $d=2$ and $N=8000$ for $d=3$) if $\fail/N\geq 10^{-4}$.
Otherwise, we estimate $\fn{\Delta}{\chi; N,\fail}$ by approximating the right-hand side of Eq. \eqref{eq:UB-} with the {\it converse Weibull distribution}:\cite{salvadori_extremes_2007,gomes_extreme_2015} 
\begin{align} \label{eq:UB-2}
 \frac{2\fail}{N} &\approx 1 - \exp\qty[-\qty(\frac{\Delta(\chi; N,\fail)-a'}{b'})^{c'}], \quad (\frac{\fail}{N} < 10^{-4}),
\end{align} 
where $a'$, $b'$, and $c'$ are fitting parameters that depend on $d$ and $\chi$; see the \href{...}{supplementary material} for details.

Figure \ref{fig:max-phi_standard} depicts the probabilistic upper bounds $\phimax^\mathrm{UB} (\chi; N,\fail)$, given in Eq. \eqref{eq:UB-}, for (a) $d=2$ and (b) $d=3$.
At a locus of $\chi$ and a desired value of $\phimax^\mathrm{UB}$, the contour value tells the ratio $\fail / N$ of the system size $N$ to the associated probability $\fail$ that cannot obtain one ground state with satisfying a prescribed value of $\phimax^\mathrm{UB}$.
For both $d=2,3$, when $\chi < 0.40$, the upper bounds on $\phimax$ rapidly decrease to zero as the ratio $\fail/N$ decreases.
Thus, it is extremely difficult or impossible ($\fail \gtrsim 1$) to obtain $\phimax>0$ (i.e., $\phimax^\mathrm{UB}\approx 0$) from the SHU ground-state packings of the standard potential \eqref{eq:pot} at large system sizes of $N\approx 10^4$.
By contrast, as $\chi$ increases beyond $0.4$, the upper bounds $\phimax^\mathrm{UB}(\chi; N, \fail)$ decrease considerably slowly with a decreasing $\fail / N$ relative to the cases of $\chi<0.4$.
Therefore, it is relatively easier to find a ground-state packing with $\phimax\sim 0.1$ - $0.2$ for the system sizes with $N\approx 10^{4}$ for $d=2$ (or $N\approx 10^2$ for $d=3$).

\begin{figure}[t h!]
\includegraphics[width=0.45\textwidth]{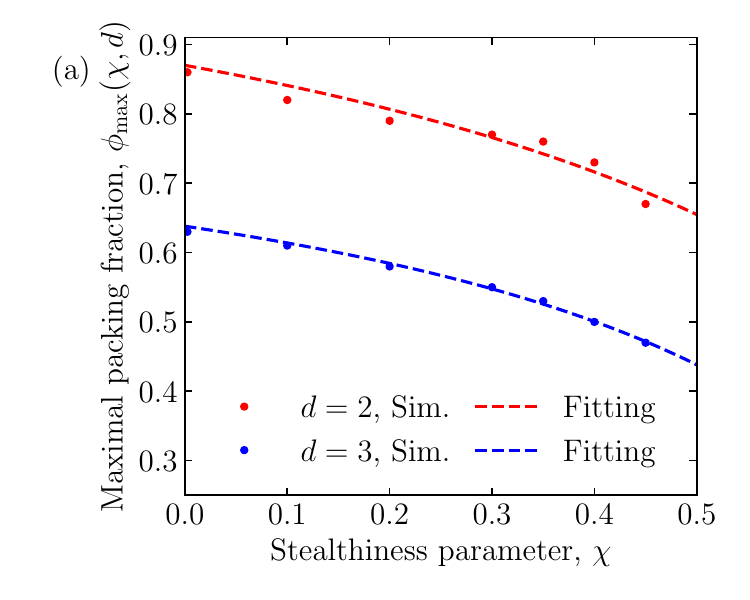}

\includegraphics[width=0.45\textwidth]{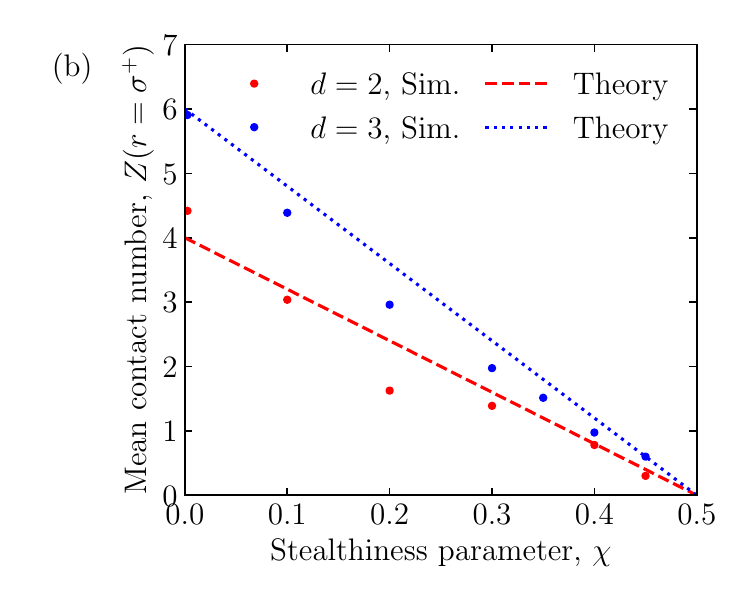}
 \caption{ 
(a) Maximal packing fraction $\phimax(\chi,d)$ as functions of stealthiness parameter $\chi$ for 2D and 3D ultradense SHU ground-state packings of the modified potential \eqref{eq:Phi}.
The simulation data (see Table \ref{tab:max-soft}) are compared to the empirical formula given in Eq. \eqref{eq:max-phi_soft} for $d=2,3$.
(b) Mean contact number per particle $\fn{Z}{r=\sigma^+}$ as functions of $\chi$ for the SHU ground-state packings in (a).
Simulation data are obtained from the ground states with $N=4000$ by counting the number of pairs whose separation distance is less than $\sigma + 0.001 \rho^{-1/d}$.
The theoretical predictions are obtained from Eq. \eqref{eq:Z-chi}. 
Note that for a given $d$, the lower and upper regions in (a) and (b) represent the SAT and UNSAT phases\cite{mezard_satisfiability_2009,franz_universality_2017} of Eq. \eqref{eq:Phi}, respectively.
\label{fig:soft_max-phi}
 }
\end{figure}

We now discuss $\phimax(\chi,d)$ for ultradense SHU ground-state packings of the modified potential \eqref{eq:Phi}.
In contrast to those without soft-core repulsions, the corresponding values of $\phimax(\chi, d)$ are identical to $\rho v_1(\sigma_{\max}/2)$ by definition, which are functions of $\chi$ for given $d$.
For $d=1$, $\phimax(\chi, d) \approx 1$ for $\chi < 1/2$, and thus, we focus on 2D and 3D cases.

\begin{figure*}[htb]
\includegraphics[width=0.8\textwidth]{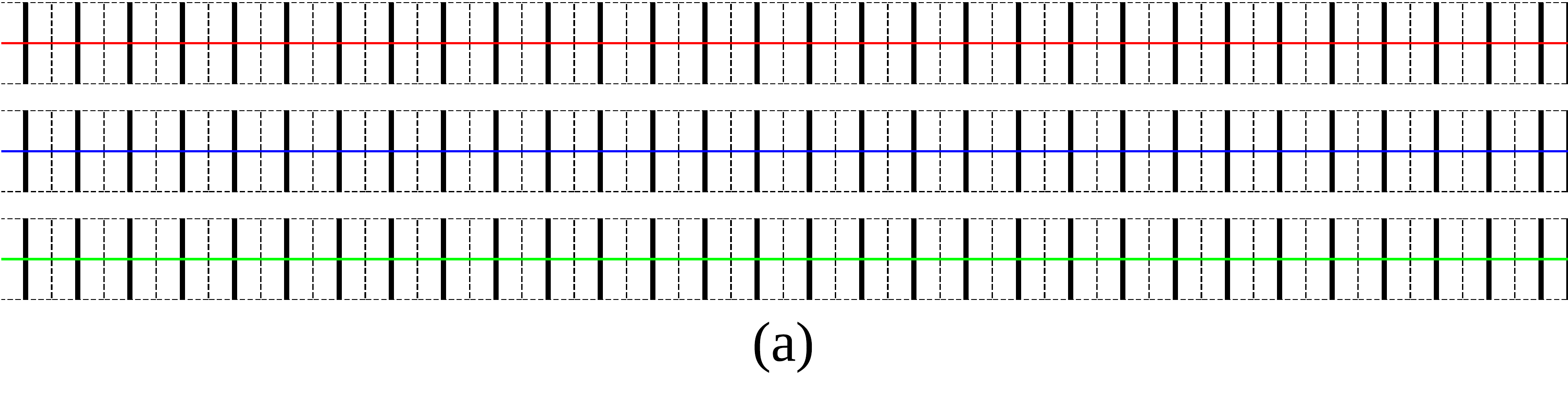}

\includegraphics[width=0.8\textwidth]{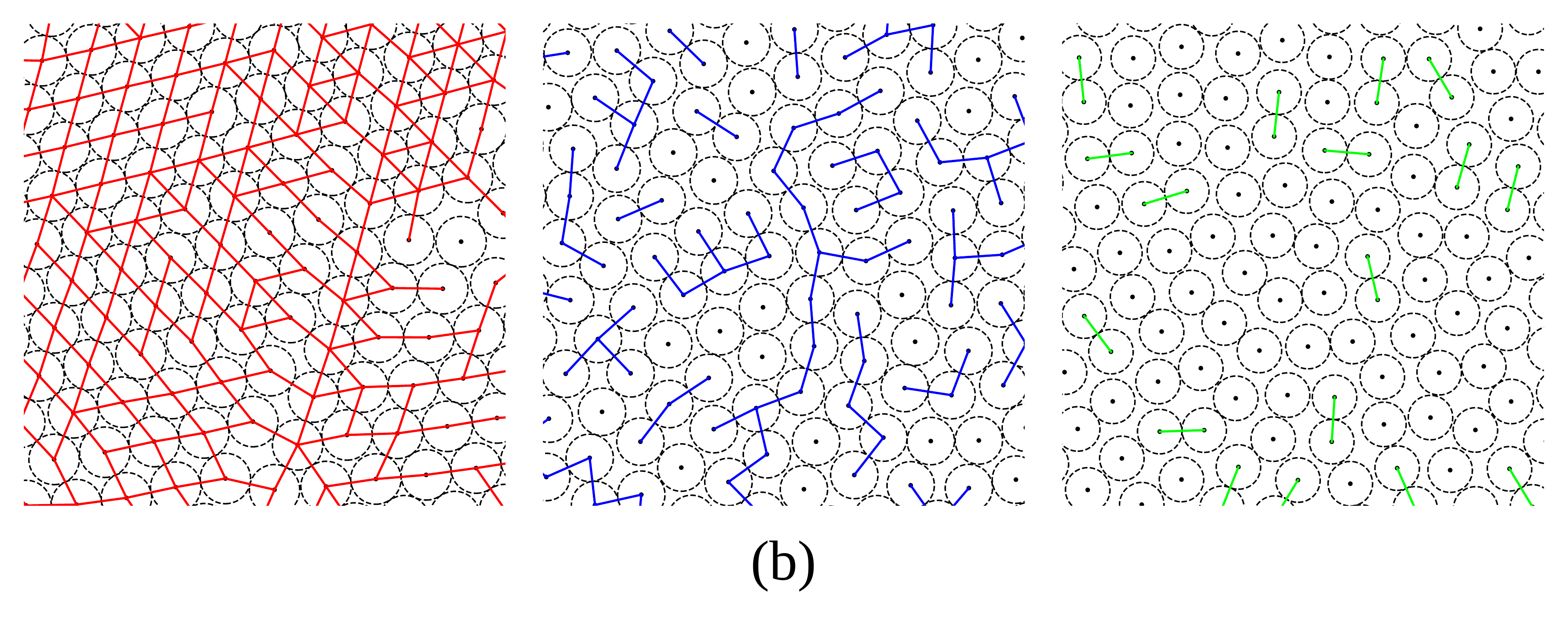}

\includegraphics[width=0.8\textwidth]{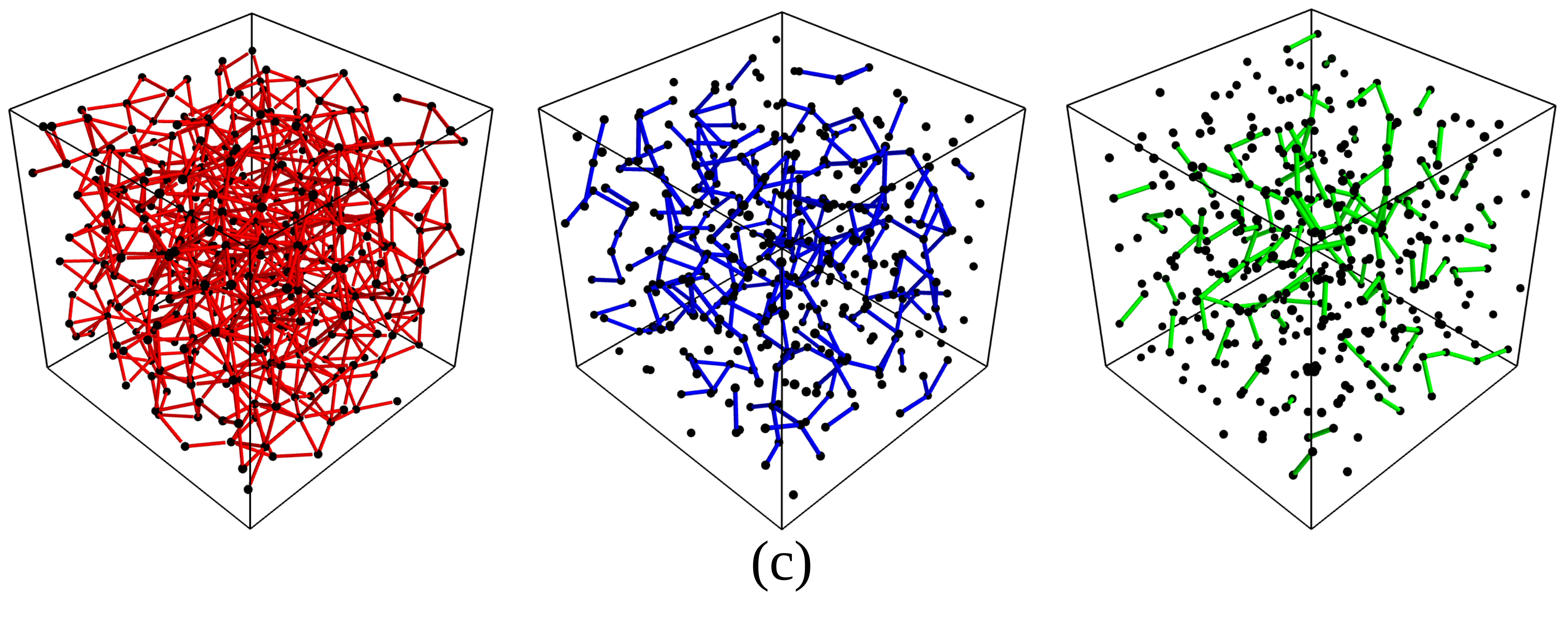}

\caption{Representative images of ultradense SHU ground-state point patterns of the modified potential \eqref{eq:Phi} with $N=400$ for (a) $d=1$, (b) $d=2$, and (c) $d=3$ at their maximal packing fractions; see Table \ref{tab:max-soft}.
The black bars in (a) and black dots in (b,c) depict the particle centroids.
In each dimension, we consider three values of $\chi=0.0025,0.35,0.45$ from the left to the right panels.
For three values of $\chi$, the bonds are drawn in red, blue, and green lines, respectively.
Bonds are drawn between pairs of centroids if they are contacting particles, meaning their separation distance $r$ obeys the inequality, $r<\sigma + 0.001\rho^{-1/d}$.
Panels (a,b) show the portions of ground states, whereas Panel (c) shows the entire ground states.
\label{fig:images}} 
\end{figure*}

\begingroup 
\squeezetable
\begin{table}
\caption{Values of the maximal packing fraction $\phimax(\chi,d)$, mean contact number per particle $Z(r=\sigma^+)$, and hyperuniformity order metric $\overline{\Lambda}$ for ultradense SHU ground-state packings of the modified potential \eqref{eq:Phi} with given values of $\chi$ for $d=1,2,3$.
We consider packings with $N=1000$ for $d=1$, and packings with $N=4000$ for $d=2,3.$
The values of $\overline{\Lambda}$ are estimated when we take the characteristic length scale $D=\sigma$.
The values in the parentheses represent standard errors.
\label{tab:max-soft}}
 \begin{ruledtabular}
 \begin{tabular}{c|d| d d d}
$d$ & \multicolumn{1}{c|}{$\chi$} & \multicolumn{1}{c}{$\phimax(\chi,d)$} &\multicolumn{1}{c}{$Z(r=\sigma^+)$} & \multicolumn{1}{c}{$\overline{\Lambda}$} \\
 \hline 
 \multirow{7}{*}{$1$}
&0.0025 &1.00 &2.0 & 0.16\overline{6}\\
&0.10 &1.00 &2.0 & 0.16\overline{6}\\
&0.20 &1.00 &2.0 & 0.16\overline{6}\\
&0.30 &1.00 &2.0 & 0.16\overline{6}\\
&0.35 &1.00 &2.0 & 0.16\overline{6}\\
&0.40 &1.00 &2.0 & 0.16\overline{6}\\
&0.45 &1.00 &2.0 & 0.16\overline{6}\\
 \hline 
 \multirow{7}{*}{$2$}
&0.0025 &0.86 &4.418(3) &0.69723(2)\\
&0.10 &0.82 &3.037(3) &0.46905(2)\\
&0.20 &0.79 &1.625(3) &0.45865(1)\\
&0.30 &0.77 &1.389(2) &0.45204(1)\\
&0.35 &0.76 &1.197(2) &0.44851(1)\\
&0.40 &0.73 &0.781(1) &0.44086(1)\\
&0.45 &0.67 &0.302(1) &0.42493(1)\\
 \hline 
 \multirow{7}{*}{$3$}
&0.0025 &0.63 &5.906(3) &0.72438(98)\\
&0.10 &0.61 &4.388(4) &0.68872(186)\\
&0.20 &0.58 &2.960(4) &0.66450(144)\\
&0.30 &0.55 &1.975(3) &0.64118(104)\\
&0.35 &0.53 &1.514(3) &0.62410(82)\\
&0.40 &0.50 &0.975(2) &0.60168(53)\\
&0.45 &0.47 &0.600(2) &0.57761(37)\\
 \end{tabular}
\end{ruledtabular}
\end{table}
\endgroup

\begin{table}[h]
\caption{Parameters $c_1(d)$ and $c_2(d)$ in the approximation formula \eqref{eq:max-phi_soft} for the maximal packing fraction $\fn{\phimax}{\chi,d}$ of SHU ground-state packings of the modified potential \eqref{eq:Phi}. 
The values of $\fn{\phi_{\max}}{0,d}$ are fixed.
\label{tab:C}
 }
\begin{tabular}{c|c c c}
 \hline 
$d$ & $\fn{\phi_{\max}}{0,d}$& $c_1(d)$& $c_2(d)$ \\
 \hline 
$2$& $0.870 $& $1.058$& $0.748 $\\ 
$3$& $0.638 $& $1.244$& $0.899 $\\
 \hline
\end{tabular}
\end{table}

As shown in Fig. \ref{fig:soft_max-phi}(a), $\phimax(\chi,d)$ for 2D and 3D ultradense SHU ground-state packings from Table \ref{tab:max-soft} counterintuitively decreases monotonically as $\chi$ decreases and approaches $1/2$, which is the trend opposite to that found for the ground states of the standard potential \eqref{eq:pot}; see Fig. \ref{fig:max-phi_standard} or Ref. \onlinecite{zhang_transport_2016}.
This trend of $\phimax(\chi,d)$ can be explained by the fact that constraints from stealthy hyperuniformity and contacting particles, in which two particles are taken to be in contact if the pair separation $r$ obeys the inequality ($r<\sigma+0.001\rho^{-1/d}$), remove degrees of freedom independently, as discussed later in this section.
Furthermore, such $\chi$-dependence on $\phimax(\chi,d)$ can be well approximated by the following [1,1] Pad\'e approximants:
\begin{align} \label{eq:max-phi_soft}
\fn{\phimax}{\chi,d} = \fn{\phimax}{0,d} \frac{1-c_1(d)\chi}{1-c_2(d)\chi},\quad (d=2,3),
\end{align}
where the $d$-dependent values of the positive parameters $c_1(d)$ and $c_2(d)$ given in Table \ref{tab:C}.
Analogously, Fig. \ref{fig:soft_max-phi}(b) also shows that the simulation data of mean contact number per particle $Z(r=\sigma^+)$, where we take $\sigma^+$, indicating the limit to $\sigma$ from above, to be $\sigma + 0.001\rho^{-1/d}$.
The reader can find values of $Z(r=\sigma^+)$ with different choices of $\sigma^+ = \sigma + \epsilon$ in the \href{...}{supplementary material}.
This mean ``contact'' number per particle is seen to monotonically decrease from $4.42$ to $0.30$ for $d=2$ and from $5.91$ to $0.60$ for $d=3$ as $\chi$ increases from $0.0025$ to $0.45$; see Table \ref{tab:max-soft} for specific values.

Such a dependence of $Z(r=\sigma^+)$ on $\chi$ is due to an interplay between two competing constraints.
Specifically, for an $N$-particle SHU ground-state packing in $\R^{d}$ with $d(N-1)$ total degrees of freedom, $2\chi \times d(N-1)$ degrees of freedom are already constrained by the stealthy hyperuniform condition; see Eq. \eqref{eq:def-chi} and Refs. \onlinecite{zhang_ground_2015,torquato_ensemble_2015}.
Thus, the maximum number of constraints from effectively contacting particles is equal to the remaining degrees of freedom, i.e., $N \fn{Z}{r=\sigma^+}/2 \leq d (N-1)(1-2\chi)$.
In the thermodynamic limit, the aforementioned constraints yield a theoretical upper bound on the mean contact number per particle $\fn{Z}{r=\sigma^+}$:
\begin{align}\label{eq:Z-chi}
 \fn{Z}{r=\sigma^+} = 2d (1-2\chi), \quad (d>1),
\end{align}
which shows good agreement with the simulation data; see Fig. \ref{fig:soft_max-phi}(b).
These predictions also explain that as $\chi$ increases, a ground-state packing has a smaller mean contact number per particle and larger void space between particles, leading to a lower value of $\phimax(\chi,d)$.
We note that Fig. \ref{fig:soft_max-phi}(a) and (b) can be interpreted as the phase diagrams of the ground state of Eq. \eqref{eq:Phi} for parameters of $(\chi, \phi)$ and $(\chi, Z(\sigma^+))$, respectively, which represent SAT-UNSAT transitions.\cite{mezard_satisfiability_2009,franz_universality_2017} 

The hyperuniformity order metric $\overline{\Lambda}$ for ultradense  SHU ground-state packings are also tabulated in Table \ref{tab:max-soft} with taking the particle diameter as the characteristic length scale, i.e., $D=\sigma$.
The value of $\overline{\Lambda}$ decreases with the degree of order at large length scales, as noted in Sec. \ref{sec:order}.
Since the 1D ground states are integer-lattice packings regardless of $\chi$, their values of $\overline{\Lambda}$ are identical to $1/6=0.16\overline{6}$ (see Ref. \citenum{torquato_local_2003}) with $D=\sigma=1/\rho.$
In contrast, for $d=2,3$, $\overline{\Lambda}$ tends to monotonically decrease with $\chi$, meaning that the ground-state packings are the least ordered at $\chi=0.0025$ but become the most ordered at $\chi=0.45$.
This trend arises partly because $\overline{\Lambda}$ is largely determined by the degree of density fluctuations at the intermediate to large length scales (associated with $S(k)$ for $k\sigma \lesssim \order{1}$) rather than those at the short length scales (associated with $S(k)$ for $k\sigma\gtrsim \order{1}$).
It follows, therefore, that the ground states with $\chi=0.0025$ have the largest values of $\overline{\Lambda}$ (e.g., $0.697$ for $d=2$ and $0.724$ for $d=3$) because of the smallest exclusion regions where $S(k)=0$.
By contrast, the ground states with a higher value of $\chi$ have larger exclusion regions and thus tend to have a lower value of $\overline{\Lambda}$.

\section{Pair Statistics of SHU Ground-State Point Configurations}
\label{sec:2pt}

Here, we present representative images and plots of pair statistics in direct and Fourier spaces for SHU point patterns with and without soft-core repulsions.
We begin by showing the representative images of ultradense SHU ground-state point patterns for $d=1,2,3$.
However, we do not include the corresponding images of the ground states without soft-core repulsions here because they were reported in Refs. \onlinecite{uche_constraints_2004, zhang_ground_2015,zhang_transport_2016}.

\begin{figure}[th!]
 {\centering
 \includegraphics[width=0.45\textwidth]{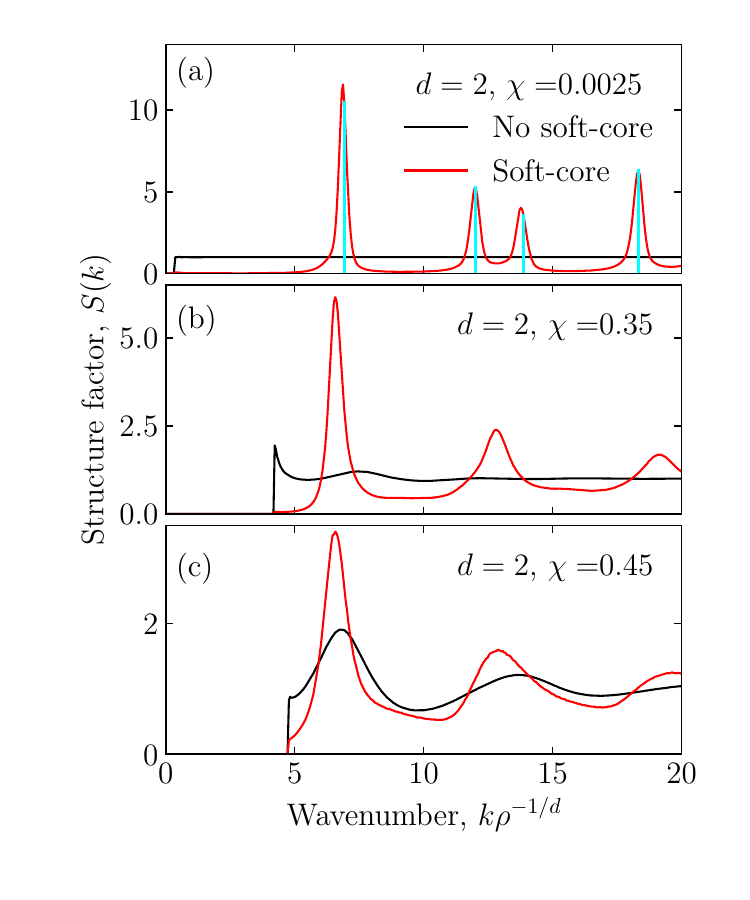}
 \vspace{-20pt}

 \includegraphics[width=0.45\textwidth]{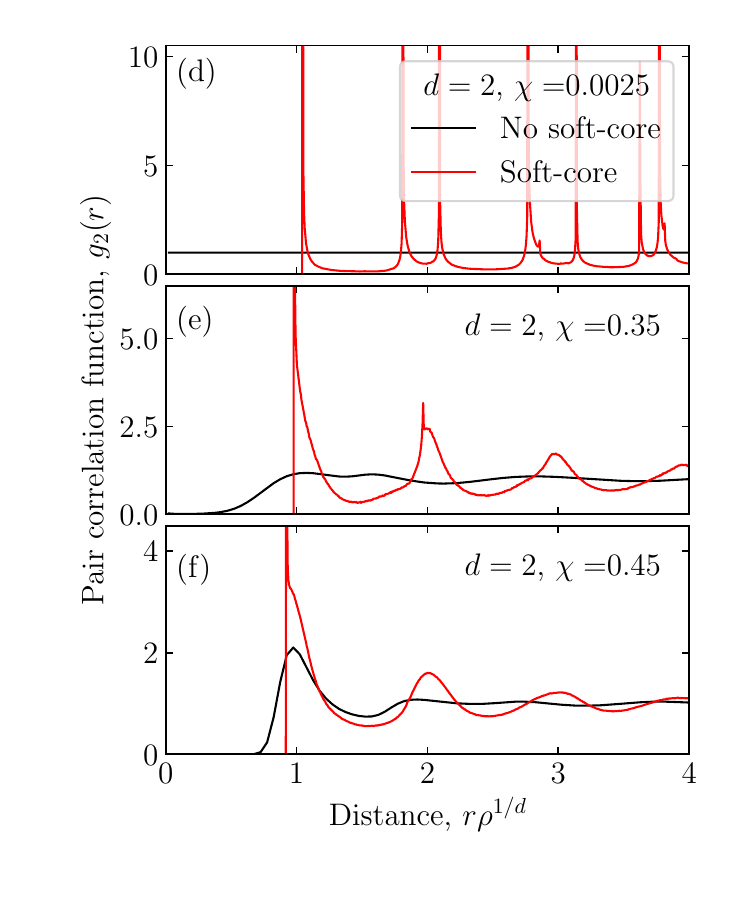}
 \vspace{-20pt}
 }
 \caption{
 Pair statistics of 2D SHU ground-state point patterns of $N=4000$ with and without soft-core repulsions for $\chi=0.0025,0.35,0.45$: (a-c) structure factor $S(k)$ and (d-f) pair correlation function $\fn{g_2}{r}$.
 The cyan lines in the topmost panel of (b) depict the Bragg peaks of the triangular lattice of $\rho=1$. 
 The analogous plots for $d=1$ are reported in the \href{...}{supplementary material}.
 \label{fig:2pt_2D}
 }
\end{figure}

Figure \ref{fig:images} shows ultradense SHU ground-state packings for $d=1,2,3$ with three different values of $\chi=0.0025,0.35,0.45$. 
It is clearly seen that 1D SHU ground-state packings are identical to the integer lattices ($\phimax(\chi,d) = 1.00$) for $\chi \leq 0.45$ in Fig. \ref{fig:images}(a).
2D and 3D ultradense SHU ground-state packings, shown in Fig. \ref{fig:images}(b,c), respectively, exhibit common $\chi$-dependences.
In the zero-$\chi$ limit (i.e., the cardinality $d_c$ of the infinitely degenerate SHU ground-state manifold set has the maximal value $d$), the ground-state packings for both $d=2$ and $d=3$ are close to `effectively' jammed states with percolating contact networks.
Specifically, the 2D ground states are highly ordered structures with large triangular coordination domains, which are the most probable outcomes for typical compression protocols to create jammed packings of identical disks under periodic boundary conditions.\cite{ohern_jamming_2003, donev_unexpected_2005, donev_configurational_2007, torquato_perspective_2018a}
The 3D counterparts are configurationally very close to the 3D nonequilibrium MRJ state prepared by rapid compression of hard-sphere systems,\cite{torquato_is_2000, donev_unexpected_2005, torquato_perspective_2018a,maher_hyperuniformity_2023} as measured by their structure factor $\fn{S}{k}$, pair correlation function $\fn{g_2}{r}$, values of packing fractions, and mean contact number per particle $Z(r=\sigma^+)\approx 6$, implying that the packing is effectively {\it isostatic}, i.e., marginally stable\cite{wyart_effects_2005}; see Ref. \onlinecite{torquato_existence_2025} for more details.

As $\chi$ increases to $0.45$, from the leftmost to the rightmost panels in Fig. \ref{fig:images}(b,c), the 2D and 3D ground states form fewer contacts, and the fraction of particles whose mean contact number per particle is greater than three gradually decreases.
Consequently, as $\chi$ increases from $0.20$ to $0.35$, linear polymer-like chains (i.e., the constituent particles have around two contacting particles) start to emerge and become an increasingly frequent motif in the ground states.
Increasing $\chi$ to $1/2$ further reduces the fraction of particles with two contacts, leading to a shorter mean chain length.
This observation is consistent with Eq. \eqref{eq:Z-chi} and Fig. \ref{fig:soft_max-phi}(b).
This trend also leads $\fn{H_P}{r;N}$ to have a broader range of support, as shown in Fig. \ref{fig:2D-SHU-soft-core}(b).
Furthermore, $\phimax(\chi,d)$ also monotonically decreases as $\chi$ increases and approaches $1/2$, as explained in Sec. \ref{sec:max}.

\begin{figure}[th!]
 {\centering
 \includegraphics[width=0.45\textwidth]{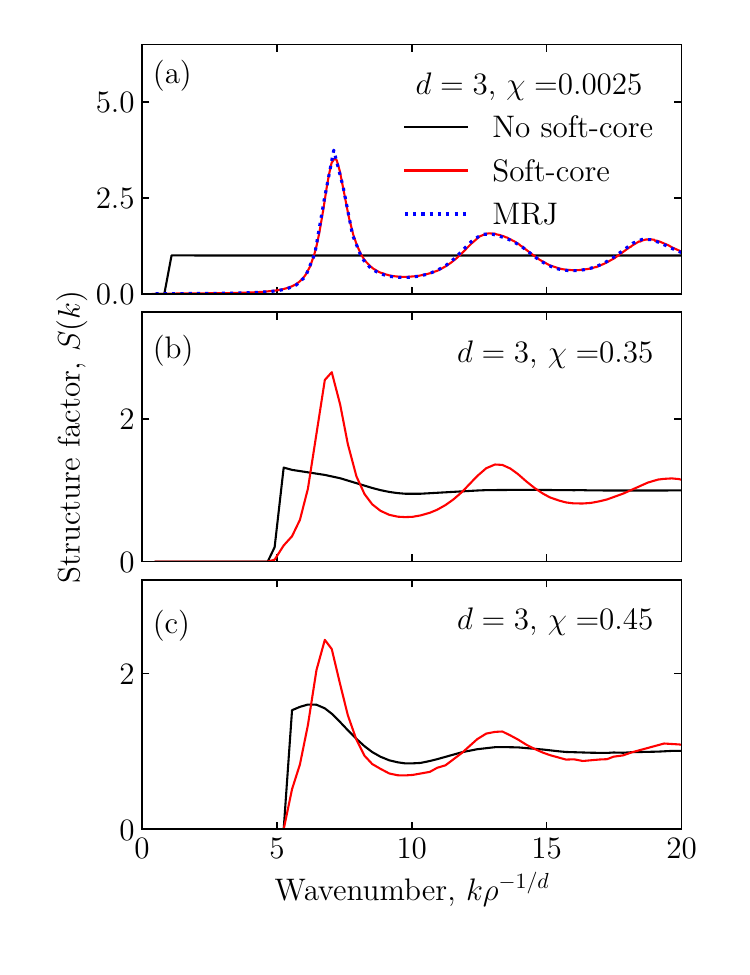}
 \vspace{-20pt}

 \includegraphics[width=0.45\textwidth]{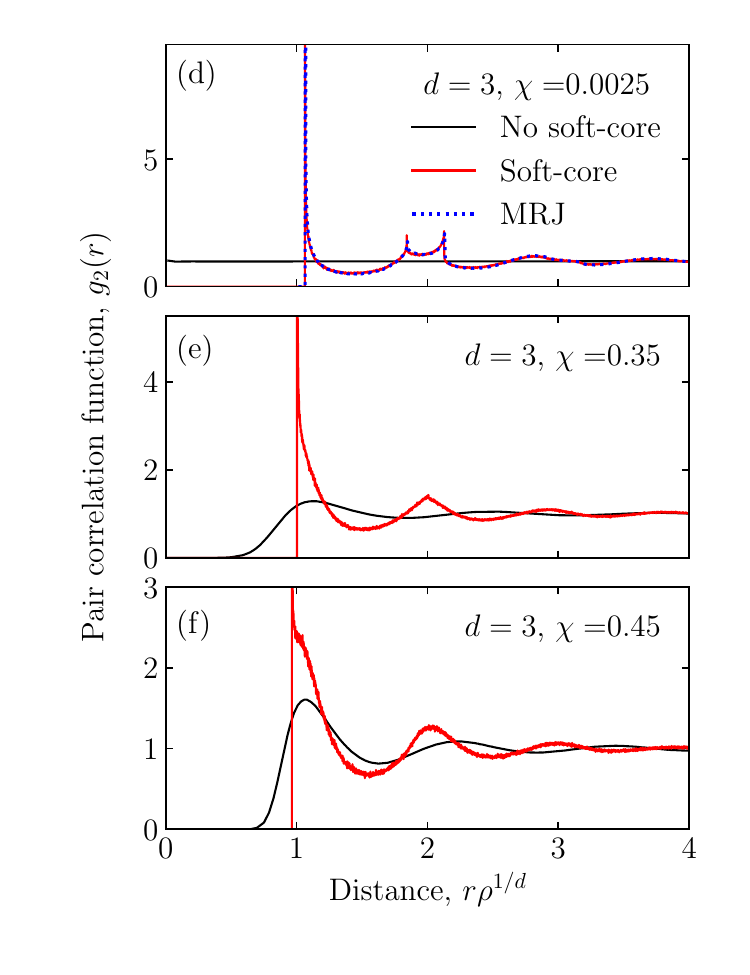}
 \vspace{-20pt}
 }
 \caption{
 Pair statistics of 3D SHU ground-state point patterns of $N=4000$ with and without soft-core repulsions for $\chi=0.0025,0.35,0.45$: (a-c) structure factor $S(k)$ and (d-f) pair correlation function $\fn{g_2}{r}$.
 Panels (a,d) also include the curves for 3D MRJ packings of $N=5000$ taken from Ref. \onlinecite{maher_hyperuniformity_2023}, which show excellent agreements with those for ultradense SHU packings.
 \label{fig:2pt_3D}
 }
\end{figure}

We now present plots of $\fn{S}{k}$ and $\fn{g_2}{r}$ for 2D and 3D SHU point patterns of the standard potential \eqref{eq:pot} and the modified one \eqref{eq:Phi} for the smallest $\chi (=0.0025)$, an intermediate $\chi (=0.35)$, and the largest $\chi (= 0.45)$; see Figs. \ref{fig:2pt_2D} and \ref{fig:2pt_3D} for $d=2$ and $3$, respectively. 
The black curves represent the ground states of the standard potential, whereas the red-dashed curves represent those of the modified potential.
The reader can find analogous plots for $d=1$ in the \href{...}{supplementary material}.
The 1D ground states without the soft-core repulsions behave approximately like an ideal gas for small $\chi$\cite{zhang_ground_2015, zhang_transport_2016, middlemas_nearestneighbor_2020} and become imperfect integer lattices of lattice spacing $2/\rho$ as $\chi$ increases up to $1/2$. 
The 1D ground states with soft-core repulsions become lattices of lattice spacing $1/\rho$ when $\chi < 1/2$.

We begin by examining 2D cases shown in Fig. \ref{fig:2pt_2D}.
When there is no soft-core repulsion, the SHU ground states are approximately like ideal gas in the zero-$\chi$ limit; see Fig. \ref{fig:2pt_2D}(a,d).
As $\chi$ increases to $0.35$ and then $0.45$, these ground states have wider exclusion regions in Fourier space [see Fig. \ref{fig:2pt_2D}(b,c)].
In the same range of $\chi$, the $\fn{g_2}{r}$'s become smaller (but nonzero) in a wider range of $r$ near the origin, as seen in Fig. \ref{fig:2pt_2D}(e,f).

By contrast, 2D ultradense SHU ground states exhibit much richer behaviors.
In the zero-$\chi$ limit, these states are highly ordered at short and intermediate length scales, approaching the densest triangular-lattice packings with some defects, characterized by the peaks of the triangular lattice in both $\fn{S}{k}$ and $\fn{g_2}{r}$, as shown in Fig. \ref{fig:2pt_2D}(a,d).
Importantly, these ultradense states resemble the densest monodisperse disk packings created via rapid-compression algorithms.\cite{ohern_jamming_2003,donev_pair_2005}
As $\chi$ increases to $0.35$, the ultradense packings become disordered at short length scales, evidenced by the disappearance of the second and third peaks of the triangular lattice disappear in both $\fn{S}{k}$ and $\fn{g_2}{r}$; see Fig. \ref{fig:2pt_2D}(b,e).
Despite such a change, $\fn{g_2}{r}$ of this ultradense state still exhibits a power-law singularity for near contacts [i.e., $\fn{g_2}{r}\sim (r-\sigma^+)^{-\gamma}$ with $\gamma\approx 0.25$ for $r\gtrsim \sigma^+$] and possesses a small sharp peak at $r=2\sigma\approx 2\rho^{-1/2}$.
This small peak indicates the presence of linear polymer-like chains consisting of three particles. 
As $\chi$ increases to $0.45$, $\fn{S}{k}$ barely changes [see Fig. \ref{fig:2pt_2D}(c)].
Since this ultradense SHU state now forms a few short linear polymer-like chains of two contacting particles, $\fn{g_2}{r}$ has a small sharp peak at $r=\sigma$ with $Z(r=\sigma^+)\approx 0.3$ and no longer has a power-law singularity for near contacts and a sharp peak at $r=2\sigma$; see Fig. \ref{fig:2pt_2D}(f).
Note that as $Z(r=\sigma^+)$ monotonically decreases with $\chi$, the discrepancies in both $S(k)$ and $\fn{g_2}{r}$ due to the soft-core repulsions become the most significant in the small-$\chi$ regime, and decrease with $\chi$.
Specifically, for $\fn{g_2}{r}$ in a range of $r<2\sigma$, there are mean relative deviations of 119\%, 69.7\%, and 31.9\% at $\chi=0.0025,0.35,0.45$, respectively; see Appendix \ref{sec:change} for the definition.

The pair statistics of 3D SHU ground states shown in Fig. \ref{fig:2pt_3D} also exhibit a $\chi$-dependence similar to that of their 2D counterparts, as discussed earlier concerning the representative images in Fig. \ref{fig:images}.
Without the soft-core repulsion, a ground state behaves approximately like an ideal gas in the zero-$\chi$ limit [see Fig. \ref{fig:2pt_3D}(a,d)].
As $\chi$ increases to $0.35$ and $0.45$, the resulting ground state has smaller but nonzero values of $\fn{g_2}{r}$ for $r\rho^{1/d}\lesssim 0.7;$ see Fig. \ref{fig:2pt_3D}(b,c).

With soft-core repulsion, Fig. \ref{fig:2pt_3D}(a,d) clearly confirm that the 3D ultradense ground state in the limit of $\chi\to0$ is configurationally nearly identical to the 3D MRJ state in pair statistics, including $\fn{S}{k}\sim k$ for $k\gtrsim K$ and $\fn{g_2}{r}$ exhibiting the split-second peak and the power-law singularity for near contacts, i.e., $\fn{g_2}{r}\sim (r-\sigma^+)^{-\gamma}$ for $r\gtrsim \sigma^+$.
Although the gap exponent $\gamma$ is not explicitly shown in this work, we demonstrated that as $N$ increases, the $\gamma$ values of the ultradense SHU packings approach $0.44$,\cite{torquato_existence_2025} which is consistent with the MRJ state.
While this ultradense ground state is clearly stealthy hyperuniform with a finite value of $N$ [see Fig. \ref{fig:2pt_3D}(a)], it becomes `nonstealthy' hyperuniform as $N$ tends to infinity with $\chi$ being the smallest value of $\chi=(N-1)^{-1}$.\footnote{For a $N$-particle system in a $d$-dimensional simple cubic fundamental cell, the smallest value of $M(K)$ is $d$, and thus $\chi = M(K)/[d(N-1)]\geq (N-1)^{-1}$.} \cite{torquato_existence_2025}
This finding is quite counterintuitive since a non-compression algorithm leads to sphere packings essentially identical to those of nonequilibrium hyperuniform MRJ packings, and since the configurations leading to $\phimax(\chi,d)=0.63$ is practically impossible to achieve from the ground states of the standard potential.
In other words, when the cardinality $d_c$ of the SHU ground-state manifold set becomes maximized in the limit $\chi\to 0$, this manifold contains the 3D MRJ packings; see Ref. \onlinecite{torquato_existence_2025}.
As $\chi$ increases to $0.35$ and $0.45$, while $\fn{S}{k}$ barely changes [see Fig. \ref{fig:2pt_3D}(b,c)], $\fn{g_2}{r}$ show a few changes in the local structures similar to those in 2D cases.
Specifically, as $\chi$ increases to $0.35$, $\fn{g_2}{r}$ still exhibits a power-law singularity for near contacts with a reduced gap exponent $\gamma\approx 0.21$, but the split-second peak around $r=2\sigma$ disappears; see Fig. \ref{fig:2pt_3D}(e).
As $\chi$ increases from $0.35$ to $0.45$, $\fn{g_2}{r}$ has a small sharp peak at $r=\sigma$ with $Z(r=\sigma^+)\approx 0.6$ and no longer has a power-law singularity for near contacts; see Fig. \ref{fig:2pt_3D}(f).
The discrepancies in both $S(k)$ and $\fn{g_2}{r}$ due to the soft-core repulsions also monotonically decrease with $\chi$.
Specifically, for $\fn{g_2}{r}$ in a range of $r<2\sigma$, there are mean relative deviations of 56.9\%, 40.0\%, and 25.9\% at $\chi=0.0025,0.35,0.45$, respectively.

For ultradense 2D and 3D SHU ground-state packings at small and large values of $\chi$ (=0.0025, 0.45), we also computed the spectral density $\spD{k}$; see Fig. \ref{fig:chik} in Appendix \ref{sec:chi}.
By construction, these packings are also stealthy hyperuniform two-phase dispersions, i.e., $\spD{k}=0$ for $k<K$.\cite{zhang_transport_2016}
It is noteworthy that the quantity $\spD{k}$ can be utilized to estimate the various physical properties of such two-phase dispersions, including the effective dynamic dielectric constant,\cite{kim_effective_2023, kim_theoretical_2024} fluid permeability, and mean survival time.\cite{torquato_predicting_2020}

\section{Conclusions and Discussion}
\label{sec:conclusion}

In this work, we have shown that SHU packings from the modified potential \eqref{eq:Phi}, which includes soft-core repulsions, achieve significantly larger packing fractions than those without these repulsions \eqref{eq:pot} because the soft-core repulsions ensure all particles are separated by at least $\sigma$; see Sec. \ref{sec:max}.
Within the disordered $\chi$ regime (i.e., $\chi<1/2$), the maximal packing fractions $\phimax(\chi,d)$ for the ultradense soft-core SHU packings are independent of the system size $N$ and at least twice as large as those without soft-core repulsions.
Specifically, the values of $\phimax(\chi,d)$ with soft-core repulsions reach up to $\phimax(\chi,d)=1.0, 0.86,0.63$ in the zero-$\chi$ limit and decrease to $\phimax(\chi,d)=1.0,0.67,0.47$ at $\chi=0.45$ for $d=1,2,3$, respectively; see Fig. \ref{fig:soft_max-phi} and Eq. \eqref{eq:max-phi_soft}.
For a given $d$, in the $\chi$-$\phi$ plane, the regions below and above the function $\phi_{\max}(\chi,d)$ represent satisfiable and unsatisfiable phases, respectively, of the soft-core SHU packings; see Fig. \ref{fig:soft_max-phi}(a).
Without soft-core repulsions, however, we showed for the first time that $\phimax$ decreases to zero on average as $N$ increases.
The local descriptors, including $\fn{P}{\rmin;N}$ and $\fn{H_P}{r;N}$, revealed that this trend comes from the fact that such a configuration is more likely to have a minimum pair distance $\rmin$ closer to zero as $N$ increases; see Fig. \ref{fig:max-phi_standard} and the approximation formula \eqref{eq:UB-2}.

Ultradense SHU ground-state packings with soft-core repulsions also exhibit considerably different morphologies from those without soft-core repulsions; see Sec. \ref{sec:2pt}.
For example, in the zero-$\chi$ limit, in which the cardinality of the SHU ground-state manifold set is maximized, the 2D and 3D ultradense SHU packings are configurationally very close to the jammed packings of identical particles created by various fast-compression protocols,\cite{ohern_jamming_2003, donev_unexpected_2005, maher_hyperuniformity_2023} as measured by their pair statistics, and packing fractions ($\phimax(\chi,d)=0.86$ for $d=2$ and $0.63$ for $d=3$).
For $d=3$, this observation is doubly counterintuitive because the sphere packings achieved via our non-compression optimization, which does not change the sizes of particles or a simulation box, are configurationally nearly identical to the nonequilibrium hyperuniform MRJ packings in various aspects, including a packing fraction of 0.63, a mean contact number $Z(\sigma^+)$ of $6$ (i.e., isostaticity), a gap exponent $\gamma$ of 0.44, and pair statistics [i.e., pair correlation functions $g_2(r)$ for all $r$ and structure factors $S(k)$ for $k>K$, including their hyperuniformity exponent of 1 for small wavenumbers].
Furthermore, $\phimax(\chi,3)=0.63$ is practically unattainable from the SHU ground states without soft-core repulsions.

The large family of 2D and 3D ultradense SHU sphere packings within the disordered
regime $(0 < \chi < 1/2)$ that we have created also possess singular structural features.
As $\chi$ increases up to $\chi=0.45$, both 2D and 3D ground-state packings become more ordered at large length scale (indicated by smaller values of the hyperuniformity order metric $\overline{\Lambda}$; see Table \ref{tab:max-soft}).
These 2D and 3D packings also have lower values of $Z(\sigma^+)~(\approx 0.46, 0.56$, respectively) and $\phimax(\chi,d)$ ($\approx 0.67, 0.47$, respectively), because the stealthy hyperuniformity constraints limit more degrees of freedom, resulting in fewer number of effectively contacting particles.
Thus, as $\chi$ increases from $0.20$ to $1/2$, this change leads to the formation of linear polymer-like chains with a decreasing mean chain length. 
Additionally, ultradense SHU ground states also have significantly different pair statistics, $\fn{S}{k}$ and $\fn{g_2}{r}$, from those without such repulsions because the repulsions enforce a nonoverlapping condition [i.e., $g_2(r)=0$ for $r<\sigma$] and induce stronger correlations in both $S(k)$ and $g_2(r)$ for intermediate wavelengths and distances, respectively; see Figs. \ref{fig:2pt_2D} and \ref{fig:2pt_3D}.
Such discrepancies in the pair correlation functions become most significant in the small-$\chi$ limit and tend to decrease as $\chi$ increases.

Our findings have important practical implications.
The spectral density $\spD{k}$ of 2D and 3D ultradense SHU packings, which we computed in Fig. \ref{fig:chik}, can be used to estimate various physical properties of these exotic two-phase dispersions.\cite{torquato_predicting_2020, kim_effective_2023, kim_theoretical_2024} 
Furthermore, the modified potential \eqref{eq:Phi} could be used in the future to create a wider range of soft-core SHU packings compared to the ones we studied here with $\sigma_{\max}(\chi) = \sigma = 2a$, where $a$ is the radius of a sphere decorating each particle.
Specifically, by choosing $\sigma$ and $a$ with $\sigma_{\max}(\chi) \geq \sigma \geq 2a > 0$, one can easily control the local structural characteristics, such as the mean contact number per particle and the pore-size distribution,\cite{torquato_new_2006, torquato_predicting_2020} without compromising long-range stealthy hyperuniformity for a given value of $\chi$.
It is noteworthy that when the radius $\sigma$ of soft-core potential is larger than the particle diameter $2a$, the resulting SHU packings no longer have pairs of contacting particles. Instead, their particles are well-separated in a fully connected matrix phase, which is a necessary requirement to attain optimal two-phase structures.\cite{torquato_multifunctional_2018}
This implies that such SHU soft-core packings may provide (nearly) optimal or novel physical properties, superior to those of SHU packings without soft-core repulsions reported in previous studies.\cite{zhang_transport_2016, froufe-perez_band_2017, torquato_multifunctional_2018, zhang_experimental_2019, christogeorgos_extraordinary_2021, tavakoli_65_2022, granchi_nearfield_2022, klatt_wave_2022, tang_hyperuniform_2023, merkel_stealthy_2023, tamraoui_hyperuniform_2023}
Indeed, \citealt{vanoni_dynamical_2025} recently showed how two dynamical physical properties (i.e., effective dynamic dielectric constant and time-dependent diffusion spreadability) of two-phase media derived from soft-core SHU packings are improved for a range of $\chi$ within the disordered regime and packing fractions $\phi$.
Our results offer a new route for discovering novel disordered SHU two-phase materials with unprecedentedly high density.

\appendix

\section{Derivation of the Probabilistic Upper Bound \eqref{eq:UB}} \label{sec:UB-derivation}

Here, we derive the probabilistic upper bound \eqref{eq:UB-}.
We begin from Eq. \eqref{eq:step2}:
\begin{align}  
\fail &= \int_0^{\Delta(\chi; N,\fail)} \fn{P}{\rmin; N} \dd{\rmin}
\nonumber \\
  &\approx
  1- \qty[1-\int_0^{\Delta(\chi; N,\fail)} \fn{H_P}{r;\infty} \dd{r}]^{N/2} \nonumber \\
  &\approx 
  1- \qty[1-\frac{N}{2}\int_0^{\Delta(\chi; N,\fail)} \fn{H_P}{r;\infty} \dd{r}]  \label{eq:UB-condition} ,
  \end{align}
  where we have used the binomial theorem in the last step.
  Algebraic simplification of Eq. \eqref{eq:UB-condition} yields the final expression given in Eq. \eqref{eq:UB-}.


  

\begin{figure}[t]
\subfloat[]{
\includegraphics[width=0.45\textwidth]{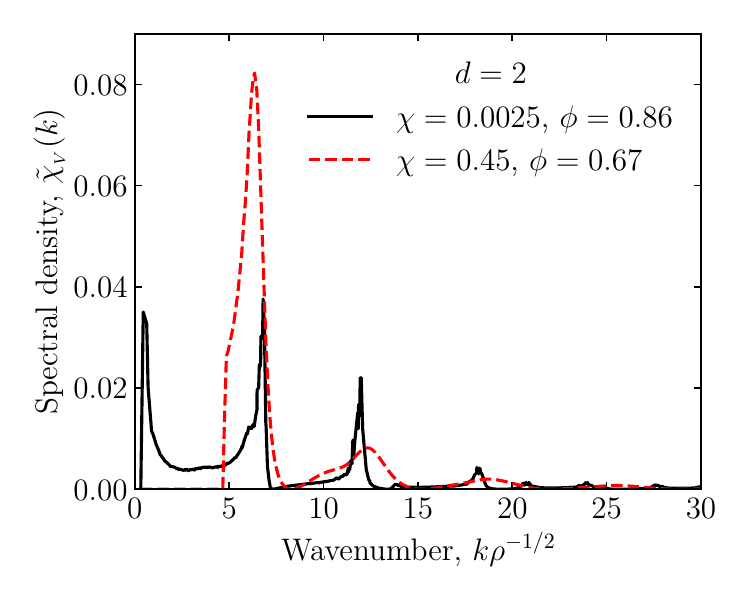}}

\subfloat[]{
\includegraphics[width=0.45\textwidth]{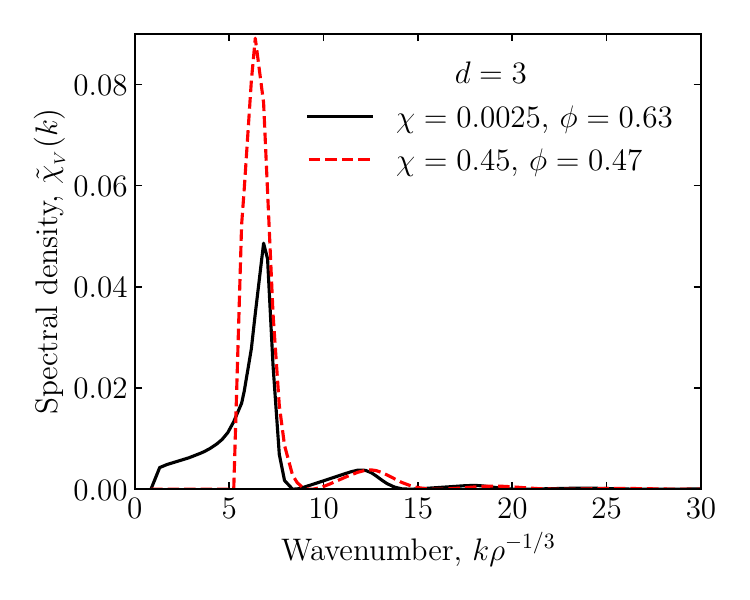}}
\caption{
Spectral density $\spD{k}$ as a function of dimensionless wavenumber $k\rho^{-1/d}$ of SHU ground-state packings of the modified potential \eqref{eq:Phi} in a matrix with $\chi=0.0025$ and $0.45$ for (a) $d=2$ and (b) $d=3$.
We consider the ground states with $N=400$.
\label{fig:chik}
}
\end{figure}

\section{Relative Change in Pair Correlation Function}
\label{sec:change}

For a prescribed value of $\chi$, we measure the discrepancy in the pair correlation function $\fn{g_2}{r}$ due to the soft-core repulsions by using the mean relative deviation, defined as the absolute value of a relative change averaged in a range of $r<2\sigma$:
\begin{align*}
 \frac{1}{v_1(2\sigma)}
 \int_0 ^{2\sigma} \abs{1-\frac{\fn{g_2}{r;\text{with soft-core}}} {\fn{g_2}{r;\text{without soft-core}}}} s_1(r) \dd{r} 
\end{align*}
where $s_1(r)$ is the surface area of a $d$-dimensional sphere of radius $r$, and $\sigma$ is the particle diameter of soft-core SHU packings.

\section{Spectral Densities of SHU Ground-State Packings} \label{sec:chi}
   
Here, we present the plots of the spectral density $\spD{k}$ of 2D and 3D soft-core SHU ground-state packings at small and large values of $\chi$ (=0.0025, 0.45).
Since these packings consist of nonoverlapping identical spheres of radius $a$ in a matrix, the spectral density $\spD{k}$ is directly related to the structure factor of the particle centers:\cite{torquato_disordered_2016}
\begin{align} \label{eq:spd}
\spD{k} = \phi \fn{\tilde{\alpha}_2}{k; a}\fn{S}{k},
\end{align}
where $\fn{\tilde{\alpha}_2}{k; a} \equiv (2\pi a/k)^d \fn{J_{d/2}}{ka}^2 / v_1(a)$, $\fn{J_\nu}{x}$ is the Bessel function of the first kind of order $\nu$, and $\phi$ is the packing fraction.
Figure \ref{fig:chik} shows the spectral density of the corresponding disk and sphere packings with $\chi=0.0025$ and $\chi=0.45$.

\section*{Supplementary Material}

The \href{...}{supplementary material} provides plots of $P(\rmin;N)$ and $H_P(r)$ for 1D and 3D systems, numerical verification of Eq. \eqref{eq:step2}, details of the probabilistic upper bound on $\phimax(\chi,d)$, and the pair statistics of 1D systems.

\begin{acknowledgments}
  The authors thank Samuel Dawley, Peter Morse, Charles Maher, Murray Skolnick, Carlo Vanoni, and Paul Steinhardt for very helpful discussions.
	This research was sponsored by the U.S. Army Research Office and was accomplished under Cooperative Agreement No. W911NF-22-2-0103.
	Simulations were performed on computational resources managed and supported by the Princeton Institute for Computational Science and Engineering (PICSciE).

\end{acknowledgments}

\section*{Data Availability Statement}

The data that support the findings of this study are available from the corresponding author upon reasonable request.

\bibliography{ref} 

\begin{thebibliography}{84}%
\makeatletter
\providecommand \@ifxundefined [1]{%
 \@ifx{#1\undefined}
}%
\providecommand \@ifnum [1]{%
 \ifnum #1\expandafter \@firstoftwo
 \else \expandafter \@secondoftwo
 \fi
}%
\providecommand \@ifx [1]{%
 \ifx #1\expandafter \@firstoftwo
 \else \expandafter \@secondoftwo
 \fi
}%
\providecommand \natexlab [1]{#1}%
\providecommand \enquote  [1]{``#1''}%
\providecommand \bibnamefont  [1]{#1}%
\providecommand \bibfnamefont [1]{#1}%
\providecommand \citenamefont [1]{#1}%
\providecommand \href@noop [0]{\@secondoftwo}%
\providecommand \href [0]{\begingroup \@sanitize@url \@href}%
\providecommand \@href[1]{\@@startlink{#1}\@@href}%
\providecommand \@@href[1]{\endgroup#1\@@endlink}%
\providecommand \@sanitize@url [0]{\catcode `\\12\catcode `\$12\catcode
  `\&12\catcode `\#12\catcode `\^12\catcode `\_12\catcode `\%12\relax}%
\providecommand \@@startlink[1]{}%
\providecommand \@@endlink[0]{}%
\providecommand \url  [0]{\begingroup\@sanitize@url \@url }%
\providecommand \@url [1]{\endgroup\@href {#1}{\urlprefix }}%
\providecommand \urlprefix  [0]{URL }%
\providecommand \Eprint [0]{\href }%
\providecommand \doibase [0]{http://dx.doi.org/}%
\providecommand \selectlanguage [0]{\@gobble}%
\providecommand \bibinfo  [0]{\@secondoftwo}%
\providecommand \bibfield  [0]{\@secondoftwo}%
\providecommand \translation [1]{[#1]}%
\providecommand \BibitemOpen [0]{}%
\providecommand \bibitemStop [0]{}%
\providecommand \bibitemNoStop [0]{.\EOS\space}%
\providecommand \EOS [0]{\spacefactor3000\relax}%
\providecommand \BibitemShut  [1]{\csname bibitem#1\endcsname}%
\let\auto@bib@innerbib\@empty
\bibitem [{\citenamefont {Torquato}\ and\ \citenamefont
  {Stillinger}(2003)}]{torquato_local_2003}%
  \BibitemOpen
  \bibfield  {author} {\bibinfo {author} {\bibfnamefont {S.}~\bibnamefont
  {Torquato}}\ and\ \bibinfo {author} {\bibfnamefont {F.}~\bibnamefont
  {Stillinger}},\ }\bibfield  {title} {\enquote {\bibinfo {title} {Local
  density fluctuations, hyperuniformity, and order metrics},}\ }\href {\doibase
  10.1103/PhysRevE.68.041113} {\bibfield  {journal} {\bibinfo  {journal} {Phys.
  Rev. E}\ }\textbf {\bibinfo {volume} {68}},\ \bibinfo {pages} {041113}
  (\bibinfo {year} {2003})}\BibitemShut {NoStop}%
\bibitem [{\citenamefont
  {Torquato}(2018{\natexlab{a}})}]{torquato_hyperuniform_2018}%
  \BibitemOpen
  \bibfield  {author} {\bibinfo {author} {\bibfnamefont {S.}~\bibnamefont
  {Torquato}},\ }\bibfield  {title} {\enquote {\bibinfo {title} {Hyperuniform
  states of matter},}\ }\href {\doibase 10.1016/j.physrep.2018.03.001}
  {\bibfield  {journal} {\bibinfo  {journal} {Phys. Rep.}\ }\textbf {\bibinfo
  {volume} {745}},\ \bibinfo {pages} {1--95} (\bibinfo {year}
  {2018}{\natexlab{a}})}\BibitemShut {NoStop}%
\bibitem [{\citenamefont {Hexner}\ and\ \citenamefont
  {Levine}(2015)}]{hexner_hyperuniformity_2015}%
  \BibitemOpen
  \bibfield  {author} {\bibinfo {author} {\bibfnamefont {D.}~\bibnamefont
  {Hexner}}\ and\ \bibinfo {author} {\bibfnamefont {D.}~\bibnamefont
  {Levine}},\ }\bibfield  {title} {\enquote {\bibinfo {title} {Hyperuniformity
  of critical absorbing states},}\ }\href {\doibase
  10.1103/PhysRevLett.114.110602} {\bibfield  {journal} {\bibinfo  {journal}
  {Phys. Rev. Lett.}\ }\textbf {\bibinfo {volume} {114}},\ \bibinfo {pages}
  {110602} (\bibinfo {year} {2015})}\BibitemShut {NoStop}%
\bibitem [{\citenamefont {Jadrich}\ \emph {et~al.}(2016)\citenamefont
  {Jadrich}, \citenamefont {Lindquist}, \citenamefont {Bollinger},\ and\
  \citenamefont {Truskett}}]{jadrich_consequences_2016}%
  \BibitemOpen
  \bibfield  {author} {\bibinfo {author} {\bibfnamefont {R.~B.}\ \bibnamefont
  {Jadrich}}, \bibinfo {author} {\bibfnamefont {B.~A.}\ \bibnamefont
  {Lindquist}}, \bibinfo {author} {\bibfnamefont {J.~A.}\ \bibnamefont
  {Bollinger}}, \ and\ \bibinfo {author} {\bibfnamefont {T.~M.}\ \bibnamefont
  {Truskett}},\ }\bibfield  {title} {\enquote {\bibinfo {title} {Consequences
  of minimising pair correlations in fluids for dynamics, thermodynamics and
  structure},}\ }\href {\doibase 10.1080/00268976.2016.1159742} {\bibfield
  {journal} {\bibinfo  {journal} {Mol. Phys.}\ } (\bibinfo {year} {2016}),\
  10.1080/00268976.2016.1159742}\BibitemShut {NoStop}%
\bibitem [{\citenamefont {Chremos}\ and\ \citenamefont
  {Douglas}(2018)}]{chremos_hidden_2018}%
  \BibitemOpen
  \bibfield  {author} {\bibinfo {author} {\bibfnamefont {A.}~\bibnamefont
  {Chremos}}\ and\ \bibinfo {author} {\bibfnamefont {J.~F.}\ \bibnamefont
  {Douglas}},\ }\bibfield  {title} {\enquote {\bibinfo {title} {Hidden
  hyperuniformity in soft polymeric materials},}\ }\href {\doibase
  10.1103/PhysRevLett.121.258002} {\bibfield  {journal} {\bibinfo  {journal}
  {Phys. Rev. Lett.}\ }\textbf {\bibinfo {volume} {121}},\ \bibinfo {pages}
  {258002} (\bibinfo {year} {2018})}\BibitemShut {NoStop}%
\bibitem [{\citenamefont {Wang}, \citenamefont {Schwarz},\ and\ \citenamefont
  {Paulsen}(2018)}]{wang_hyperuniformity_2018}%
  \BibitemOpen
  \bibfield  {author} {\bibinfo {author} {\bibfnamefont {J.}~\bibnamefont
  {Wang}}, \bibinfo {author} {\bibfnamefont {J.~M.}\ \bibnamefont {Schwarz}}, \
  and\ \bibinfo {author} {\bibfnamefont {J.~D.}\ \bibnamefont {Paulsen}},\
  }\bibfield  {title} {\enquote {\bibinfo {title} {Hyperuniformity with no fine
  tuning in sheared sedimenting suspensions},}\ }\href {\doibase
  10.1038/s41467-018-05195-4} {\bibfield  {journal} {\bibinfo  {journal} {Nat.
  Commun.}\ }\textbf {\bibinfo {volume} {9}},\ \bibinfo {pages} {2836}
  (\bibinfo {year} {2018})}\BibitemShut {NoStop}%
\bibitem [{\citenamefont {Zhuravlyov}\ \emph {et~al.}(2022)\citenamefont
  {Zhuravlyov}, \citenamefont {Goree}, \citenamefont {Douglas}, \citenamefont
  {Elvati},\ and\ \citenamefont {Violi}}]{zhuravlyov_comparison_2022}%
  \BibitemOpen
  \bibfield  {author} {\bibinfo {author} {\bibfnamefont {V.}~\bibnamefont
  {Zhuravlyov}}, \bibinfo {author} {\bibfnamefont {J.}~\bibnamefont {Goree}},
  \bibinfo {author} {\bibfnamefont {J.~F.}\ \bibnamefont {Douglas}}, \bibinfo
  {author} {\bibfnamefont {P.}~\bibnamefont {Elvati}}, \ and\ \bibinfo {author}
  {\bibfnamefont {A.}~\bibnamefont {Violi}},\ }\bibfield  {title} {\enquote
  {\bibinfo {title} {Comparison of the static structure factor at long
  wavelengths for a dusty plasma liquid and other liquids},}\ }\href {\doibase
  10.1103/PhysRevE.106.055212} {\bibfield  {journal} {\bibinfo  {journal}
  {Phys. Rev. E}\ }\textbf {\bibinfo {volume} {106}},\ \bibinfo {pages}
  {055212} (\bibinfo {year} {2022})}\BibitemShut {NoStop}%
\bibitem [{\citenamefont {Oppenheimer}\ \emph {et~al.}(2022)\citenamefont
  {Oppenheimer}, \citenamefont {Stein}, \citenamefont {Zion},\ and\
  \citenamefont {Shelley}}]{oppenheimer_hyperuniformity_2022}%
  \BibitemOpen
  \bibfield  {author} {\bibinfo {author} {\bibfnamefont {N.}~\bibnamefont
  {Oppenheimer}}, \bibinfo {author} {\bibfnamefont {D.~B.}\ \bibnamefont
  {Stein}}, \bibinfo {author} {\bibfnamefont {M.~Y.~B.}\ \bibnamefont {Zion}},
  \ and\ \bibinfo {author} {\bibfnamefont {M.~J.}\ \bibnamefont {Shelley}},\
  }\bibfield  {title} {\enquote {\bibinfo {title} {Hyperuniformity and phase
  enrichment in vortex and rotor assemblies},}\ }\href {\doibase
  10.1038/s41467-022-28375-9} {\bibfield  {journal} {\bibinfo  {journal} {Nat
  Commun}\ }\textbf {\bibinfo {volume} {13}},\ \bibinfo {pages} {804} (\bibinfo
  {year} {2022})}\BibitemShut {NoStop}%
\bibitem [{\citenamefont {Onishi}\ and\ \citenamefont
  {Fu}(2024)}]{onishi_topological_2024}%
  \BibitemOpen
  \bibfield  {author} {\bibinfo {author} {\bibfnamefont {Y.}~\bibnamefont
  {Onishi}}\ and\ \bibinfo {author} {\bibfnamefont {L.}~\bibnamefont {Fu}},\
  }\bibfield  {title} {\enquote {\bibinfo {title} {Topological bound on the
  structure factor},}\ }\href {\doibase 10.1103/PhysRevLett.133.206602}
  {\bibfield  {journal} {\bibinfo  {journal} {Phys. Rev. Lett.}\ }\textbf
  {\bibinfo {volume} {133}},\ \bibinfo {pages} {206602} (\bibinfo {year}
  {2024})}\BibitemShut {NoStop}%
\bibitem [{\citenamefont {Torquato}(2022)}]{torquato_extraordinary_2022}%
  \BibitemOpen
  \bibfield  {author} {\bibinfo {author} {\bibfnamefont {S.}~\bibnamefont
  {Torquato}},\ }\bibfield  {title} {\enquote {\bibinfo {title} {Extraordinary
  disordered hyperuniform multifunctional composites},}\ }\href {\doibase
  10.1177/00219983221116432} {\bibfield  {journal} {\bibinfo  {journal} {J.
  Comp. Mater.}\ }\textbf {\bibinfo {volume} {56}},\ \bibinfo {pages}
  {3635--3649} (\bibinfo {year} {2022})}\BibitemShut {NoStop}%
\bibitem [{\citenamefont {Uche}, \citenamefont {Stillinger},\ and\
  \citenamefont {Torquato}(2004)}]{uche_constraints_2004}%
  \BibitemOpen
  \bibfield  {author} {\bibinfo {author} {\bibfnamefont {O.}~\bibnamefont
  {Uche}}, \bibinfo {author} {\bibfnamefont {F.}~\bibnamefont {Stillinger}}, \
  and\ \bibinfo {author} {\bibfnamefont {S.}~\bibnamefont {Torquato}},\
  }\bibfield  {title} {\enquote {\bibinfo {title} {Constraints on collective
  density variables: Two dimensions},}\ }\href {\doibase
  10.1103/PhysRevE.70.046122} {\bibfield  {journal} {\bibinfo  {journal} {Phys.
  Rev. E}\ }\textbf {\bibinfo {volume} {70}},\ \bibinfo {pages} {046122}
  (\bibinfo {year} {2004})}\BibitemShut {NoStop}%
\bibitem [{\citenamefont {Zhang}, \citenamefont {Stillinger},\ and\
  \citenamefont {Torquato}(2015)}]{zhang_ground_2015}%
  \BibitemOpen
  \bibfield  {author} {\bibinfo {author} {\bibfnamefont {G.}~\bibnamefont
  {Zhang}}, \bibinfo {author} {\bibfnamefont {F.}~\bibnamefont {Stillinger}}, \
  and\ \bibinfo {author} {\bibfnamefont {S.}~\bibnamefont {Torquato}},\
  }\bibfield  {title} {\enquote {\bibinfo {title} {Ground states of stealthy
  hyperuniform potentials: I. entropically favored configurations},}\ }\href
  {\doibase 10.1103/PhysRevE.92.022119} {\bibfield  {journal} {\bibinfo
  {journal} {Phys. Rev. E}\ }\textbf {\bibinfo {volume} {92}},\ \bibinfo
  {pages} {022119} (\bibinfo {year} {2015})}\BibitemShut {NoStop}%
\bibitem [{\citenamefont {Torquato}, \citenamefont {Zhang},\ and\ \citenamefont
  {Stillinger}(2015)}]{torquato_ensemble_2015}%
  \BibitemOpen
  \bibfield  {author} {\bibinfo {author} {\bibfnamefont {S.}~\bibnamefont
  {Torquato}}, \bibinfo {author} {\bibfnamefont {G.}~\bibnamefont {Zhang}}, \
  and\ \bibinfo {author} {\bibfnamefont {F.~H.}\ \bibnamefont {Stillinger}},\
  }\bibfield  {title} {\enquote {\bibinfo {title} {Ensemble theory for stealthy
  hyperuniform disordered ground states},}\ }\href {\doibase
  10.1103/PhysRevX.5.021020} {\bibfield  {journal} {\bibinfo  {journal} {Phys.
  Rev. X}\ }\textbf {\bibinfo {volume} {5}},\ \bibinfo {pages} {021020}
  (\bibinfo {year} {2015})}\BibitemShut {NoStop}%
\bibitem [{\citenamefont {Batten}, \citenamefont {Stillinger},\ and\
  \citenamefont {Torquato}(2008)}]{batten_classical_2008}%
  \BibitemOpen
  \bibfield  {author} {\bibinfo {author} {\bibfnamefont {R.}~\bibnamefont
  {Batten}}, \bibinfo {author} {\bibfnamefont {F.}~\bibnamefont {Stillinger}},
  \ and\ \bibinfo {author} {\bibfnamefont {S.}~\bibnamefont {Torquato}},\
  }\bibfield  {title} {\enquote {\bibinfo {title} {Classical disordered ground
  states: Super-ideal gases and stealth and equi-luminous materials},}\ }\href
  {\doibase 10.1063/1.2961314} {\bibfield  {journal} {\bibinfo  {journal} {J.
  Appl. Phys.}\ }\textbf {\bibinfo {volume} {104}},\ \bibinfo {pages} {033504}
  (\bibinfo {year} {2008})}\BibitemShut {NoStop}%
\bibitem [{\citenamefont {Florescu}, \citenamefont {Torquato},\ and\
  \citenamefont {Steinhardt}(2009)}]{florescu_designer_2009}%
  \BibitemOpen
  \bibfield  {author} {\bibinfo {author} {\bibfnamefont {M.}~\bibnamefont
  {Florescu}}, \bibinfo {author} {\bibfnamefont {S.}~\bibnamefont {Torquato}},
  \ and\ \bibinfo {author} {\bibfnamefont {P.}~\bibnamefont {Steinhardt}},\
  }\bibfield  {title} {\enquote {\bibinfo {title} {Designer disordered
  materials with large, complete photonic band gaps},}\ }\href {\doibase
  10.1073/pnas.0907744106} {\bibfield  {journal} {\bibinfo  {journal} {Proc.
  Natl. Acad. Sci. U.S.A.}\ }\textbf {\bibinfo {volume} {106}},\ \bibinfo
  {pages} {20658--20663} (\bibinfo {year} {2009})}\BibitemShut {NoStop}%
\bibitem [{\citenamefont {Aeby}\ \emph {et~al.}(2022)\citenamefont {Aeby},
  \citenamefont {Aubry}, \citenamefont {{Froufe-P{\'e}rez}},\ and\
  \citenamefont {Scheffold}}]{aeby_fabrication_2022}%
  \BibitemOpen
  \bibfield  {author} {\bibinfo {author} {\bibfnamefont {S.}~\bibnamefont
  {Aeby}}, \bibinfo {author} {\bibfnamefont {G.~J.}\ \bibnamefont {Aubry}},
  \bibinfo {author} {\bibfnamefont {L.~S.}\ \bibnamefont {{Froufe-P{\'e}rez}}},
  \ and\ \bibinfo {author} {\bibfnamefont {F.}~\bibnamefont {Scheffold}},\
  }\bibfield  {title} {\enquote {\bibinfo {title} {Fabrication of hyperuniform
  dielectric networks via heat-induced shrinkage reveals a bandgap at telecom
  wavelengths},}\ }\href {\doibase 10.1002/adom.202200232} {\bibfield
  {journal} {\bibinfo  {journal} {Adv. Opt. Mater.}\ }\textbf {\bibinfo
  {volume} {2022}},\ \bibinfo {pages} {2200232} (\bibinfo {year}
  {2022})}\BibitemShut {NoStop}%
\bibitem [{\citenamefont {Torquato}\ and\ \citenamefont
  {Chen}(2018)}]{torquato_multifunctional_2018}%
  \BibitemOpen
  \bibfield  {author} {\bibinfo {author} {\bibfnamefont {S.}~\bibnamefont
  {Torquato}}\ and\ \bibinfo {author} {\bibfnamefont {D.}~\bibnamefont
  {Chen}},\ }\bibfield  {title} {\enquote {\bibinfo {title} {Multifunctional
  hyperuniform cellular networks: optimality, anisotropy and disorder},}\
  }\href {\doibase 10.1088/2399-7532/aaca91} {\bibfield  {journal} {\bibinfo
  {journal} {Multifunct. Mater.}\ }\textbf {\bibinfo {volume} {1}},\ \bibinfo
  {pages} {015001} (\bibinfo {year} {2018})}\BibitemShut {NoStop}%
\bibitem [{\citenamefont {Zhou}\ \emph {et~al.}(2020)\citenamefont {Zhou},
  \citenamefont {Tong}, \citenamefont {Sun},\ and\ \citenamefont
  {Tsang}}]{zhou_ultrabroadband_2020}%
  \BibitemOpen
  \bibfield  {author} {\bibinfo {author} {\bibfnamefont {W.}~\bibnamefont
  {Zhou}}, \bibinfo {author} {\bibfnamefont {Y.}~\bibnamefont {Tong}}, \bibinfo
  {author} {\bibfnamefont {X.}~\bibnamefont {Sun}}, \ and\ \bibinfo {author}
  {\bibfnamefont {H.~K.}\ \bibnamefont {Tsang}},\ }\bibfield  {title} {\enquote
  {\bibinfo {title} {Ultra-broadband hyperuniform disordered silicon photonic
  polarizers},}\ }\href {\doibase 10.1109/JSTQE.2019.2938069} {\bibfield
  {journal} {\bibinfo  {journal} {IEEE J. Sel. Top. Quantum Electron.}\
  }\textbf {\bibinfo {volume} {26}},\ \bibinfo {pages} {1--9} (\bibinfo {year}
  {2020})}\BibitemShut {NoStop}%
\bibitem [{\citenamefont {Klatt}, \citenamefont {Steinhardt},\ and\
  \citenamefont {Torquato}(2022)}]{klatt_wave_2022}%
  \BibitemOpen
  \bibfield  {author} {\bibinfo {author} {\bibfnamefont {M.~A.}\ \bibnamefont
  {Klatt}}, \bibinfo {author} {\bibfnamefont {P.~J.}\ \bibnamefont
  {Steinhardt}}, \ and\ \bibinfo {author} {\bibfnamefont {S.}~\bibnamefont
  {Torquato}},\ }\bibfield  {title} {\enquote {\bibinfo {title} {Wave
  propagation and band tails of two-dimensional disordered systems in the
  thermodynamic limit},}\ }\href {\doibase 10.1073/pnas.2213633119} {\bibfield
  {journal} {\bibinfo  {journal} {Proc. Natl. Acad. Sci. U.S.A.}\ }\textbf
  {\bibinfo {volume} {119}},\ \bibinfo {pages} {e2213633119} (\bibinfo {year}
  {2022})}\BibitemShut {NoStop}%
\bibitem [{\citenamefont {Granchi}\ \emph {et~al.}(2022)\citenamefont
  {Granchi}, \citenamefont {Spalding}, \citenamefont {Lodde}, \citenamefont
  {Petruzzella}, \citenamefont {Otten}, \citenamefont {Fiore}, \citenamefont
  {Intonti}, \citenamefont {Sapienza}, \citenamefont {Florescu},\ and\
  \citenamefont {Gurioli}}]{granchi_nearfield_2022}%
  \BibitemOpen
  \bibfield  {author} {\bibinfo {author} {\bibfnamefont {N.}~\bibnamefont
  {Granchi}}, \bibinfo {author} {\bibfnamefont {R.}~\bibnamefont {Spalding}},
  \bibinfo {author} {\bibfnamefont {M.}~\bibnamefont {Lodde}}, \bibinfo
  {author} {\bibfnamefont {M.}~\bibnamefont {Petruzzella}}, \bibinfo {author}
  {\bibfnamefont {F.~W.}\ \bibnamefont {Otten}}, \bibinfo {author}
  {\bibfnamefont {A.}~\bibnamefont {Fiore}}, \bibinfo {author} {\bibfnamefont
  {F.}~\bibnamefont {Intonti}}, \bibinfo {author} {\bibfnamefont
  {R.}~\bibnamefont {Sapienza}}, \bibinfo {author} {\bibfnamefont
  {M.}~\bibnamefont {Florescu}}, \ and\ \bibinfo {author} {\bibfnamefont
  {M.}~\bibnamefont {Gurioli}},\ }\bibfield  {title} {\enquote {\bibinfo
  {title} {Near-field investigation of luminescent hyperuniform disordered
  materials},}\ }\href {\doibase 10.1002/adom.202102565} {\bibfield  {journal}
  {\bibinfo  {journal} {Adv. Opt. Mater.}\ }\textbf {\bibinfo {volume} {10}},\
  \bibinfo {pages} {2102565} (\bibinfo {year} {2022})}\BibitemShut {NoStop}%
\bibitem [{\citenamefont {Zhang}, \citenamefont {Stillinger},\ and\
  \citenamefont {Torquato}(2016{\natexlab{a}})}]{zhang_transport_2016}%
  \BibitemOpen
  \bibfield  {author} {\bibinfo {author} {\bibfnamefont {G.}~\bibnamefont
  {Zhang}}, \bibinfo {author} {\bibfnamefont {F.}~\bibnamefont {Stillinger}}, \
  and\ \bibinfo {author} {\bibfnamefont {S.}~\bibnamefont {Torquato}},\
  }\bibfield  {title} {\enquote {\bibinfo {title} {Transport, geometrical, and
  topological properties of stealthy disordered hyperuniform two-phase
  systems},}\ }\href {\doibase 10.1063/1.4972862} {\bibfield  {journal}
  {\bibinfo  {journal} {J. Chem. Phys.}\ }\textbf {\bibinfo {volume} {145}},\
  \bibinfo {pages} {244109} (\bibinfo {year} {2016}{\natexlab{a}})}\BibitemShut
  {NoStop}%
\bibitem [{\citenamefont {Kim}\ and\ \citenamefont
  {Torquato}(2020)}]{kim_multifunctional_2020}%
  \BibitemOpen
  \bibfield  {author} {\bibinfo {author} {\bibfnamefont {J.}~\bibnamefont
  {Kim}}\ and\ \bibinfo {author} {\bibfnamefont {S.}~\bibnamefont {Torquato}},\
  }\bibfield  {title} {\enquote {\bibinfo {title} {Multifunctional composites
  for elastic and electromagnetic wave propagation},}\ }\href {\doibase
  10.1073/pnas.1914086117} {\bibfield  {journal} {\bibinfo  {journal} {Proc.
  Natl. Acad. Sci. U.S.A.}\ }\textbf {\bibinfo {volume} {117}},\ \bibinfo
  {pages} {8764--8774} (\bibinfo {year} {2020})},\ \Eprint
  {http://arxiv.org/abs/1908.06662} {arXiv:1908.06662} \BibitemShut {NoStop}%
\bibitem [{\citenamefont {Romero-Garc\'{i}a}\ \emph {et~al.}(2019)\citenamefont
  {Romero-Garc\'{i}a}, \citenamefont {Lamothe}, \citenamefont {Theocharis},
  \citenamefont {Richoux},\ and\ \citenamefont
  {Garc\'{i}a-Raffi}}]{romero-garcia_stealth_2019}%
  \BibitemOpen
  \bibfield  {author} {\bibinfo {author} {\bibfnamefont {V.}~\bibnamefont
  {Romero-Garc\'{i}a}}, \bibinfo {author} {\bibfnamefont {N.}~\bibnamefont
  {Lamothe}}, \bibinfo {author} {\bibfnamefont {G.}~\bibnamefont {Theocharis}},
  \bibinfo {author} {\bibfnamefont {O.}~\bibnamefont {Richoux}}, \ and\
  \bibinfo {author} {\bibfnamefont {L.~M.}\ \bibnamefont {Garc\'{i}a-Raffi}},\
  }\bibfield  {title} {\enquote {\bibinfo {title} {Stealth acoustic
  materials},}\ }\href {\doibase 10.1103/PhysRevApplied.11.054076} {\bibfield
  {journal} {\bibinfo  {journal} {Phys. Rev. Applied}\ }\textbf {\bibinfo
  {volume} {11}},\ \bibinfo {pages} {054076} (\bibinfo {year}
  {2019})}\BibitemShut {NoStop}%
\bibitem [{\citenamefont {Kim}\ and\ \citenamefont
  {Torquato}(2023)}]{kim_effective_2023}%
  \BibitemOpen
  \bibfield  {author} {\bibinfo {author} {\bibfnamefont {J.}~\bibnamefont
  {Kim}}\ and\ \bibinfo {author} {\bibfnamefont {S.}~\bibnamefont {Torquato}},\
  }\bibfield  {title} {\enquote {\bibinfo {title} {Effective electromagnetic
  wave properties of disordered stealthy hyperuniform layered media beyond the
  quasistatic regime},}\ }\href {\doibase 10.1364/optica.489797} {\bibfield
  {journal} {\bibinfo  {journal} {Optica}\ }\textbf {\bibinfo {volume} {10}},\
  \bibinfo {pages} {965--972} (\bibinfo {year} {2023})}\BibitemShut {NoStop}%
\bibitem [{\citenamefont {{Froufe-P{\'e}rez}}\ \emph
  {et~al.}(2023)\citenamefont {{Froufe-P{\'e}rez}}, \citenamefont {Aubry},
  \citenamefont {Scheffold},\ and\ \citenamefont
  {Magkiriadou}}]{froufe-perez_bandgap_2023}%
  \BibitemOpen
  \bibfield  {author} {\bibinfo {author} {\bibfnamefont {L.~S.}\ \bibnamefont
  {{Froufe-P{\'e}rez}}}, \bibinfo {author} {\bibfnamefont {G.~J.}\ \bibnamefont
  {Aubry}}, \bibinfo {author} {\bibfnamefont {F.}~\bibnamefont {Scheffold}}, \
  and\ \bibinfo {author} {\bibfnamefont {S.}~\bibnamefont {Magkiriadou}},\
  }\bibfield  {title} {\enquote {\bibinfo {title} {Bandgap fluctuations and
  robustness in two-dimensional hyperuniform dielectric materials},}\ }\href
  {\doibase 10.1364/OE.484232} {\bibfield  {journal} {\bibinfo  {journal}
  {Optics Express}\ }\textbf {\bibinfo {volume} {31}},\ \bibinfo {pages}
  {18509--18515} (\bibinfo {year} {2023})}\BibitemShut {NoStop}%
\bibitem [{\citenamefont {Alha{\"i}tz}, \citenamefont {Conoir},\ and\
  \citenamefont {{Valier-Brasier}}(2023)}]{alhaitz_experimental_2023}%
  \BibitemOpen
  \bibfield  {author} {\bibinfo {author} {\bibfnamefont {L.}~\bibnamefont
  {Alha{\"i}tz}}, \bibinfo {author} {\bibfnamefont {J.-M.}\ \bibnamefont
  {Conoir}}, \ and\ \bibinfo {author} {\bibfnamefont {T.}~\bibnamefont
  {{Valier-Brasier}}},\ }\bibfield  {title} {\enquote {\bibinfo {title}
  {Experimental evidence of isotropic transparency and complete band gap
  formation for ultrasound propagation in stealthy hyperuniform media},}\
  }\href {\doibase 10.1103/PhysRevE.108.065001} {\bibfield  {journal} {\bibinfo
   {journal} {Phy. Rev. E}\ }\textbf {\bibinfo {volume} {108}},\ \bibinfo
  {pages} {065001} (\bibinfo {year} {2023})}\BibitemShut {NoStop}%
\bibitem [{\citenamefont {Bigourdan}, \citenamefont {Pierrat},\ and\
  \citenamefont {Carminati}(2019)}]{bigourdan_enhanced_2019}%
  \BibitemOpen
  \bibfield  {author} {\bibinfo {author} {\bibfnamefont {F.}~\bibnamefont
  {Bigourdan}}, \bibinfo {author} {\bibfnamefont {R.}~\bibnamefont {Pierrat}},
  \ and\ \bibinfo {author} {\bibfnamefont {R.}~\bibnamefont {Carminati}},\
  }\bibfield  {title} {\enquote {\bibinfo {title} {Enhanced absorption of waves
  in stealth hyperuniform disordered media},}\ }\href {\doibase
  10.1364/OE.27.008666} {\bibfield  {journal} {\bibinfo  {journal} {Opt.
  Express}\ }\textbf {\bibinfo {volume} {27}},\ \bibinfo {pages} {8666--8682}
  (\bibinfo {year} {2019})}\BibitemShut {NoStop}%
\bibitem [{\citenamefont {{Froufe-P{\'e}rez}}\ \emph
  {et~al.}(2017)\citenamefont {{Froufe-P{\'e}rez}}, \citenamefont {Engel},
  \citenamefont {S{\'a}enz},\ and\ \citenamefont
  {Scheffold}}]{froufe-perez_band_2017}%
  \BibitemOpen
  \bibfield  {author} {\bibinfo {author} {\bibfnamefont {L.}~\bibnamefont
  {{Froufe-P{\'e}rez}}}, \bibinfo {author} {\bibfnamefont {M.}~\bibnamefont
  {Engel}}, \bibinfo {author} {\bibfnamefont {J.}~\bibnamefont {S{\'a}enz}}, \
  and\ \bibinfo {author} {\bibfnamefont {F.}~\bibnamefont {Scheffold}},\
  }\bibfield  {title} {\enquote {\bibinfo {title} {Band gap formation and
  anderson localization in disordered photonic materials with structural
  correlations},}\ }\href {\doibase 10.1073/pnas.1705130114} {\bibfield
  {journal} {\bibinfo  {journal} {Proc. Natl. Acad. Sci. U.S.A.}\ }\textbf
  {\bibinfo {volume} {114}},\ \bibinfo {pages} {9570--9574} (\bibinfo {year}
  {2017})}\BibitemShut {NoStop}%
\bibitem [{\citenamefont {Sgrignuoli}, \citenamefont {Torquato},\ and\
  \citenamefont {Dal~Negro}(2022)}]{sgrignuoli_subdiffusive_2022}%
  \BibitemOpen
  \bibfield  {author} {\bibinfo {author} {\bibfnamefont {F.}~\bibnamefont
  {Sgrignuoli}}, \bibinfo {author} {\bibfnamefont {S.}~\bibnamefont
  {Torquato}}, \ and\ \bibinfo {author} {\bibfnamefont {L.}~\bibnamefont
  {Dal~Negro}},\ }\bibfield  {title} {\enquote {\bibinfo {title} {Subdiffusive
  wave transport and weak localization transition in three-dimensional stealthy
  hyperuniform disordered systems},}\ }\href {\doibase
  10.1103/PhysRevB.105.064204} {\bibfield  {journal} {\bibinfo  {journal}
  {Phys. Rev. B}\ }\textbf {\bibinfo {volume} {105}},\ \bibinfo {pages}
  {064204} (\bibinfo {year} {2022})}\BibitemShut {NoStop}%
\bibitem [{\citenamefont {Tavakoli}\ \emph {et~al.}(2022)\citenamefont
  {Tavakoli}, \citenamefont {Spalding}, \citenamefont {Lambertz}, \citenamefont
  {Koppejan}, \citenamefont {Gkantzounis}, \citenamefont {Wan}, \citenamefont
  {R{\"o}hrich}, \citenamefont {Kontoleta}, \citenamefont {Koenderink},
  \citenamefont {Sapienza}, \citenamefont {Florescu},\ and\ \citenamefont
  {{Alarcon-Llado}}}]{tavakoli_65_2022}%
  \BibitemOpen
  \bibfield  {author} {\bibinfo {author} {\bibfnamefont {N.}~\bibnamefont
  {Tavakoli}}, \bibinfo {author} {\bibfnamefont {R.}~\bibnamefont {Spalding}},
  \bibinfo {author} {\bibfnamefont {A.}~\bibnamefont {Lambertz}}, \bibinfo
  {author} {\bibfnamefont {P.}~\bibnamefont {Koppejan}}, \bibinfo {author}
  {\bibfnamefont {G.}~\bibnamefont {Gkantzounis}}, \bibinfo {author}
  {\bibfnamefont {C.}~\bibnamefont {Wan}}, \bibinfo {author} {\bibfnamefont
  {R.}~\bibnamefont {R{\"o}hrich}}, \bibinfo {author} {\bibfnamefont
  {E.}~\bibnamefont {Kontoleta}}, \bibinfo {author} {\bibfnamefont {A.~F.}\
  \bibnamefont {Koenderink}}, \bibinfo {author} {\bibfnamefont
  {R.}~\bibnamefont {Sapienza}}, \bibinfo {author} {\bibfnamefont
  {M.}~\bibnamefont {Florescu}}, \ and\ \bibinfo {author} {\bibfnamefont
  {E.}~\bibnamefont {{Alarcon-Llado}}},\ }\bibfield  {title} {\enquote
  {\bibinfo {title} {Over 65\% sunlight absorption in a 1 {$\mu$}m si slab with
  hyperuniform texture},}\ }\href {\doibase 10.1021/acsphotonics.1c01668}
  {\bibfield  {journal} {\bibinfo  {journal} {ACS Photonics}\ }\textbf
  {\bibinfo {volume} {9}},\ \bibinfo {pages} {1206--1217} (\bibinfo {year}
  {2022})}\BibitemShut {NoStop}%
\bibitem [{\citenamefont {Merkel}\ \emph {et~al.}(2023)\citenamefont {Merkel},
  \citenamefont {Stappers}, \citenamefont {Ray}, \citenamefont {Denz},\ and\
  \citenamefont {Imbrock}}]{merkel_stealthy_2023}%
  \BibitemOpen
  \bibfield  {author} {\bibinfo {author} {\bibfnamefont {M.}~\bibnamefont
  {Merkel}}, \bibinfo {author} {\bibfnamefont {M.}~\bibnamefont {Stappers}},
  \bibinfo {author} {\bibfnamefont {D.}~\bibnamefont {Ray}}, \bibinfo {author}
  {\bibfnamefont {C.}~\bibnamefont {Denz}}, \ and\ \bibinfo {author}
  {\bibfnamefont {J.}~\bibnamefont {Imbrock}},\ }\bibfield  {title} {\enquote
  {\bibinfo {title} {Stealthy hyperuniform surface structures for efficiency
  enhancement of organic solar cells},}\ }\href {\doibase
  10.1002/adpr.202300256} {\bibfield  {journal} {\bibinfo  {journal} {Adv.
  Photonics Res.}\ }\textbf {\bibinfo {volume} {n/a}},\ \bibinfo {pages}
  {2300256} (\bibinfo {year} {2023})}\BibitemShut {NoStop}%
\bibitem [{\citenamefont {Gkantzounis}, \citenamefont {Amoah},\ and\
  \citenamefont {Florescu}(2017)}]{gkantzounis_hyperuniform_2017}%
  \BibitemOpen
  \bibfield  {author} {\bibinfo {author} {\bibfnamefont {G.}~\bibnamefont
  {Gkantzounis}}, \bibinfo {author} {\bibfnamefont {T.}~\bibnamefont {Amoah}},
  \ and\ \bibinfo {author} {\bibfnamefont {M.}~\bibnamefont {Florescu}},\
  }\bibfield  {title} {\enquote {\bibinfo {title} {Hyperuniform disordered
  phononic structures},}\ }\href {\doibase 10.1103/PhysRevB.95.094120}
  {\bibfield  {journal} {\bibinfo  {journal} {Phys. Rev. B}\ }\textbf {\bibinfo
  {volume} {95}},\ \bibinfo {pages} {094120} (\bibinfo {year}
  {2017})}\BibitemShut {NoStop}%
\bibitem [{\citenamefont {Gkantzounis}\ and\ \citenamefont
  {Florescu}(2017)}]{gkantzounis_freeform_2017}%
  \BibitemOpen
  \bibfield  {author} {\bibinfo {author} {\bibfnamefont {G.}~\bibnamefont
  {Gkantzounis}}\ and\ \bibinfo {author} {\bibfnamefont {M.}~\bibnamefont
  {Florescu}},\ }\bibfield  {title} {\enquote {\bibinfo {title} {Freeform
  phononic waveguides},}\ }\href {\doibase 10.3390/cryst7120353} {\bibfield
  {journal} {\bibinfo  {journal} {Crystals}\ }\textbf {\bibinfo {volume} {7}},\
  \bibinfo {pages} {353} (\bibinfo {year} {2017})}\BibitemShut {NoStop}%
\bibitem [{\citenamefont {{Romero-Garc{\'i}a}}\ \emph
  {et~al.}(2021)\citenamefont {{Romero-Garc{\'i}a}}, \citenamefont
  {Ch{\'e}ron}, \citenamefont {Kuznetsova}, \citenamefont {Groby},
  \citenamefont {F{\'e}lix}, \citenamefont {Pagneux},\ and\ \citenamefont
  {{Garcia-Raffi}}}]{romero-garcia_wave_2021}%
  \BibitemOpen
  \bibfield  {author} {\bibinfo {author} {\bibfnamefont {V.}~\bibnamefont
  {{Romero-Garc{\'i}a}}}, \bibinfo {author} {\bibfnamefont
  {{\'E}.}~\bibnamefont {Ch{\'e}ron}}, \bibinfo {author} {\bibfnamefont
  {S.}~\bibnamefont {Kuznetsova}}, \bibinfo {author} {\bibfnamefont {J.-P.}\
  \bibnamefont {Groby}}, \bibinfo {author} {\bibfnamefont {S.}~\bibnamefont
  {F{\'e}lix}}, \bibinfo {author} {\bibfnamefont {V.}~\bibnamefont {Pagneux}},
  \ and\ \bibinfo {author} {\bibfnamefont {L.~M.}\ \bibnamefont
  {{Garcia-Raffi}}},\ }\bibfield  {title} {\enquote {\bibinfo {title} {Wave
  transport in 1d stealthy hyperuniform phononic materials made of non-resonant
  and resonant scatterers},}\ }\href {\doibase 10.1063/5.0059928} {\bibfield
  {journal} {\bibinfo  {journal} {APL Materials}\ }\textbf {\bibinfo {volume}
  {9}},\ \bibinfo {pages} {101101} (\bibinfo {year} {2021})}\BibitemShut
  {NoStop}%
\bibitem [{\citenamefont {Rohfritsch}\ \emph {et~al.}(2020)\citenamefont
  {Rohfritsch}, \citenamefont {Conoir}, \citenamefont {{Valier-Brasier}},\ and\
  \citenamefont {Marchiano}}]{rohfritsch_impact_2020}%
  \BibitemOpen
  \bibfield  {author} {\bibinfo {author} {\bibfnamefont {A.}~\bibnamefont
  {Rohfritsch}}, \bibinfo {author} {\bibfnamefont {J.-M.}\ \bibnamefont
  {Conoir}}, \bibinfo {author} {\bibfnamefont {T.}~\bibnamefont
  {{Valier-Brasier}}}, \ and\ \bibinfo {author} {\bibfnamefont
  {R.}~\bibnamefont {Marchiano}},\ }\bibfield  {title} {\enquote {\bibinfo
  {title} {Impact of particle size and multiple scattering on the propagation
  of waves in stealthy-hyperuniform media},}\ }\href {\doibase
  10.1103/PhysRevE.102.053001} {\bibfield  {journal} {\bibinfo  {journal}
  {Phys. Rev. E}\ }\textbf {\bibinfo {volume} {102}},\ \bibinfo {pages}
  {053001} (\bibinfo {year} {2020})}\BibitemShut {NoStop}%
\bibitem [{\citenamefont {Ch\'{e}ron}\ \emph {et~al.}(2022)\citenamefont
  {Ch\'{e}ron}, \citenamefont {Groby}, \citenamefont {Pagneux}, \citenamefont
  {Félix},\ and\ \citenamefont
  {Romero-Garc\'{i}a}}]{cheron_experimental_2022}%
  \BibitemOpen
  \bibfield  {author} {\bibinfo {author} {\bibfnamefont {E.}~\bibnamefont
  {Ch\'{e}ron}}, \bibinfo {author} {\bibfnamefont {J.-P.}\ \bibnamefont
  {Groby}}, \bibinfo {author} {\bibfnamefont {V.}~\bibnamefont {Pagneux}},
  \bibinfo {author} {\bibfnamefont {S.}~\bibnamefont {Félix}}, \ and\ \bibinfo
  {author} {\bibfnamefont {V.}~\bibnamefont {Romero-Garc\'{i}a}},\ }\bibfield
  {title} {\enquote {\bibinfo {title} {Experimental characterization of
  rigid-scatterer hyperuniform distributions for audible acoustics},}\ }\href
  {\doibase 10.1103/PhysRevB.106.064206} {\bibfield  {journal} {\bibinfo
  {journal} {Phys. Rev. B}\ }\textbf {\bibinfo {volume} {106}},\ \bibinfo
  {pages} {064206} (\bibinfo {year} {2022})}\BibitemShut {NoStop}%
\bibitem [{\citenamefont {Zhuang}\ \emph {et~al.}(2024)\citenamefont {Zhuang},
  \citenamefont {Chen}, \citenamefont {Liu}, \citenamefont {Keeney},
  \citenamefont {Zhang},\ and\ \citenamefont {Jiao}}]{zhuang_vibrational_2024}%
  \BibitemOpen
  \bibfield  {author} {\bibinfo {author} {\bibfnamefont {H.}~\bibnamefont
  {Zhuang}}, \bibinfo {author} {\bibfnamefont {D.}~\bibnamefont {Chen}},
  \bibinfo {author} {\bibfnamefont {L.}~\bibnamefont {Liu}}, \bibinfo {author}
  {\bibfnamefont {D.}~\bibnamefont {Keeney}}, \bibinfo {author} {\bibfnamefont
  {G.}~\bibnamefont {Zhang}}, \ and\ \bibinfo {author} {\bibfnamefont
  {Y.}~\bibnamefont {Jiao}},\ }\bibfield  {title} {\enquote {\bibinfo {title}
  {Vibrational properties of disordered stealthy hyperuniform 1d atomic
  chains},}\ }\href {\doibase 10.1088/1361-648X/ad3b5c} {\bibfield  {journal}
  {\bibinfo  {journal} {J. Phys.: Condens. Matter}\ }\textbf {\bibinfo {volume}
  {36}},\ \bibinfo {pages} {285703} (\bibinfo {year} {2024})}\BibitemShut
  {NoStop}%
\bibitem [{\citenamefont {Castro-Lopez}\ \emph {et~al.}(2017)\citenamefont
  {Castro-Lopez}, \citenamefont {Gaio}, \citenamefont {Sellers}, \citenamefont
  {Gkantzounis}, \citenamefont {Florescu},\ and\ \citenamefont
  {Sapienza}}]{castro-lopez_reciprocal_2017}%
  \BibitemOpen
  \bibfield  {author} {\bibinfo {author} {\bibfnamefont {M.}~\bibnamefont
  {Castro-Lopez}}, \bibinfo {author} {\bibfnamefont {M.}~\bibnamefont {Gaio}},
  \bibinfo {author} {\bibfnamefont {S.}~\bibnamefont {Sellers}}, \bibinfo
  {author} {\bibfnamefont {G.}~\bibnamefont {Gkantzounis}}, \bibinfo {author}
  {\bibfnamefont {M.}~\bibnamefont {Florescu}}, \ and\ \bibinfo {author}
  {\bibfnamefont {R.}~\bibnamefont {Sapienza}},\ }\bibfield  {title} {\enquote
  {\bibinfo {title} {Reciprocal space engineering with hyperuniform gold
  disordered surfaces},}\ }\href {\doibase 10.1063/1.4983990} {\bibfield
  {journal} {\bibinfo  {journal} {APL Photonics}\ }\textbf {\bibinfo {volume}
  {2}},\ \bibinfo {pages} {061302} (\bibinfo {year} {2017})}\BibitemShut
  {NoStop}%
\bibitem [{\citenamefont {Gorsky}\ \emph {et~al.}(2019)\citenamefont {Gorsky},
  \citenamefont {Britton}, \citenamefont {Chen}, \citenamefont {Montaner},
  \citenamefont {Lenef}, \citenamefont {Raukas},\ and\ \citenamefont
  {Dal~Negro}}]{gorsky_engineered_2019}%
  \BibitemOpen
  \bibfield  {author} {\bibinfo {author} {\bibfnamefont {S.}~\bibnamefont
  {Gorsky}}, \bibinfo {author} {\bibfnamefont {W.~A.}\ \bibnamefont {Britton}},
  \bibinfo {author} {\bibfnamefont {Y.}~\bibnamefont {Chen}}, \bibinfo {author}
  {\bibfnamefont {J.}~\bibnamefont {Montaner}}, \bibinfo {author}
  {\bibfnamefont {A.}~\bibnamefont {Lenef}}, \bibinfo {author} {\bibfnamefont
  {M.}~\bibnamefont {Raukas}}, \ and\ \bibinfo {author} {\bibfnamefont
  {L.}~\bibnamefont {Dal~Negro}},\ }\bibfield  {title} {\enquote {\bibinfo
  {title} {Engineered hyperuniformity for directional light extraction},}\
  }\href {\doibase 10.1063/1.5124302} {\bibfield  {journal} {\bibinfo
  {journal} {APL Photonics}\ }\textbf {\bibinfo {volume} {4}},\ \bibinfo
  {pages} {110801} (\bibinfo {year} {2019})}\BibitemShut {NoStop}%
\bibitem [{\citenamefont {Degl\'Innocenti}\ \emph {et~al.}(2016)\citenamefont
  {Degl\'Innocenti}, \citenamefont {Shah}, \citenamefont {Masini},
  \citenamefont {Ronzani}, \citenamefont {Pitanti}, \citenamefont {Ren},
  \citenamefont {Jessop}, \citenamefont {Tredicucci}, \citenamefont {Beere},\
  and\ \citenamefont {Ritchie}}]{deglinnocenti_hyperuniform_2016}%
  \BibitemOpen
  \bibfield  {author} {\bibinfo {author} {\bibfnamefont {R.}~\bibnamefont
  {Degl\'Innocenti}}, \bibinfo {author} {\bibfnamefont {Y.~D.}\ \bibnamefont
  {Shah}}, \bibinfo {author} {\bibfnamefont {L.}~\bibnamefont {Masini}},
  \bibinfo {author} {\bibfnamefont {A.}~\bibnamefont {Ronzani}}, \bibinfo
  {author} {\bibfnamefont {A.}~\bibnamefont {Pitanti}}, \bibinfo {author}
  {\bibfnamefont {Y.}~\bibnamefont {Ren}}, \bibinfo {author} {\bibfnamefont
  {D.~S.}\ \bibnamefont {Jessop}}, \bibinfo {author} {\bibfnamefont
  {A.}~\bibnamefont {Tredicucci}}, \bibinfo {author} {\bibfnamefont {H.~E.}\
  \bibnamefont {Beere}}, \ and\ \bibinfo {author} {\bibfnamefont {D.~A.}\
  \bibnamefont {Ritchie}},\ }\bibfield  {title} {\enquote {\bibinfo {title}
  {Hyperuniform disordered terahertz quantum cascade laser},}\ }\href {\doibase
  10.1038/srep19325} {\bibfield  {journal} {\bibinfo  {journal} {Sci Rep}\
  }\textbf {\bibinfo {volume} {6}},\ \bibinfo {pages} {19325} (\bibinfo {year}
  {2016})},\ \Eprint {http://arxiv.org/abs/26758959} {26758959} \BibitemShut
  {NoStop}%
\bibitem [{\citenamefont {Zhang}\ \emph {et~al.}(2019)\citenamefont {Zhang},
  \citenamefont {Chu}, \citenamefont {Giddens}, \citenamefont {Wu},\ and\
  \citenamefont {Hao}}]{zhang_experimental_2019}%
  \BibitemOpen
  \bibfield  {author} {\bibinfo {author} {\bibfnamefont {H.}~\bibnamefont
  {Zhang}}, \bibinfo {author} {\bibfnamefont {H.}~\bibnamefont {Chu}}, \bibinfo
  {author} {\bibfnamefont {H.}~\bibnamefont {Giddens}}, \bibinfo {author}
  {\bibfnamefont {W.}~\bibnamefont {Wu}}, \ and\ \bibinfo {author}
  {\bibfnamefont {Y.}~\bibnamefont {Hao}},\ }\bibfield  {title} {\enquote
  {\bibinfo {title} {Experimental demonstration of luneburg lens based on
  hyperuniform disordered media},}\ }\href {\doibase 10.1063/1.5055295}
  {\bibfield  {journal} {\bibinfo  {journal} {Appl. Phys. Lett.}\ }\textbf
  {\bibinfo {volume} {114}},\ \bibinfo {pages} {053507} (\bibinfo {year}
  {2019})}\BibitemShut {NoStop}%
\bibitem [{\citenamefont {Christogeorgos}\ \emph {et~al.}(2021)\citenamefont
  {Christogeorgos}, \citenamefont {Zhang}, \citenamefont {Cheng},\ and\
  \citenamefont {Hao}}]{christogeorgos_extraordinary_2021}%
  \BibitemOpen
  \bibfield  {author} {\bibinfo {author} {\bibfnamefont {O.}~\bibnamefont
  {Christogeorgos}}, \bibinfo {author} {\bibfnamefont {H.}~\bibnamefont
  {Zhang}}, \bibinfo {author} {\bibfnamefont {Q.}~\bibnamefont {Cheng}}, \ and\
  \bibinfo {author} {\bibfnamefont {Y.}~\bibnamefont {Hao}},\ }\bibfield
  {title} {\enquote {\bibinfo {title} {Extraordinary directive emission and
  scanning from an array of radiation sources with hyperuniform disorder},}\
  }\href {\doibase 10.1103/PhysRevApplied.15.014062} {\bibfield  {journal}
  {\bibinfo  {journal} {Phys. Rev. Appl.}\ }\textbf {\bibinfo {volume} {15}},\
  \bibinfo {pages} {014062} (\bibinfo {year} {2021})}\BibitemShut {NoStop}%
\bibitem [{\citenamefont {Tang}\ \emph {et~al.}(2023)\citenamefont {Tang},
  \citenamefont {Wang}, \citenamefont {Wang}, \citenamefont {Gao},
  \citenamefont {Li}, \citenamefont {Liang}, \citenamefont {Sebbah},
  \citenamefont {Zhang},\ and\ \citenamefont {Shi}}]{tang_hyperuniform_2023}%
  \BibitemOpen
  \bibfield  {author} {\bibinfo {author} {\bibfnamefont {K.}~\bibnamefont
  {Tang}}, \bibinfo {author} {\bibfnamefont {Y.}~\bibnamefont {Wang}}, \bibinfo
  {author} {\bibfnamefont {S.}~\bibnamefont {Wang}}, \bibinfo {author}
  {\bibfnamefont {D.}~\bibnamefont {Gao}}, \bibinfo {author} {\bibfnamefont
  {H.}~\bibnamefont {Li}}, \bibinfo {author} {\bibfnamefont {X.}~\bibnamefont
  {Liang}}, \bibinfo {author} {\bibfnamefont {P.}~\bibnamefont {Sebbah}},
  \bibinfo {author} {\bibfnamefont {J.}~\bibnamefont {Zhang}}, \ and\ \bibinfo
  {author} {\bibfnamefont {J.}~\bibnamefont {Shi}},\ }\bibfield  {title}
  {\enquote {\bibinfo {title} {Hyperuniform disordered parametric loudspeaker
  array},}\ }\href {\doibase 10.1103/PhysRevApplied.19.054035} {\bibfield
  {journal} {\bibinfo  {journal} {Phys. Rev. Appl.}\ }\textbf {\bibinfo
  {volume} {19}},\ \bibinfo {pages} {054035} (\bibinfo {year}
  {2023})}\BibitemShut {NoStop}%
\bibitem [{\citenamefont {Zhang}\ \emph {et~al.}(2021)\citenamefont {Zhang},
  \citenamefont {Cheng}, \citenamefont {Chu}, \citenamefont {Christogeorgos},
  \citenamefont {Wu},\ and\ \citenamefont {Hao}}]{zhang_hyperuniform_2021}%
  \BibitemOpen
  \bibfield  {author} {\bibinfo {author} {\bibfnamefont {H.}~\bibnamefont
  {Zhang}}, \bibinfo {author} {\bibfnamefont {Q.}~\bibnamefont {Cheng}},
  \bibinfo {author} {\bibfnamefont {H.}~\bibnamefont {Chu}}, \bibinfo {author}
  {\bibfnamefont {O.}~\bibnamefont {Christogeorgos}}, \bibinfo {author}
  {\bibfnamefont {W.}~\bibnamefont {Wu}}, \ and\ \bibinfo {author}
  {\bibfnamefont {Y.}~\bibnamefont {Hao}},\ }\bibfield  {title} {\enquote
  {\bibinfo {title} {Hyperuniform disordered distribution metasurface for
  scattering reduction},}\ }\href {\doibase 10.1063/5.0041911} {\bibfield
  {journal} {\bibinfo  {journal} {Appl. Phys. Lett.}\ }\textbf {\bibinfo
  {volume} {118}},\ \bibinfo {pages} {101601} (\bibinfo {year}
  {2021})}\BibitemShut {NoStop}%
\bibitem [{\citenamefont {Kuznetsova}\ \emph {et~al.}(2021)\citenamefont
  {Kuznetsova}, \citenamefont {Groby}, \citenamefont {Garcia-Raffi},\ and\
  \citenamefont {Romero-Garc\'{i}a}}]{kuznetsova_stealth_2021}%
  \BibitemOpen
  \bibfield  {author} {\bibinfo {author} {\bibfnamefont {S.}~\bibnamefont
  {Kuznetsova}}, \bibinfo {author} {\bibfnamefont {J.~P.}\ \bibnamefont
  {Groby}}, \bibinfo {author} {\bibfnamefont {L.~M.}\ \bibnamefont
  {Garcia-Raffi}}, \ and\ \bibinfo {author} {\bibfnamefont {V.}~\bibnamefont
  {Romero-Garc\'{i}a}},\ }\bibfield  {title} {\enquote {\bibinfo {title}
  {Stealth and equiluminous materials for scattering cancellation and wave
  diffusion},}\ }\href {\doibase 10.1080/17455030.2021.1948630} {\bibfield
  {journal} {\bibinfo  {journal} {Waves Random Complex Media}\ }\textbf
  {\bibinfo {volume} {0}},\ \bibinfo {pages} {1--19} (\bibinfo {year}
  {2021})}\BibitemShut {NoStop}%
\bibitem [{\citenamefont {Tamraoui}, \citenamefont {Roux},\ and\ \citenamefont
  {Liebgott}(2023)}]{tamraoui_hyperuniform_2023}%
  \BibitemOpen
  \bibfield  {author} {\bibinfo {author} {\bibfnamefont {M.}~\bibnamefont
  {Tamraoui}}, \bibinfo {author} {\bibfnamefont {E.}~\bibnamefont {Roux}}, \
  and\ \bibinfo {author} {\bibfnamefont {H.}~\bibnamefont {Liebgott}},\
  }\bibfield  {title} {\enquote {\bibinfo {title} {Hyperuniform disordered
  sparse array for 3d ultrasound imaging},}\ }in\ \href {\doibase
  10.1109/IUS51837.2023.10308368} {\emph {\bibinfo {booktitle} {2023 IEEE
  International Ultrasonics Symposium (IUS)}}}\ (\bibinfo {year} {2023})\ pp.\
  \bibinfo {pages} {1--4}\BibitemShut {NoStop}%
\bibitem [{\citenamefont {Zhang}, \citenamefont {Stillinger},\ and\
  \citenamefont {Torquato}(2017)}]{zhang_can_2017}%
  \BibitemOpen
  \bibfield  {author} {\bibinfo {author} {\bibfnamefont {G.}~\bibnamefont
  {Zhang}}, \bibinfo {author} {\bibfnamefont {F.~H.}\ \bibnamefont
  {Stillinger}}, \ and\ \bibinfo {author} {\bibfnamefont {S.}~\bibnamefont
  {Torquato}},\ }\bibfield  {title} {\enquote {\bibinfo {title} {Can exotic
  disordered ``stealthy" particle configurations tolerate arbitrarily large
  holes?}}\ }\href {\doibase 10.1039/c7sm01028a} {\bibfield  {journal}
  {\bibinfo  {journal} {Soft Matter}\ }\textbf {\bibinfo {volume} {13}},\
  \bibinfo {pages} {6197--6207} (\bibinfo {year} {2017})}\BibitemShut {NoStop}%
\bibitem [{\citenamefont {Ghosh}\ and\ \citenamefont
  {Lebowitz}(2018)}]{ghosh_generalized_2018}%
  \BibitemOpen
  \bibfield  {author} {\bibinfo {author} {\bibfnamefont {S.}~\bibnamefont
  {Ghosh}}\ and\ \bibinfo {author} {\bibfnamefont {J.~L.}\ \bibnamefont
  {Lebowitz}},\ }\bibfield  {title} {\enquote {\bibinfo {title} {Generalized
  stealthy hyperuniform processes: maximal rigidity and the bounded holes
  conjecture},}\ }\href {\doibase 10.1007/s00220-018-3226-5} {\bibfield
  {journal} {\bibinfo  {journal} {Commun. Math. Phys.}\ }\textbf {\bibinfo
  {volume} {363}},\ \bibinfo {pages} {97--110} (\bibinfo {year}
  {2018})}\BibitemShut {NoStop}%
\bibitem [{\citenamefont {Martis}\ \emph {et~al.}(2013)\citenamefont {Martis},
  \citenamefont {Marcotte}, \citenamefont {Stillinger},\ and\ \citenamefont
  {Torquato}}]{martis_exotic_2013}%
  \BibitemOpen
  \bibfield  {author} {\bibinfo {author} {\bibfnamefont {S.}~\bibnamefont
  {Martis}}, \bibinfo {author} {\bibfnamefont {{\'E}.}~\bibnamefont
  {Marcotte}}, \bibinfo {author} {\bibfnamefont {F.~H.}\ \bibnamefont
  {Stillinger}}, \ and\ \bibinfo {author} {\bibfnamefont {S.}~\bibnamefont
  {Torquato}},\ }\bibfield  {title} {\enquote {\bibinfo {title} {Exotic ground
  states of directional pair potentials via collective-density variables},}\
  }\href {\doibase 10.1007/s10955-012-0619-2} {\bibfield  {journal} {\bibinfo
  {journal} {J. Stat. Phys.}\ }\textbf {\bibinfo {volume} {150}},\ \bibinfo
  {pages} {414--431} (\bibinfo {year} {2013})}\BibitemShut {NoStop}%
\bibitem [{\citenamefont
  {Torquato}(2016{\natexlab{a}})}]{torquato_hyperuniformity_2016}%
  \BibitemOpen
  \bibfield  {author} {\bibinfo {author} {\bibfnamefont {S.}~\bibnamefont
  {Torquato}},\ }\bibfield  {title} {\enquote {\bibinfo {title}
  {Hyperuniformity and its generalizations},}\ }\href {\doibase
  10.1103/PhysRevE.94.022122} {\bibfield  {journal} {\bibinfo  {journal} {Phys.
  Rev. E}\ }\textbf {\bibinfo {volume} {94}},\ \bibinfo {pages} {022122}
  (\bibinfo {year} {2016}{\natexlab{a}})}\BibitemShut {NoStop}%
\bibitem [{Note1()}]{Note1}%
  \BibitemOpen
  \bibinfo {note} {The total energy expression (\ref {eq:pot}) actually also
  contains the sum $\DOTSB \sum@ \slimits@ _{\protect \bf k} {\protect \tilde
  v}({\protect \bf k})$, but this is a structure-independent constant that is
  set to zero here without loss of generality.}\BibitemShut {Stop}%
\bibitem [{Note2()}]{Note2}%
  \BibitemOpen
  \bibinfo {note} {More generally, stealthy configurations are ground states
  that minimize $S(k)$ to be zero at other sets of wave vectors, not
  necessarily in a connected set around the origin and hence nonhyperuniform,
  specific examples of which were investigated in Ref. \protect \rev@citealpnum
  {batten_classical_2008}. The collective-coordinate optimization technique has
  been used to target more general forms of the structure factor for a
  prescribed set of wave vectors such that $S({\protect \bf k})$ is not
  minimized to be zero in this set (e.g., power-law forms and positive
  constants).\cite {batten_classical_2008, zhang_perfect_2016} There, the
  resulting configurations are the ground states of interacting many-particle
  systems with 2-,3- and 4-body interactions.}\BibitemShut {Stop}%
\bibitem [{\citenamefont {Morse}\ \emph {et~al.}(2023)\citenamefont {Morse},
  \citenamefont {Kim}, \citenamefont {Steinhardt},\ and\ \citenamefont
  {Torquato}}]{morse_generating_2023}%
  \BibitemOpen
  \bibfield  {author} {\bibinfo {author} {\bibfnamefont {P.~K.}\ \bibnamefont
  {Morse}}, \bibinfo {author} {\bibfnamefont {J.}~\bibnamefont {Kim}}, \bibinfo
  {author} {\bibfnamefont {P.~J.}\ \bibnamefont {Steinhardt}}, \ and\ \bibinfo
  {author} {\bibfnamefont {S.}~\bibnamefont {Torquato}},\ }\bibfield  {title}
  {\enquote {\bibinfo {title} {Generating large disordered stealthy
  hyperuniform systems with ultrahigh accuracy to determine their physical
  properties},}\ }\href {\doibase 10.1103/PhysRevResearch.5.033190} {\bibfield
  {journal} {\bibinfo  {journal} {Phys. Rev. Research}\ }\textbf {\bibinfo
  {volume} {5}},\ \bibinfo {pages} {033190} (\bibinfo {year}
  {2023})}\BibitemShut {NoStop}%
\bibitem [{\citenamefont {Shih}, \citenamefont {Casiulis},\ and\ \citenamefont
  {Martiniani}(2024)}]{shih_fast_2024}%
  \BibitemOpen
  \bibfield  {author} {\bibinfo {author} {\bibfnamefont {A.}~\bibnamefont
  {Shih}}, \bibinfo {author} {\bibfnamefont {M.}~\bibnamefont {Casiulis}}, \
  and\ \bibinfo {author} {\bibfnamefont {S.}~\bibnamefont {Martiniani}},\
  }\bibfield  {title} {\enquote {\bibinfo {title} {Fast generation of
  spectrally shaped disorder},}\ }\href {\doibase 10.1103/PhysRevE.110.034122}
  {\bibfield  {journal} {\bibinfo  {journal} {Phys. Rev. E}\ }\textbf {\bibinfo
  {volume} {110}},\ \bibinfo {pages} {034122} (\bibinfo {year}
  {2024})}\BibitemShut {NoStop}%
\bibitem [{Note3()}]{Note3}%
  \BibitemOpen
  \bibinfo {note} {Reference \protect \citenum {morse_generating_2023} reports
  techniques that reduce the deviations from zero within the exclusion region
  (distance to exact stealthiness) by a factor of approximately $10^{30}$
  compared to the previous ones described in Refs. \cite
  {uche_constraints_2004,batten_classical_2008,martis_exotic_2013}, even for
  much larger system sizes.}\BibitemShut {Stop}%
\bibitem [{\citenamefont {Middlemas}\ and\ \citenamefont
  {Torquato}(2020)}]{middlemas_nearestneighbor_2020}%
  \BibitemOpen
  \bibfield  {author} {\bibinfo {author} {\bibfnamefont {T.~M.}\ \bibnamefont
  {Middlemas}}\ and\ \bibinfo {author} {\bibfnamefont {S.}~\bibnamefont
  {Torquato}},\ }\bibfield  {title} {\enquote {\bibinfo {title}
  {Nearest-neighbor functions for disordered stealthy hyperuniform
  many-particle systems},}\ }\href {\doibase 10.1088/1742-5468/abb8cb}
  {\bibfield  {journal} {\bibinfo  {journal} {J. Stat. Mech.}\ }\textbf
  {\bibinfo {volume} {2020}},\ \bibinfo {pages} {103302} (\bibinfo {year}
  {2020})}\BibitemShut {NoStop}%
\bibitem [{\citenamefont {Kim}\ and\ \citenamefont
  {Torquato}(2024)}]{kim_theoretical_2024}%
  \BibitemOpen
  \bibfield  {author} {\bibinfo {author} {\bibfnamefont {J.}~\bibnamefont
  {Kim}}\ and\ \bibinfo {author} {\bibfnamefont {S.}~\bibnamefont {Torquato}},\
  }\bibfield  {title} {\enquote {\bibinfo {title} {Theoretical prediction of
  the effective dynamic dielectric constant of disordered hyperuniform
  anisotropic composites beyond the long-wavelength regime},}\ }\href {\doibase
  10.1364/OME.507918} {\bibfield  {journal} {\bibinfo  {journal} {Opt. Mater.
  Express}\ }\textbf {\bibinfo {volume} {14}},\ \bibinfo {pages} {194}
  (\bibinfo {year} {2024})}\BibitemShut {NoStop}%
\bibitem [{\citenamefont {Torquato}, \citenamefont {Truskett},\ and\
  \citenamefont {Debenedetti}(2000)}]{torquato_is_2000}%
  \BibitemOpen
  \bibfield  {author} {\bibinfo {author} {\bibfnamefont {S.}~\bibnamefont
  {Torquato}}, \bibinfo {author} {\bibfnamefont {T.~M.}\ \bibnamefont
  {Truskett}}, \ and\ \bibinfo {author} {\bibfnamefont {P.~G.}\ \bibnamefont
  {Debenedetti}},\ }\bibfield  {title} {\enquote {\bibinfo {title} {Is random
  close packing of spheres well defined?}}\ }\href {\doibase
  10.1103/physrevlett.84.2064} {\bibfield  {journal} {\bibinfo  {journal}
  {Phys. Rev. Lett.}\ }\textbf {\bibinfo {volume} {84}},\ \bibinfo {pages}
  {2064--2067} (\bibinfo {year} {2000})}\BibitemShut {NoStop}%
\bibitem [{\citenamefont {Donev}, \citenamefont {Stillinger},\ and\
  \citenamefont {Torquato}(2005)}]{donev_unexpected_2005}%
  \BibitemOpen
  \bibfield  {author} {\bibinfo {author} {\bibfnamefont {A.}~\bibnamefont
  {Donev}}, \bibinfo {author} {\bibfnamefont {F.}~\bibnamefont {Stillinger}}, \
  and\ \bibinfo {author} {\bibfnamefont {S.}~\bibnamefont {Torquato}},\
  }\bibfield  {title} {\enquote {\bibinfo {title} {Unexpected density
  fluctuations in jammed disordered sphere packings},}\ }\href {\doibase
  10.1103/PhysRevLett.95.090604} {\bibfield  {journal} {\bibinfo  {journal}
  {Phys. Rev. Lett.}\ }\textbf {\bibinfo {volume} {95}},\ \bibinfo {pages}
  {090604} (\bibinfo {year} {2005})}\BibitemShut {NoStop}%
\bibitem [{\citenamefont {Maher}, \citenamefont {Jiao},\ and\ \citenamefont
  {Torquato}(2023)}]{maher_hyperuniformity_2023}%
  \BibitemOpen
  \bibfield  {author} {\bibinfo {author} {\bibfnamefont {C.~E.}\ \bibnamefont
  {Maher}}, \bibinfo {author} {\bibfnamefont {Y.}~\bibnamefont {Jiao}}, \ and\
  \bibinfo {author} {\bibfnamefont {S.}~\bibnamefont {Torquato}},\ }\bibfield
  {title} {\enquote {\bibinfo {title} {Hyperuniformity of maximally random
  jammed packings of hyperspheres across spatial dimensions},}\ }\href
  {\doibase 10.1103/PhysRevE.108.064602} {\bibfield  {journal} {\bibinfo
  {journal} {Phys. Rev. E}\ }\textbf {\bibinfo {volume} {108}},\ \bibinfo
  {pages} {064602} (\bibinfo {year} {2023})}\BibitemShut {NoStop}%
\bibitem [{\citenamefont {Torquato}\ and\ \citenamefont
  {Kim}(2025)}]{torquato_existence_2025}%
  \BibitemOpen
  \bibfield  {author} {\bibinfo {author} {\bibfnamefont {S.}~\bibnamefont
  {Torquato}}\ and\ \bibinfo {author} {\bibfnamefont {J.}~\bibnamefont {Kim}},\
  }\href {\doibase 10.48550/arXiv.2504.10310} {\enquote {\bibinfo {title}
  {Existence of nonequilibrium glasses in the degenerate stealthy hyperuniform
  ground-state manifold},}\ } (\bibinfo {year} {2025}),\ \Eprint
  {http://arxiv.org/abs/2504.10310} {arXiv:2504.10310 [cond-mat]} \BibitemShut
  {NoStop}%
\bibitem [{\citenamefont {Torquato}(2020)}]{torquato_predicting_2020}%
  \BibitemOpen
  \bibfield  {author} {\bibinfo {author} {\bibfnamefont {S.}~\bibnamefont
  {Torquato}},\ }\bibfield  {title} {\enquote {\bibinfo {title} {Predicting
  transport characteristics of hyperuniform porous media via rigorous
  microstructure-property relations},}\ }\href {\doibase
  10.1016/j.advwatres.2020.103565} {\bibfield  {journal} {\bibinfo  {journal}
  {Adv. Water Resour.}\ }\textbf {\bibinfo {volume} {140}},\ \bibinfo {pages}
  {103565} (\bibinfo {year} {2020})}\BibitemShut {NoStop}%
\bibitem [{\citenamefont {Torquato}\ and\ \citenamefont
  {Stillinger}(2006)}]{torquato_new_2006}%
  \BibitemOpen
  \bibfield  {author} {\bibinfo {author} {\bibfnamefont {S.}~\bibnamefont
  {Torquato}}\ and\ \bibinfo {author} {\bibfnamefont {F.~H.}\ \bibnamefont
  {Stillinger}},\ }\bibfield  {title} {\enquote {\bibinfo {title} {New
  conjectural lower bounds on the optimal density of sphere packings},}\ }\href
  {\doibase 10.1080/10586458.2006.10128964} {\bibfield  {journal} {\bibinfo
  {journal} {Exper. Math.}\ }\textbf {\bibinfo {volume} {15}},\ \bibinfo
  {pages} {307--331} (\bibinfo {year} {2006})}\BibitemShut {NoStop}%
\bibitem [{\citenamefont {Salvadori}(2007)}]{salvadori_extremes_2007}%
  \BibitemOpen
  \bibinfo {editor} {\bibfnamefont {G.}~\bibnamefont {Salvadori}},\ ed.,\
  \href@noop {} {\emph {\bibinfo {title} {Extremes in nature: an approach using
  copulas}}},\ \bibinfo {series} {Water science and technology library}\
  No.~\bibinfo {number} {56}\ (\bibinfo  {publisher} {Springer},\ \bibinfo
  {address} {Dordrecht},\ \bibinfo {year} {2007})\BibitemShut {NoStop}%
\bibitem [{\citenamefont {Gomes}\ and\ \citenamefont
  {Guillou}(2015)}]{gomes_extreme_2015}%
  \BibitemOpen
  \bibfield  {author} {\bibinfo {author} {\bibfnamefont {M.~I.}\ \bibnamefont
  {Gomes}}\ and\ \bibinfo {author} {\bibfnamefont {A.}~\bibnamefont
  {Guillou}},\ }\bibfield  {title} {\enquote {\bibinfo {title} {Extreme value
  theory and statistics of univariate extremes: A review},}\ }\href {\doibase
  10.1111/insr.12058} {\bibfield  {journal} {\bibinfo  {journal} {Int. Stat.
  Rev.}\ }\textbf {\bibinfo {volume} {83}},\ \bibinfo {pages} {263--292}
  (\bibinfo {year} {2015})}\BibitemShut {NoStop}%
\bibitem [{\citenamefont {Zachary}\ and\ \citenamefont
  {Torquato}(2009)}]{zachary_hyperuniformity_2009}%
  \BibitemOpen
  \bibfield  {author} {\bibinfo {author} {\bibfnamefont {C.}~\bibnamefont
  {Zachary}}\ and\ \bibinfo {author} {\bibfnamefont {S.}~\bibnamefont
  {Torquato}},\ }\bibfield  {title} {\enquote {\bibinfo {title}
  {Hyperuniformity in point patterns and two-phase random heterogeneous
  media},}\ }\href {\doibase 10.1088/1742-5468/2009/12/P12015} {\bibfield
  {journal} {\bibinfo  {journal} {J. Stat. Mech: Theory Exp.}\ }\textbf
  {\bibinfo {volume} {2009}},\ \bibinfo {pages} {P12015} (\bibinfo {year}
  {2009})}\BibitemShut {NoStop}%
\bibitem [{\citenamefont {Torquato}(2021)}]{torquato_structural_2021b}%
  \BibitemOpen
  \bibfield  {author} {\bibinfo {author} {\bibfnamefont {S.}~\bibnamefont
  {Torquato}},\ }\bibfield  {title} {\enquote {\bibinfo {title} {Structural
  characterization of many-particle systems on approach to hyperuniform
  states},}\ }\href {\doibase 10.1103/PhysRevE.103.052126} {\bibfield
  {journal} {\bibinfo  {journal} {Phys. Rev. E}\ }\textbf {\bibinfo {volume}
  {103}},\ \bibinfo {pages} {052126} (\bibinfo {year} {2021})}\BibitemShut
  {NoStop}%
\bibitem [{\citenamefont {O'Hern}\ \emph {et~al.}(2003)\citenamefont {O'Hern},
  \citenamefont {Silbert}, \citenamefont {Liu},\ and\ \citenamefont
  {Nagel}}]{ohern_jamming_2003}%
  \BibitemOpen
  \bibfield  {author} {\bibinfo {author} {\bibfnamefont {C.~S.}\ \bibnamefont
  {O'Hern}}, \bibinfo {author} {\bibfnamefont {L.~E.}\ \bibnamefont {Silbert}},
  \bibinfo {author} {\bibfnamefont {A.~J.}\ \bibnamefont {Liu}}, \ and\
  \bibinfo {author} {\bibfnamefont {S.~R.}\ \bibnamefont {Nagel}},\ }\bibfield
  {title} {\enquote {\bibinfo {title} {Jamming at zero temperature and zero
  applied stress: The epitome of disorder},}\ }\href {\doibase
  10.1103/PhysRevE.68.011306} {\bibfield  {journal} {\bibinfo  {journal} {Phys.
  Rev. E}\ }\textbf {\bibinfo {volume} {68}},\ \bibinfo {pages} {011306}
  (\bibinfo {year} {2003})}\BibitemShut {NoStop}%
\bibitem [{\citenamefont {Charbonneau}\ \emph {et~al.}(2012)\citenamefont
  {Charbonneau}, \citenamefont {Corwin}, \citenamefont {Parisi},\ and\
  \citenamefont {Zamponi}}]{charbonneau_universal_2012}%
  \BibitemOpen
  \bibfield  {author} {\bibinfo {author} {\bibfnamefont {P.}~\bibnamefont
  {Charbonneau}}, \bibinfo {author} {\bibfnamefont {E.~I.}\ \bibnamefont
  {Corwin}}, \bibinfo {author} {\bibfnamefont {G.}~\bibnamefont {Parisi}}, \
  and\ \bibinfo {author} {\bibfnamefont {F.}~\bibnamefont {Zamponi}},\
  }\bibfield  {title} {\enquote {\bibinfo {title} {Universal microstructure and
  mechanical stability of jammed packings},}\ }\href {\doibase
  10.1103/PhysRevLett.109.205501} {\bibfield  {journal} {\bibinfo  {journal}
  {Phys. Rev. Lett.}\ }\textbf {\bibinfo {volume} {109}},\ \bibinfo {pages}
  {205501} (\bibinfo {year} {2012})}\BibitemShut {NoStop}%
\bibitem [{\citenamefont {Jin}\ and\ \citenamefont
  {Yoshino}(2021)}]{jin_jamming_21}%
  \BibitemOpen
  \bibfield  {author} {\bibinfo {author} {\bibfnamefont {Y.}~\bibnamefont
  {Jin}}\ and\ \bibinfo {author} {\bibfnamefont {H.}~\bibnamefont {Yoshino}},\
  }\bibfield  {title} {\enquote {\bibinfo {title} {A jamming plane of sphere
  packings},}\ }\href {\doibase 10.1073/pnas.2021794118} {\bibfield  {journal}
  {\bibinfo  {journal} {Proc. Natl. Acad. Sci. U.S.A.}\ }\textbf {\bibinfo
  {volume} {118}},\ \bibinfo {pages} {e2021794118} (\bibinfo {year}
  {2021})}\BibitemShut {NoStop}%
\bibitem [{\citenamefont {M{\'e}zard}\ and\ \citenamefont
  {Montanari}(2009)}]{mezard_satisfiability_2009}%
  \BibitemOpen
  \bibfield  {author} {\bibinfo {author} {\bibfnamefont {M.}~\bibnamefont
  {M{\'e}zard}}\ and\ \bibinfo {author} {\bibfnamefont {A.}~\bibnamefont
  {Montanari}},\ }\bibfield  {title} {\enquote {\bibinfo {title}
  {Satisfiability},}\ }in\ \href {\doibase
  10.1093/acprof:oso/9780198570837.003.0010} {\emph {\bibinfo {booktitle}
  {Information, Physics, and Computation}}},\ \bibinfo {editor} {edited by\
  \bibinfo {editor} {\bibfnamefont {M.}~\bibnamefont {M{\'e}zard}}\ and\
  \bibinfo {editor} {\bibfnamefont {A.}~\bibnamefont {Montanari}}}\ (\bibinfo
  {publisher} {Oxford University Press},\ \bibinfo {year} {2009})\ p.~\bibinfo
  {pages} {0}\BibitemShut {NoStop}%
\bibitem [{\citenamefont {Franz}\ \emph {et~al.}(2017)\citenamefont {Franz},
  \citenamefont {Parisi}, \citenamefont {Sevelev}, \citenamefont {Urbani},\
  and\ \citenamefont {Zamponi}}]{franz_universality_2017}%
  \BibitemOpen
  \bibfield  {author} {\bibinfo {author} {\bibfnamefont {S.}~\bibnamefont
  {Franz}}, \bibinfo {author} {\bibfnamefont {G.}~\bibnamefont {Parisi}},
  \bibinfo {author} {\bibfnamefont {M.}~\bibnamefont {Sevelev}}, \bibinfo
  {author} {\bibfnamefont {P.}~\bibnamefont {Urbani}}, \ and\ \bibinfo {author}
  {\bibfnamefont {F.}~\bibnamefont {Zamponi}},\ }\bibfield  {title} {\enquote
  {\bibinfo {title} {Universality of the sat-unsat (jamming) threshold in
  non-convex continuous constraint satisfaction problems},}\ }\href {\doibase
  10.21468/SciPostPhys.2.3.019} {\bibfield  {journal} {\bibinfo  {journal}
  {SciPost Phys.}\ }\textbf {\bibinfo {volume} {2}},\ \bibinfo {pages} {019}
  (\bibinfo {year} {2017})}\BibitemShut {NoStop}%
\bibitem [{Note4()}]{Note4}%
  \BibitemOpen
  \bibinfo {note} {For example, on an Intel(R) Xeon(R) CPU (E5-2680, 2.40 GHz),
  it takes around 15 core-hours to generate one 3D SHU ground-state packing
  with $\chi =0.45$, $\phi =0.47$, and $N=4000$. It takes around 0.1 core-hours
  on the same CPU to generate one 3D SHU ground-state packing with $\chi
  =0.0025$, $\phi =0.63$, and $N=4000$}\BibitemShut {NoStop}%
\bibitem [{\citenamefont {Nocedal}(1980)}]{nocedal_updating_1980}%
  \BibitemOpen
  \bibfield  {author} {\bibinfo {author} {\bibfnamefont {J.}~\bibnamefont
  {Nocedal}},\ }\bibfield  {title} {\enquote {\bibinfo {title} {Updating
  quasi-newton matrices with limited storage},}\ }\href {\doibase
  10.1090/S0025-5718-1980-0572855-7} {\bibfield  {journal} {\bibinfo  {journal}
  {Math. Comput.}\ }\textbf {\bibinfo {volume} {35}},\ \bibinfo {pages}
  {773--782} (\bibinfo {year} {1980})}\BibitemShut {NoStop}%
\bibitem [{\citenamefont {Liu}\ and\ \citenamefont
  {Nocedal}(1989)}]{liu_limited_1989}%
  \BibitemOpen
  \bibfield  {author} {\bibinfo {author} {\bibfnamefont {D.~C.}\ \bibnamefont
  {Liu}}\ and\ \bibinfo {author} {\bibfnamefont {J.}~\bibnamefont {Nocedal}},\
  }\bibfield  {title} {\enquote {\bibinfo {title} {On the limited memory bfgs
  method for large scale optimization},}\ }\href {\doibase 10.1007/BF01589116}
  {\bibfield  {journal} {\bibinfo  {journal} {Math. Program.}\ }\textbf
  {\bibinfo {volume} {45}},\ \bibinfo {pages} {503--528} (\bibinfo {year}
  {1989})}\BibitemShut {NoStop}%
\bibitem [{Note5()}]{Note5}%
  \BibitemOpen
  \bibinfo {note} {This threshold is close to zero energy of the potentials
  \protect \textup {\hbox {\mathsurround \z@ \protect \normalfont
  (\ignorespaces \ref {eq:pot}\unskip \@@italiccorr )}} and \protect \textup
  {\hbox {\mathsurround \z@ \protect \normalfont (\ignorespaces \ref
  {eq:Phi}\unskip \@@italiccorr )}} within the double precision of the
  machine}\BibitemShut {NoStop}%
\bibitem [{\citenamefont {Donev}, \citenamefont {Stillinger},\ and\
  \citenamefont {Torquato}(2007)}]{donev_configurational_2007}%
  \BibitemOpen
  \bibfield  {author} {\bibinfo {author} {\bibfnamefont {A.}~\bibnamefont
  {Donev}}, \bibinfo {author} {\bibfnamefont {F.~H.}\ \bibnamefont
  {Stillinger}}, \ and\ \bibinfo {author} {\bibfnamefont {S.}~\bibnamefont
  {Torquato}},\ }\bibfield  {title} {\enquote {\bibinfo {title}
  {Configurational entropy of binary hard-disk glasses: Nonexistence of an
  ideal glass transition},}\ }\href {\doibase 10.1063/1.2775928} {\bibfield
  {journal} {\bibinfo  {journal} {J. Chem. Phys.}\ }\textbf {\bibinfo {volume}
  {127}},\ \bibinfo {pages} {124509} (\bibinfo {year} {2007})}\BibitemShut
  {NoStop}%
\bibitem [{\citenamefont
  {Torquato}(2018{\natexlab{b}})}]{torquato_perspective_2018a}%
  \BibitemOpen
  \bibfield  {author} {\bibinfo {author} {\bibfnamefont {S.}~\bibnamefont
  {Torquato}},\ }\bibfield  {title} {\enquote {\bibinfo {title} {Perspective:
  Basic understanding of condensed phases of matter via packing models},}\
  }\href {\doibase 10.1063/1.5036657} {\bibfield  {journal} {\bibinfo
  {journal} {J. Chem. Phys.}\ }\textbf {\bibinfo {volume} {149}},\ \bibinfo
  {pages} {020901} (\bibinfo {year} {2018}{\natexlab{b}})}\BibitemShut
  {NoStop}%
\bibitem [{\citenamefont {Wyart}\ \emph {et~al.}(2005)\citenamefont {Wyart},
  \citenamefont {Silbert}, \citenamefont {Nagel},\ and\ \citenamefont
  {Witten}}]{wyart_effects_2005}%
  \BibitemOpen
  \bibfield  {author} {\bibinfo {author} {\bibfnamefont {M.}~\bibnamefont
  {Wyart}}, \bibinfo {author} {\bibfnamefont {L.~E.}\ \bibnamefont {Silbert}},
  \bibinfo {author} {\bibfnamefont {S.~R.}\ \bibnamefont {Nagel}}, \ and\
  \bibinfo {author} {\bibfnamefont {T.~A.}\ \bibnamefont {Witten}},\ }\bibfield
   {title} {\enquote {\bibinfo {title} {Effects of compression on the
  vibrational modes of marginally jammed solids},}\ }\href {\doibase
  10.1103/PhysRevE.72.051306} {\bibfield  {journal} {\bibinfo  {journal} {Phys.
  Rev. E}\ }\textbf {\bibinfo {volume} {72}},\ \bibinfo {pages} {051306}
  (\bibinfo {year} {2005})}\BibitemShut {NoStop}%
\bibitem [{\citenamefont {Donev}, \citenamefont {Torquato},\ and\ \citenamefont
  {Stillinger}(2005)}]{donev_pair_2005}%
  \BibitemOpen
  \bibfield  {author} {\bibinfo {author} {\bibfnamefont {A.}~\bibnamefont
  {Donev}}, \bibinfo {author} {\bibfnamefont {S.}~\bibnamefont {Torquato}}, \
  and\ \bibinfo {author} {\bibfnamefont {F.~H.}\ \bibnamefont {Stillinger}},\
  }\bibfield  {title} {\enquote {\bibinfo {title} {Pair correlation function
  characteristics of nearly jammed disordered and ordered hard-sphere
  packings},}\ }\href {\doibase 10.1103/PhysRevE.71.011105} {\bibfield
  {journal} {\bibinfo  {journal} {Phys. Rev. E}\ }\textbf {\bibinfo {volume}
  {71}},\ \bibinfo {pages} {011105} (\bibinfo {year} {2005})}\BibitemShut
  {NoStop}%
\bibitem [{Note6()}]{Note6}%
  \BibitemOpen
  \bibinfo {note} {For a $N$-particle system in a $d$-dimensional simple cubic
  fundamental cell, the smallest value of $M(K)$ is $d$, and thus $\chi =
  M(K)/[d(N-1)]\geq (N-1)^{-1}$.}\BibitemShut {Stop}%
\bibitem [{\citenamefont {Vanoni}\ \emph {et~al.}(2025)\citenamefont {Vanoni},
  \citenamefont {Kim}, \citenamefont {Steinhardt},\ and\ \citenamefont
  {Torquato}}]{vanoni_dynamical_2025}%
  \BibitemOpen
  \bibfield  {author} {\bibinfo {author} {\bibfnamefont {C.}~\bibnamefont
  {Vanoni}}, \bibinfo {author} {\bibfnamefont {J.}~\bibnamefont {Kim}},
  \bibinfo {author} {\bibfnamefont {P.~J.}\ \bibnamefont {Steinhardt}}, \ and\
  \bibinfo {author} {\bibfnamefont {S.}~\bibnamefont {Torquato}},\ }\href
  {\doibase 10.48550/arXiv.2503.24297} {\enquote {\bibinfo {title} {Dynamical
  properties of particulate composites derived from ultradense stealthy
  hyperuniform sphere packings},}\ } (\bibinfo {year} {2025}),\ \Eprint
  {http://arxiv.org/abs/2503.24297} {2503.24297} \BibitemShut {NoStop}%
\bibitem [{\citenamefont
  {Torquato}(2016{\natexlab{b}})}]{torquato_disordered_2016}%
  \BibitemOpen
  \bibfield  {author} {\bibinfo {author} {\bibfnamefont {S.}~\bibnamefont
  {Torquato}},\ }\bibfield  {title} {\enquote {\bibinfo {title} {Disordered
  hyperuniform heterogeneous materials},}\ }\href {\doibase
  10.1088/0953-8984/28/41/414012} {\bibfield  {journal} {\bibinfo  {journal}
  {J. Phys.: Condens. Matter}\ }\textbf {\bibinfo {volume} {28}},\ \bibinfo
  {pages} {414012} (\bibinfo {year} {2016}{\natexlab{b}})}\BibitemShut
  {NoStop}%
\bibitem [{\citenamefont {Zhang}, \citenamefont {Stillinger},\ and\
  \citenamefont {Torquato}(2016{\natexlab{b}})}]{zhang_perfect_2016}%
  \BibitemOpen
  \bibfield  {author} {\bibinfo {author} {\bibfnamefont {G.}~\bibnamefont
  {Zhang}}, \bibinfo {author} {\bibfnamefont {F.~H.}\ \bibnamefont
  {Stillinger}}, \ and\ \bibinfo {author} {\bibfnamefont {S.}~\bibnamefont
  {Torquato}},\ }\bibfield  {title} {\enquote {\bibinfo {title} {The perfect
  glass paradigm: Disordered hyperuniform glasses down to absolute zero},}\
  }\href {\doibase 10.1038/srep36963} {\bibfield  {journal} {\bibinfo
  {journal} {Sci. Rep.}\ }\textbf {\bibinfo {volume} {6}},\ \bibinfo {pages}
  {36963} (\bibinfo {year} {2016}{\natexlab{b}})}\BibitemShut {NoStop}%
\end{thebibliography}%


\end{document}